\documentclass[preprint,final,3p,times]{elsarticle}
\usepackage[utf8]{inputenc}
\usepackage{graphicx}
\usepackage{amsmath,amsfonts,amssymb}
\usepackage{caption}
\usepackage{subcaption}
\usepackage[stretch=10,shrink=10,tracking=true]{microtype}
\usepackage{siunitx}

\makeatletter
\let\cl@chapter\undefined
\makeatletter
\usepackage{hyperref}
\usepackage{tikz}

\newcommand{\diff}[2]{\frac{\partial #1}{\partial #2}}
\newcommand{\Pe}{\mathit{Pe}}

\newcommand{\rhodesign}{\rho}
\newcommand{\rhoreg}{\Tilde{\rho}}
\newcommand{\rhoshad}{\check{\Tilde{\rho}}_s}
\newcommand{\rhoshadsub}[1]{\check{\Tilde{\rho}}_{#1}}
\newcommand{\rhoaggl}{\check{\Tilde{\rho}}}
\newcommand{\rhophys}{\bar{\check{\Tilde{\rho}}}}
\newcommand{\rhophysel}{\rhophys^e}
\newcommand{\rhoinit}{\rho_{init}}
\newcommand{\ADsysmat}{\mathbf{A}^s}
\newcommand{\milldir}{\theta_\mathit{milling}}
\newcommand{\scale}{s}

\usepackage[mathlines]{lineno} 
\let\oldequation\equation
\let\oldendequation\endequation
\let\oldalign\align
\let\oldendalign\endalign

\renewenvironment{equation}
  {\linenomathNonumbers\oldequation}
  {\oldendequation\endlinenomath}
\renewenvironment{align}
  {\linenomathNonumbers\oldalign}
  {\oldendalign\endlinenomath}

\usepackage[noabbrev]{cleveref}
\renewcommand{\cref}{\Cref}

\usetikzlibrary{positioning}
\usetikzlibrary{shapes,arrows}
\tikzstyle{block} = [rectangle, draw, fill=blue!20, text width=10em, text centered, rounded corners, minimum height=3em]
\tikzstyle{s} = [text width=3em, text centered, minimum height=3em]
\tikzstyle{line} = [draw, -latex']

\title{An Advection-Diffusion based Filter for Machinable Designs in Topology Optimization}
\author{Lukas C. Høghøj\corref{c}}\ead{luch@mek.dtu.dk}
\author{Erik A. Träff}
\cortext[c]{Corresponding author: luch@mek.dtu.dk}
\address{Department of Mechanical Engineering, Section for Solid Mechanics, Technical University of Denmark, Kgs. Lyngby, Denmark}
\date{January 2021}

\begin{document}
\begin{frontmatter}
\begin{abstract}
This paper introduces a simple formulation for topology optimization problems ensuring manufacturability by machining. The method distinguishes itself from existing methods by using the advection-diffusion equation with Robin boundary conditions to perform a filtering of the design variables. The proposed approach is less computationally expensive than the traditional methods used. Furthermore, the approach is easy to implement on unstructured meshes and in a distributed memory setting. Finally, the proposed approach can be performed with few to no continuation steps in any system parameters. Applications are demonstrated with topology optimization on unstructured meshes with up to 64 million elements and up to 29 milling tool directions.

\end{abstract}
\end{frontmatter}
\section{Introduction}
Topology optimization (TO) is a widely adopted method for structural optimization \cite{Bendsoee2003,Sigmund2013}. The main advantage of TO is its ability to generate structures of arbitrary topology regardless of the prescribed initial conditions, i.e. little to no prior knowledge of the optimal structure is required. Due to this property, topology optimization is often used in the initial conceptualization of new designs.
If a manufacturing method, such as milling, is imposed on the designer when conceptualizing a part, the main advantage of topology optimization can become a disadvantage. The freedom for TO to generate arbitrary topologies can result in structures with features, which are impossible to manufacture with traditional machining techniques. This can be characterized as a mismatch between the constraints placed on the designer and the constraints used in the TO formulation.

Extending the topology optimization approach to ensure manufacturability is an ongoing field of research. A wide variety of methods for topology optimization which ensures manufacturability by additive manufacturing have been proposed during recent years \cite{Gaynor2016a,Langelaar2016,Mass2017,Qian2017,Zhang2019}. The additive manufacturing approaches usually constrain the allowed overhang of the structure in order to ensure self-support during the manufacturing process. Other types of manufacturing are also considered in the literature, such as \citet{Wang2020}, which introduces a method for ensuring topologies that can be extruded, by utilizing the so-called PDE-filter \cite{Lazarov2011} with high anisotropy. 

Several approaches for manufacturable TO using milling have been suggested. \citet{Gersborg2011} proposes a method to consider an explicitly castable or millable design by using one design variable for each row or column of a structured grid. This approach is contrasted by \citet{Guest2012}, which proposes a similar method, but retains all design variables and uses cumulative summation along rows or columns as a filter to ensure that a design can be cast or milled. The approach by cumulative summation is extended to arbitrary directions by \citet{Langelaar2019}, which maps the densities to a structured grid aligned with the milling direction in order to perform the summation. Finally, \citet{Hur2020} recently proposed a milling filter, which uses a modified variant of the advection-diffusion equation to perform the cumulative summation for the level-set formulation, not quite unlike the proposed method. In \citet{Hur2020} Dirichlet boundary conditions are used for the flow problem, which introduce the need for domain padding in order to correctly capture non-zero physical densities on the boundary of the design domain.

The approaches which use a cumulative sum provide a high degree of control allowing tool shapes to be taken into account \cite{Langelaar2019}, but the cumulative summation is also computationally very expensive to evaluate. If the sum operation is precomputed, the resulting matrix will contain many non-zero values and result in very high memory usage, which hinders application on large-scale systems. Furthermore the assembly of such a matrix is a non-trivial task, especially on distributed memory systems. This paper presents an approach to perform the milling tool emulation, by solving the advection-diffusion equation with Robin boundary conditions. The proposed approach is conceptually simple and scales well, facilitating the solution of high resolution 3D problems. Like the cumulative summation approach, this method can be employed without any continuation scheme, and with little to no tuning for problem specific parameters. For now, however, the approach is limited to milling considerations without explicit tool shapes.

This paper is organized as follows; \Cref{sec:formulation} introduces the formulation necessary for the advection-diffusion filtering step; \Cref{sec:optimization} introduces the optimization problem used for numerical examples; \Cref{sec:efficiency} discusses the computational efficiency of the presented methodology; \Cref{sec:results} presents two and three dimensional machinable results, with machining in one or multiple directions and \Cref{sec:conclusion} provides a discussion of the presented methodology and a conclusion.

\section{Formulation}
\label{sec:formulation}
\begin{figure}[htb]
    \centering
    \def\svgwidth{.6\textwidth}
\begingroup%
  \makeatletter%
  \providecommand\color[2][]{%
    \errmessage{(Inkscape) Color is used for the text in Inkscape, but the package 'color.sty' is not loaded}%
    \renewcommand\color[2][]{}%
  }%
  \providecommand\transparent[1]{%
    \errmessage{(Inkscape) Transparency is used (non-zero) for the text in Inkscape, but the package 'transparent.sty' is not loaded}%
    \renewcommand\transparent[1]{}%
  }%
  \providecommand\rotatebox[2]{#2}%
  \newcommand*\fsize{\dimexpr\f@size pt\relax}%
  \newcommand*\lineheight[1]{\fontsize{\fsize}{#1\fsize}\selectfont}%
  \ifx\svgwidth\undefined%
    \setlength{\unitlength}{375.53869061bp}%
    \ifx\svgscale\undefined%
      \relax%
    \else%
      \setlength{\unitlength}{\unitlength * \real{\svgscale}}%
    \fi%
  \else%
    \setlength{\unitlength}{\svgwidth}%
  \fi%
  \global\let\svgwidth\undefined%
  \global\let\svgscale\undefined%
  \makeatother%
  \begin{picture}(1,1.10287772)%
    \lineheight{1}%
    \setlength\tabcolsep{0pt}%
    \put(-0.20656182,1.23629163){\makebox(0,0)[lt]{\lineheight{1.75}\smash{\begin{tabular}[t]{l}\end{tabular}}}}%
    \put(0,0){\includegraphics[width=\unitlength,page=1]{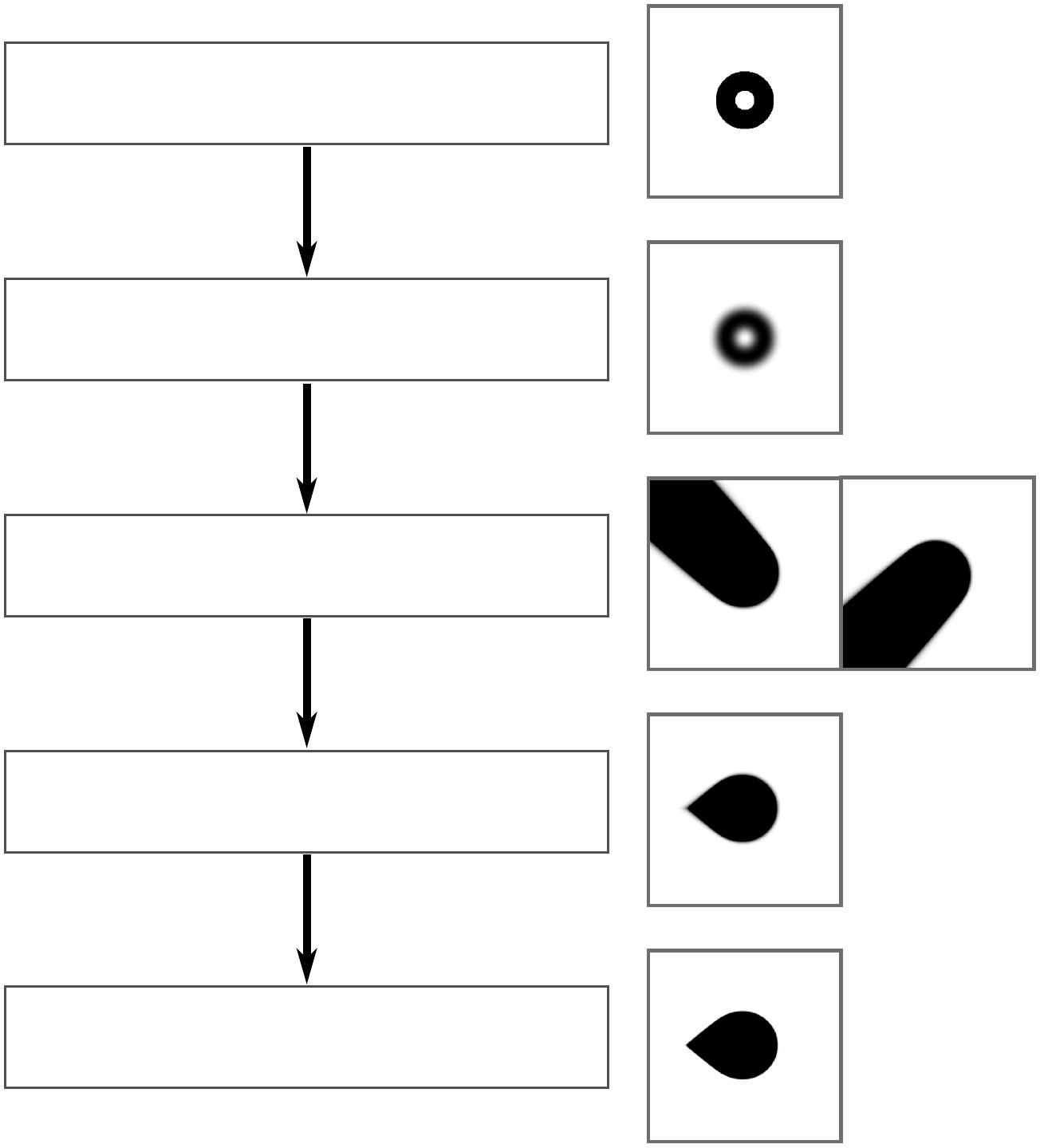}}%
    \put(0.01897518,1.00137201){\makebox(0,0)[lt]{\lineheight{1.75}\smash{\begin{tabular}[t]{l}0: Design variable $\rhodesign$\end{tabular}}}}%
    \put(0.01710287,0.77446057){\makebox(0,0)[lt]{\lineheight{1.75}\smash{\begin{tabular}[t]{l}1: Density filter $\rhoreg$\end{tabular}}}}%
    \put(0.01866312,0.54752644){\makebox(0,0)[lt]{\lineheight{1.75}\smash{\begin{tabular}[t]{l}2: Shadowing steps $\rhoshadsub{1}$, $\rhoshadsub{2}$, .. \end{tabular}}}}%
    \put(0.0185383,0.32060279){\makebox(0,0)[lt]{\lineheight{1.75}\smash{\begin{tabular}[t]{l}3: Agglomorating fields $\rhoaggl$\end{tabular}}}}%
    \put(0.0197033,0.09368633){\makebox(0,0)[lt]{\lineheight{1.75}\smash{\begin{tabular}[t]{l}4: Heaviside projection $\rhophys$\end{tabular}}}}%
  \end{picture}%
\endgroup%
    \caption{Flowchart and visualization of the composition of filters to realize the milling filter. Two tool directions are used for the milling filter, resulting in two distinct shadowed fields.}
    \label{fig:flowchart}
\end{figure}
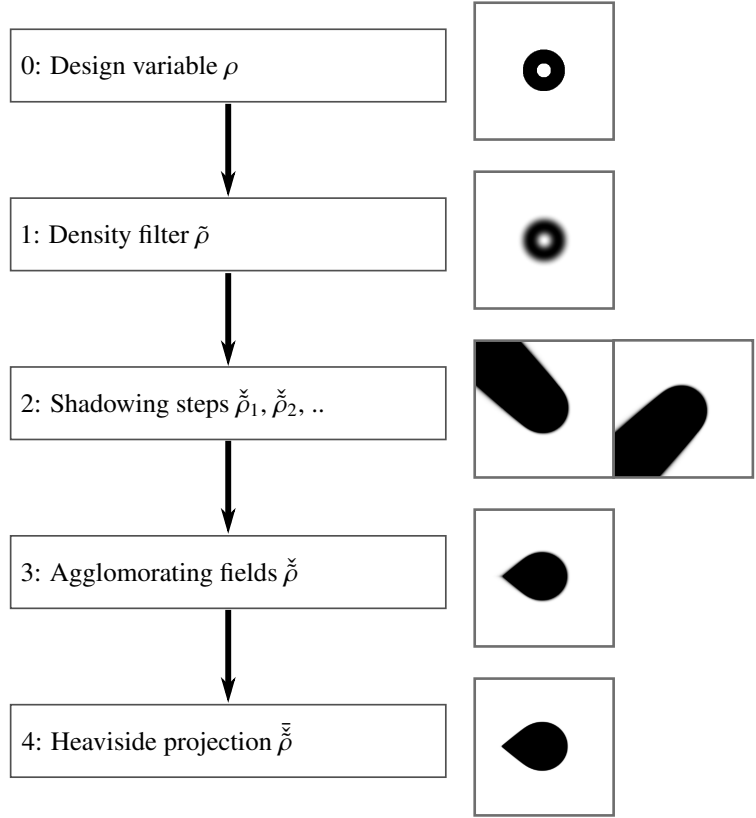

The milling filter is based on the approach presented by \citet{Langelaar2019}, with the notable difference that the cumulative sums have been replaced by solving the advection-diffusion equation, with a dominating advective term. The filter relates the design variables to a 'physical' density field, which guarantees that the resulting void regions are reachable by a tool direction from outside the design domain. This relation is performed through a series of filters, of which most are already common in many topology optimization approaches. A brief overview of the steps used to compose the complete milling filter is presented in \cref{fig:flowchart}. The initial design variable is filtered in step 1, yielding the regularized design field, $\rhoreg$, as discussed in \cref{sec:reg}. In step 2, the shadowing operations are performed, resulting in the shadowed intermediate design fields, $\rhoshad$, as discussed in \cref{sec:shadow}. Note that a shadowing step is performed for each prescribed tool direction. The shadowed intermediate design fields are then agglomerated in step 3, resulting in the intermediate agglomerated variable $\rhoaggl$, as discussed in \cref{sec:aggl}. Finally, in step 4, the agglomerated variable is projected, as discussed in \cref{sec:proj}. The sensitivity analysis of the filter is included in \cref{sec:sens} for completeness.

\subsection{Step 1: Smoothing of design field}
\label{sec:reg}
A density filter is initially applied to the design field. This can be performed either by a classical convolution approach \cite{bendsoe}, or by employing the so-called PDE-filter which solves a modified Poisson equation to perform density filtering \cite{Lazarov2011,Lazarov2016}. The reader is referred to the cited articles for details on the respective filters and their sensitivity analysis.

The density filter is introduced to regularize the design field. This is especially useful when evaluating a black and white field, where the gradients will be zero in most of the void region (due to the SIMP interpolation) and in parts of the solid region (due to the agglomeration of shadowing steps). By including the density filter, non-zero gradient values will be smoothed into the domain with zero sensitivities.

In this work, the convolution approach is used for the 2D examples in \cref{sec:2Dexamples}, due to the widespread use of this approach for 2D codes, such as \cite{top88,Ferrari2020}. The PDE-filter approach \cite{Lazarov2011,Lazarov2016} is used in the 3D examples, due to the increased performance and reduced memory footprint compared to the convolution approach.

\subsection{Step 2: Shadowing by advection-diffusion}
\label{sec:shadow}
The advection-diffusion equation describes the transport of a scalar field, it is known from e.g. heat transfer problems. The steady state advection-diffusion equation is given as:
\begin{equation}
\label{eq:ad}
    \Pe \; u_i \; \diff{T}{x_i}-\diff{^2T}{x_i^2}=q
\end{equation}
where $T$ is the transported field, $u_i$ the prescribed advective field, $q$ the normalized source term and $\Pe$ the Peclet number which is the ratio between the advective and diffusive transport.

In the presented shadowing methodology, a regularized design field $\rhoreg$ is used as a source term, and the obtained transported field $\rhoshad$ is shadowed in the direction of the constant advection term $u_i^s$ of unity norm $||u_i^s||=1$, corresponding to the machining direction, where $s$ refers to the shadowing angle index, since multiple shadowing angles may be applied. For numerical reasons, the equation is normalized by the Peclet number. Thus, the equation solved using the filter notation becomes
\begin{equation}
\label{eq:ad_concrete}
    u_i^s \; \diff{\rhoshad}{x_i}-\frac{1}{\Pe}\diff{^2\rhoshad}{x_i^2}=\scale\rhoreg \quad \text{in} \;\; \Omega
\end{equation}
where $\scale$ is a factor scaling the source term on an element basis. The factor $\scale$ is chosen such that the solution can go from $0$ (void) to $1$ (solid) over a single element and is further discussed in \cref{sec:peclet}. The scaling is applied on the source rather than on the solution, as unstructured meshes are considered. 

Solutions to \cref{eq:ad} are known to become numerically unstable in cases with a high Peclet number. However, while paying a price in solution accuracy, the Finite Volume formulation using an upwind difference scheme on the advection term is known to be numerically stable for high Peclet numbers. The derivations of the used finite volume schemes are presented in \ref{app:FVMimplementation}.

\subsubsection{Boundary conditions}
\begin{figure}[htb]
    \centering
    \includegraphics[width=0.5\linewidth]{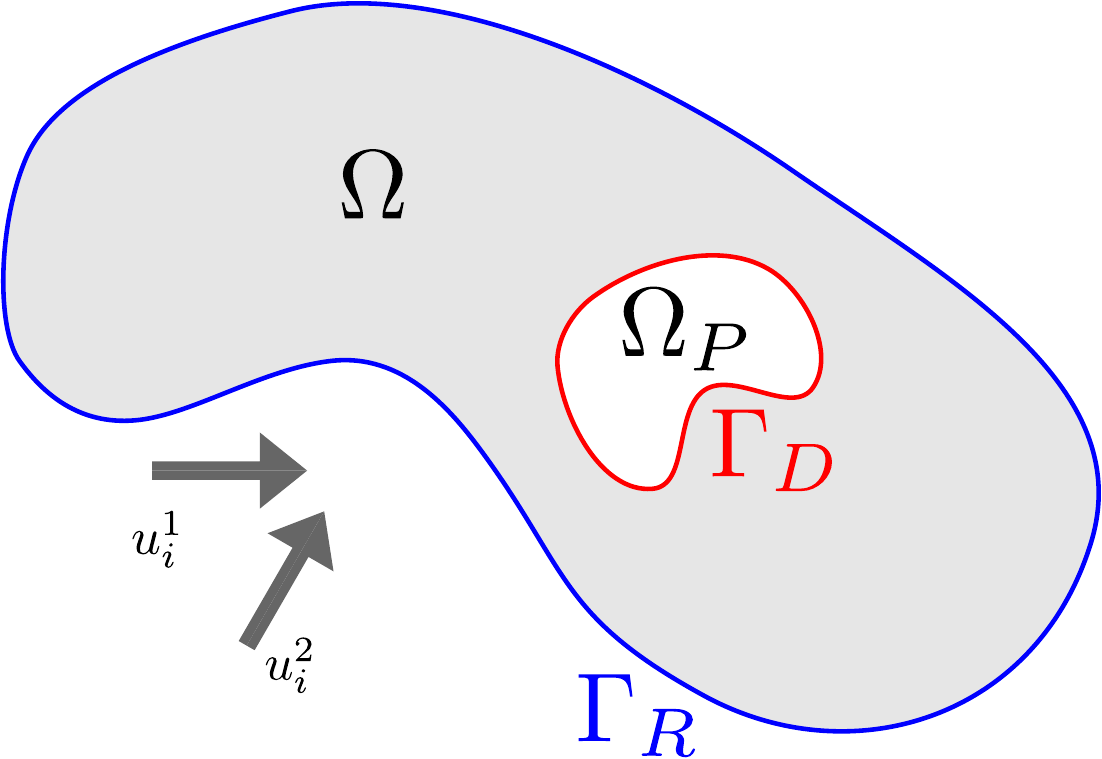}
    \caption{Overview of the employed boundary conditions on the optimization domain $\Omega$ with shadowing directions $\theta_1$ and $\theta_2$. The passive solid domain $\Omega_P$ has Dirichlet-type boundary conditions along its boundary $\Gamma_D$ to the optimization domain. Free boundaries of the optimization domain, $\Gamma_R$, have Robin-type boundary conditions imposed. }
    \label{fig:kartoffel}
\end{figure}

The advection-diffusion equation is only solved in the optimization domain. Hence, boundary conditions are imposed on free boundaries, as well as on the boundaries between the optimization- and passive solid domains - as illustrated in \cref{fig:kartoffel}.

The employed boundary conditions on the free domain boundaries, $\Gamma_R$, are of Robin type on boundary surfaces:
\begin{equation}
\label{eq:robinBC}
    \rhoshad+\mathbf{n}\frac{1}{\scale\Pe}\nabla\rhoshad=0\quad \text{on} \;\; \Gamma_R
\end{equation}
where $\mathbf{n}$ is the outward pointing surface normal at the boundary.

On the domain boundaries adjacent to solid passive domains, $\Gamma_D$, Dirichlet boundary conditions are introduced to obtain the shadow of the passive domain in the corresponding milling direction. This ensures manufacturability of the full design, including the passive domains.
\begin{equation}
    \rhoshad=1\quad \text{on} \;\; \Gamma_D
\end{equation}
The implementation presented in the present work is based on unstructured meshes, hence passive void elements are not required. If one wants to use the presented methodology with the use of passive void elements, the authors suggest to ensure that the embedded design domain is machinable with the used tool directions.

Further details on the boundary conditions and their implementations are given in \ref{app:FVMimplementation}.

\subsubsection{Choice of Peclet number and scaling factor}
\label{sec:peclet}
The $\scale$ parameter, that is used for scaling the source term in the advection-diffusion equation, \cref{eq:ad_concrete}, is set such that a $\rhoreg=1$ value permits the shadowed variable to go from 0 to 1 over one cell face (normal to the shadowing direction). Assuming a very large Peclet number, and $|u_i|=1$, this parameter should be set as:
\begin{equation}
    \scale=\frac{1}{h_e}
\end{equation}
where $h_e$ is the average element side length. 

\begin{figure}[htb]
    \centering
    \begin{subfigure}[t]{0.23\linewidth}
        \includegraphics[trim=40 25 15 15,clip,width=\linewidth]{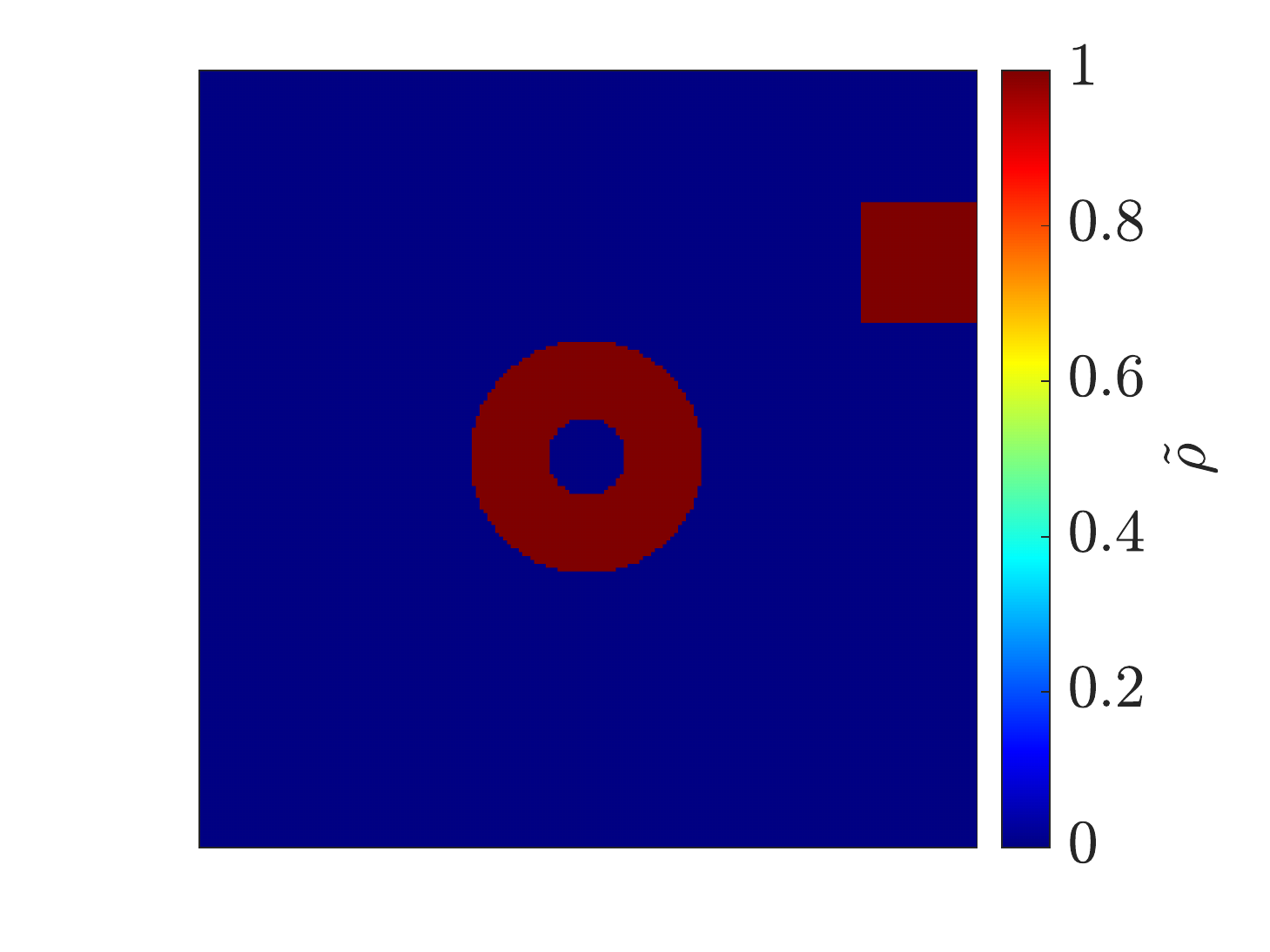}
        \caption{$\rhoreg$}
        \label{fig:peinf1}
    \end{subfigure}
    \hfill
    \begin{subfigure}[t]{0.23\linewidth}
        \includegraphics[trim=40 25 15 15,clip,width=\linewidth]{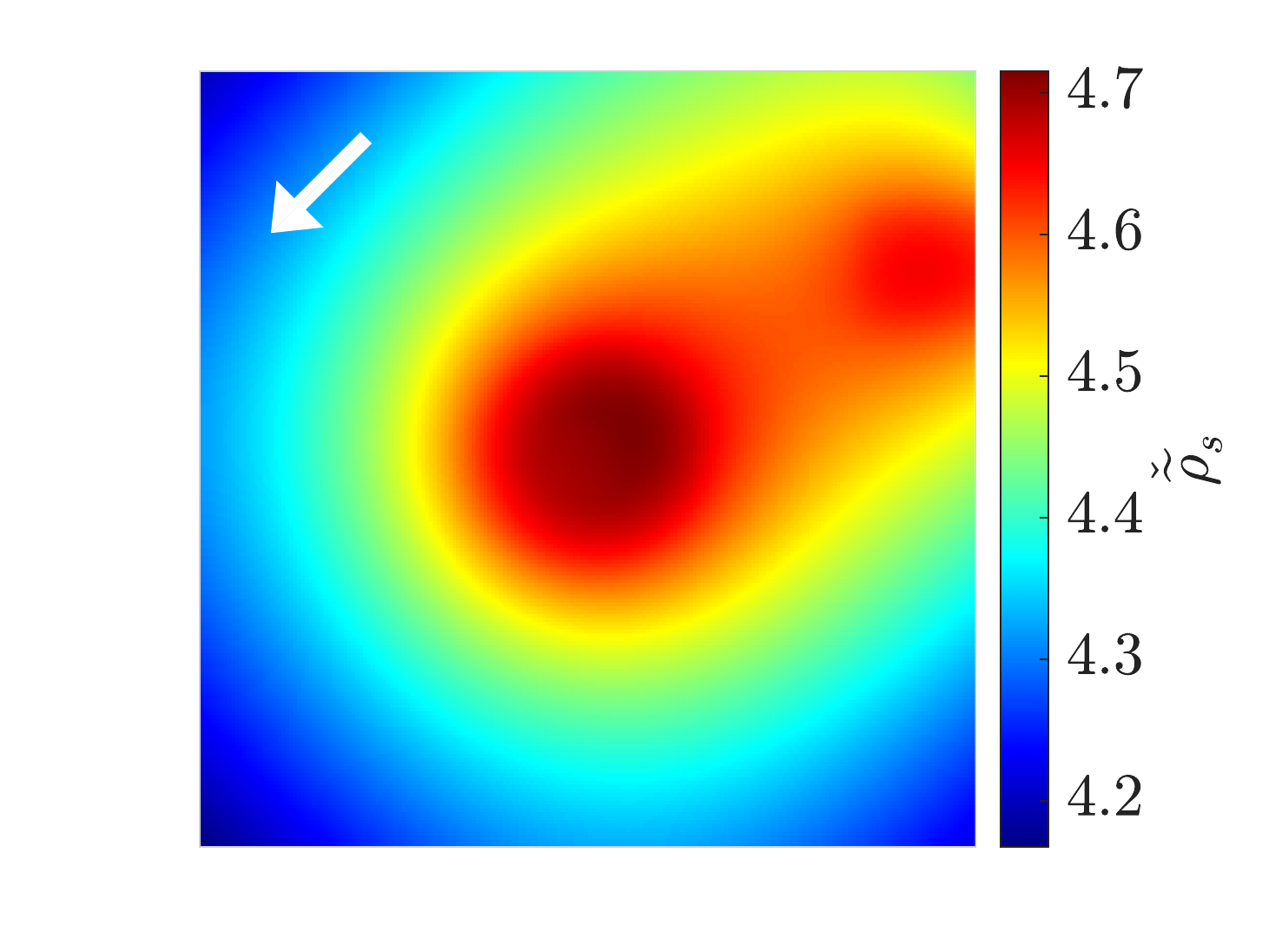}
        \caption{$\Pe=0.1$}
        \label{fig:peinf2}
    \end{subfigure}
    \hfill
    \begin{subfigure}[t]{0.23\linewidth}
        \includegraphics[trim=40 25 15 15,clip,width=\linewidth]{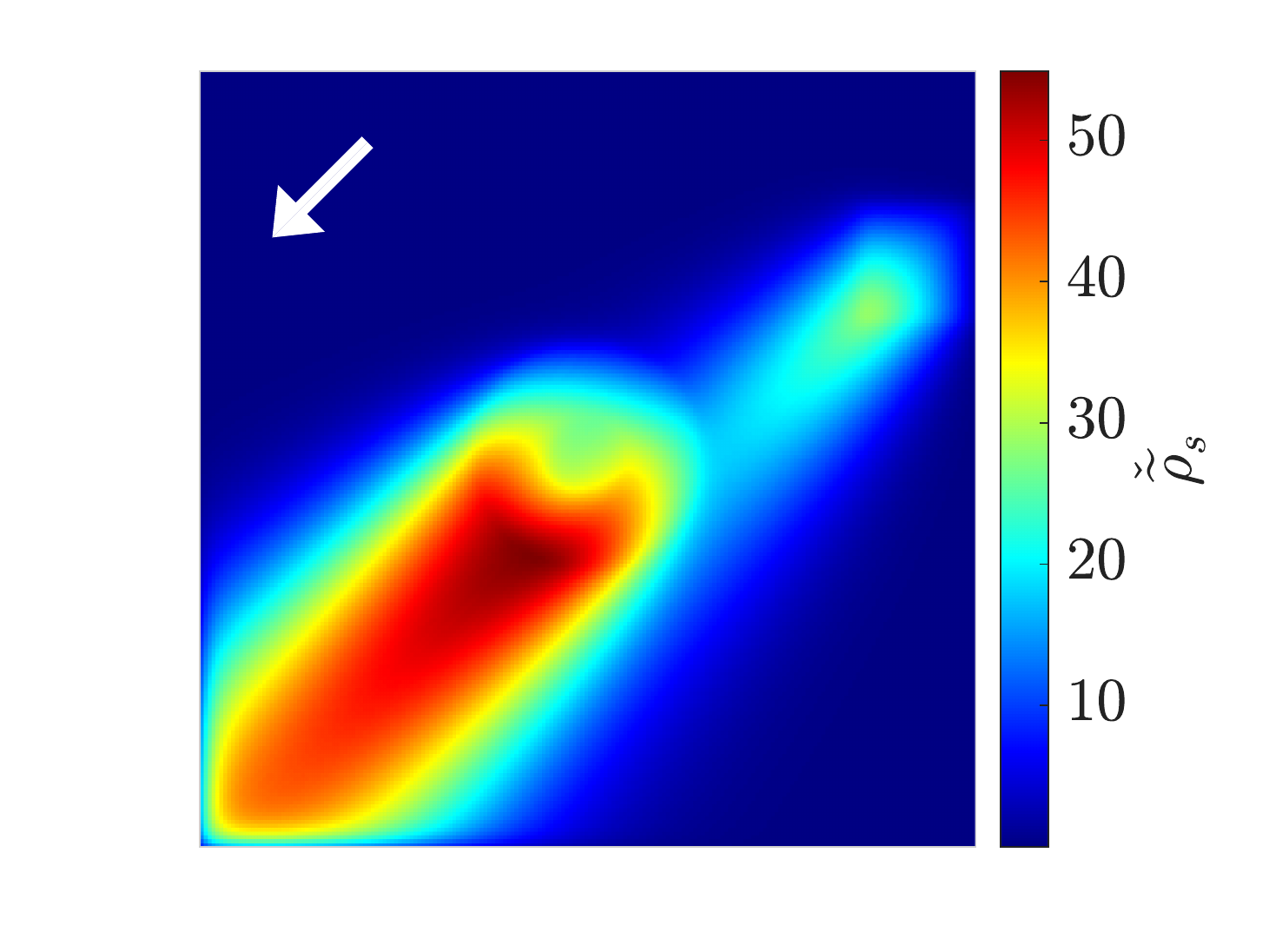}
        \caption{$\Pe=100$}
        \label{fig:peinf3}
    \end{subfigure}
    \hfill
    \begin{subfigure}[t]{0.23\linewidth}
        \includegraphics[trim=40 25 15 15,clip,width=\linewidth]{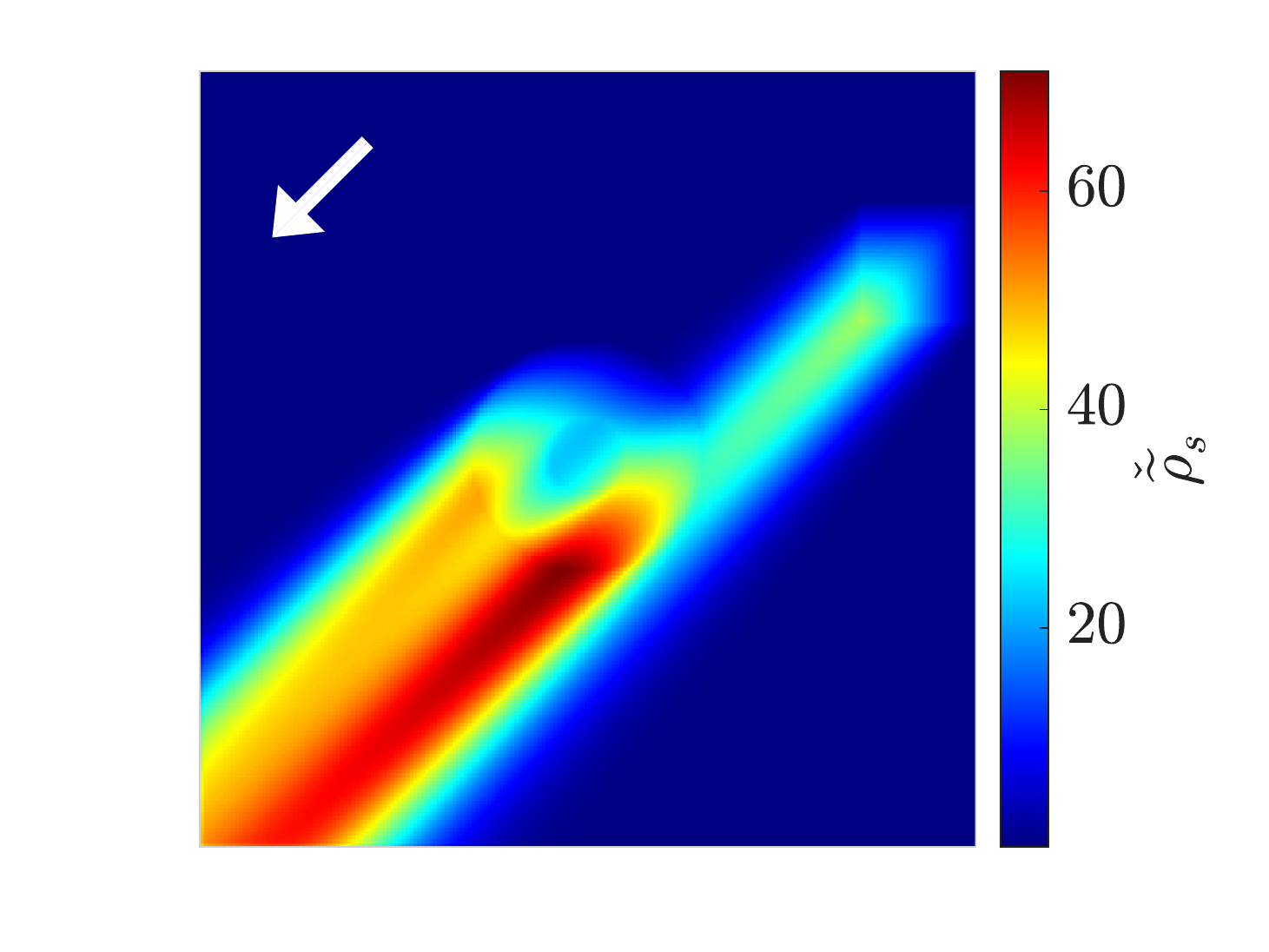}
        \caption{$\Pe=10^4$}
        \label{fig:peinf4}
    \end{subfigure}
    \caption{Shadowing step performed on the field (a) with different Peclet numbers. Note the different color-bars for the respective cases.}
    \label{fig:peinfluence}
\end{figure}

The choice of Peclet number is critical to the effect of the advection-diffusion filtering step on the obtained results. As straight walls are machinable, the conic (diffusive) effects of the filter are to be minimized. The desired effect of the shadowing is hence to obtain features parallel to the specified direction, $u_i$, with as little diffusion as possible. This can be achieved by using a high Peclet number. When setting $\Pe\gg1$, the problem becomes advection dominated. This is also seen in \cref{fig:peinfluence}, where the field from \cref{fig:peinf1} is shadowed in a diagonal direction at three different Peclet numbers. With $\Pe=0.1$, as seen in \cref{fig:peinf2}, the diffusive effect is clearly dominating. When raising the Peclet number to $\Pe=100$, as demonstrated in \cref{fig:peinf3}, the regime is advection dominated, however diffusion still has a significant effect. In the case with $\Pe=10^4$, seen in \cref{fig:peinf4}, information is seen to almost exclusively propagate downstream. However, inside the shadowed region, some diffusive effects are still observed. As the values inside these regions are $\rhoshad>>1$, this is not an issue as they are thresholded using the smooth Heaviside projection, introduced in \cref{sec:proj}.

It should be noted that the Peclet number is defined based on a characteristic length scale of the considered domain. This implies that when larger domains are considered, a lower Peclet number will suffice to achieve similar results. For reference, \cref{fig:peinfluence} is computed on a domain of unit side-lengths. A rule of thumb to choose the Peclet number based on a characteristic domain length $l_c$ is
\begin{equation}
    \Pe_\text{thumb} = \frac{10^4}{l_c}.
\end{equation}

The constant $10^4$ is by no means a fixed rule, as acceptable results have also been found using constant factors of $10^3$, $10^5$, and $10^6$. We do however advise ensuring that the used Peclet number satisfies $\Pe \; l_c > 10^3$.

\subsection{Step 3: Field agglomeration}
\label{sec:aggl}
The resulting fields of the shadowing for each tool direction $\rhoshad$ are agglomerated to a single field using the $p$-mean, which provides a differentiable approximation to the $\min$ operator.
\begin{equation}
\label{eq:agglomoration}
    \rhoaggl^e = \left(\frac{1}{n_s}\sum_{s=1}^{n_s}(\rhoshad^e)^p\right)^\frac{1}{p} \approx \min_s \rhoshad^e
\end{equation}
Here $p < 0$ denotes an agglomeration parameter. The $p$-mean becomes a more accurate approximation of the $\min$ operator, but also more non-linear, as $p \rightarrow -\infty$. In practice $p=-4$ is commonly used throughout this article, as it was found that the field projection step reduces the effect of a low accuracy approximation of the min operator. When a large number of milling directions are used it becomes necessary to use a lower value of $p$, in order to ensure a sufficiently good approximation of the $\min$ operator. Other field agglomeration approaches are possible as discussed in detail in \cite{Kennedy2015}. For instance, \citet{Langelaar2019} suggested using the KS functional for the agglomeration. For large problems, however, where $\rhoshad$ can take very large values, due to the high number of elements along one axis, this resulted in numerical issues with the KS function, and the $p$-mean was found to be more robust in these cases.

\subsection{Step 4: Field projection}
\label{sec:proj}
The aggregated fields are projected using the smoothened Heaviside filter \cite{Wang2011} in order to bound the final density field between 0 and 1:
\begin{equation}
\label{eq:heavi}
    \rhophys^e =\frac{\tanh(\beta\eta)+\tanh(\beta(\rhoaggl^e-\eta))}{\tanh(\beta\eta)+\tanh(\beta(1-\eta))}
\end{equation}
where $\beta$ is the projection sharpness and $\eta$ the projection threshold. A desirable side effect is that a black-and-white design is also obtained. Throughout this article the value $\eta=0.5$ is used, while $\beta$ is chosen to either \num{8} or \num{10} depending on the required projection sharpness.

\section{Optimization formulation}
\label{sec:optimization}
The general optimization problem is formulated on the design variables $\rhodesign$. $\rhophys$ is used to denote the 'physical' density field, and is obtained by performing all milling filter steps described in \cref{sec:formulation} on the design field $\rhodesign$. The posed optimization problem is minimization of the compliance with a global volume constraint:
\begin{equation}
    \label{eq:optimization}
    \begin{split}
        \min_{\rhodesign\in\mathbb{R}^N}\quad& c(\rhodesign)=\mathbf{u}^\top\mathbf{K}(\rhophys)\mathbf{u}\\
        \mathrm{s.t.} \quad & g(\rhodesign)=\frac{V(\rhophys)}{V^*}-1\leq0\\
        & \mathbf{K}(\rhophys)\mathbf{u}-\mathbf{p}=0\\
        & 0 \leq \rhodesign_i \leq 1 \quad i = 1\dots N
    \end{split}
\end{equation}
where $\mathbf{K}(\rhophys)$ is the stiffness matrix, $\mathbf{p}$ the load vector and $\mathbf{u}$ the resulting displacement vector. The volume constraint is enforced based on the volume of the current design, $V(\rhophys)$, and the maximum allowable volume, $V^*$.

The Solid Isotropic Material Penalization (SIMP) interpolation scheme is used to map the element projected design variable $\rhophysel$ to the corresponding Young's modulus \cite{bendsoe}:
\begin{equation}
    E(\rhophysel) = E_\mathit{min}+(\rhophysel)^p \left( E_\mathit{max} - E_\mathit{min} \right)
\end{equation}
Where $E_\mathit{min}$ and $E_\mathit{max}$ are the lower and upper values of the Young's modulus, respectively and $p$ is the SIMP penalization parameter.

\subsection{Parameters for the Method of Moving Asymptotes}
The optimization problem is solved using the Method of Moving Asymptotes \cite{Svanberg1987} implemented by \cite{Aage2013} with non standard parameters. These parameters are necessary as the milling filter projection does not work for low projection sharpnesses, as this results in $\rhophys$ not being bounded correctly. This requires the usage of a constant and high $\beta$ value. When optimizing with a constant projection sharpness, the problem becomes highly sensitive and the asymptotes in the MMA algorithm hence need to be tightened, as discussed by \citet{Guest2011}.

Therefore, to eliminate oscillatory behavior the initial asymptotes are tightened by setting the initial asympote spacing, \texttt{asyinit}, 
\begin{equation}
    s_0=\frac{0.5}{2\beta+1}.
\end{equation}

Furthermore the parameter to widen the asymptotes, \texttt{asyincr}, is set to 1.05 and the parameter to tighten them, \texttt{asydecr}, to $0.65$ instead of 1.2 and 0.7, respectively, in the standard distribution of MMA. The outer mover limit was set to $0.1$ per design iteration.

A scaling parameter is applied on the objective function, such that its value is $10$ in the first optimization iteration. When the scaled objective function gets below $0.1$ the scaling parameter is updated by a factor of $10$. Likewise the volume constraint is used in a scaled formulation, as shown in \cref{eq:optimization}. This ensures a good numerical conditioning on the optimization problem.

\subsection{Sensitivity analysis}
\label{sec:sens}
The sensitivities of the objective- and constraint functions are obtained with respect to the final projected element variable, $\rhophysel$. The sensitivities of a function $f$ with respect to the projected variable, $\diff{f}{\rhophys}$ are projected back to the design variable, $\rhodesign$, by use of the chainrule:
\begin{equation}
    \label{eq:chainrule}
    \frac{df}{d\rhodesign} = \frac{df}{d\rhophys} \diff{\rhophys}{\rhoaggl} \sum_{s=1}^{n_{s}} \left[ \diff{\rhoaggl}{\rhoshad} \diff{\rhoshad}{\rhoreg} \right] \diff{\rhoreg}{\rhodesign}.
\end{equation}
where the therm $\diff{\rhoshad}{\rhoreg}$, represents the chainrule term of the shadowing step. To correct the filter for the advection-diffusion equations the discretized milling filtering step is considered in tensor notation:
\begin{equation}
    A^s_{i,j}\rhoshadsub{s,j}=\rhoreg_i
\end{equation}
differentiating the expression with respect to the regularized field $\rhoreg_i$ and multiplying with the sensitivities $\diff{f}{\rhoshadsub{s,j}}$ with respect to the shadow yields
\begin{equation}
    A^s_{i,j}\left.\diff{f}{\rhoreg_i}\right|_s=\diff{f}{\rhoshadsub{s,j}}
\end{equation}
which in matrix notation yields to solving the transposed filtering equation:
\begin{equation}
{\ADsysmat}^\intercal\left.\diff{f}{\rhoreg}\right|_s=\diff{f}{\rhoshad}
\end{equation}
where the partial derivatives $\left.\diff{f}{\rhoreg}\right|_s$ need to be summed for all shadowing steps to obtain the full sensitivity with respect to the regularized field, as seen in \cref{eq:chainrule}. More details on the sensitivity analysis and chain-rule terms are outlined in \ref{app:sensitivity}.

\section{Computational Efficiency}
\label{sec:efficiency}
The computational cost of introducing the proposed milling filter depends directly on both the desired number of milling directions and the number of constraints used in the optimization formulation. For every milling direction an advection-diffusion equation needs to be solved at every design iteration. Likewise, an adjoint system of equations needs to be solved for every tool direction for the objective function and for every constraint every design iteration. When also accounting for the used density filter this results in a total of $(n_\text{constraints} + 2)(n_\text{tools} + 1)$ linear system solves for filtering, where $n_\text{constraints}$ denotes the number of constraints and $n_\text{tools}$ denotes the number of tool directions. 

The worst case presented in this article uses one constraint and \num{29} tool directions, resulting in \num{90} auxiliary linear systems solutions every design iteration. While many auxiliary systems might need to be solved, the time required to solve the finite volume problems is significantly lower than the time required to solve the linear elasticity state equation, since they are scalar problems and are hence much cheaper than the vectorial state problem. Furthermore, the system matrix of the advection-diffusion problem does not change between design iterations, amortizing the computational cost of constructing preconditioners, as these can be stored throughout the optimization process. The cost of solving the advection-diffusion equation depends on a large set of parameters, where some of the most significant are; the used discretization of the equations, the number of elements in the mesh, and the used method to solve the resulting system of equations. 

The used method for solving the resulting system of equations is also of great importance for the efficiency. If the system is sufficiently small, e.g. less than one million elements, a direct solution technique is the most viable strategy. As the operator does not change with the design iterations, only one LU factorization is required for every tool direction for the entirety of the optimization problem. This is the approach used in the 2D examples from \cref{sec:2Dexamples}. 

If the number of elements grows too large, iterative solvers are required. For the presented 3D examples the Flexible Generalized Minimal RESidual \cite{Saad1993} method was used, with additive Schwartz preconditioning, which in turn uses incomplete LU to approximate the local solutions. All of these solvers are implemented by the Portable, Extensible Toolkit for Scientific Computation \cite{Balay2019}. More advanced preconditioners, such as the multigrid method employed in solving the linear elastic state equation, can also be implemented in order to improve the computational efficiency of solving the advection-diffusion equation. This was not deemed necessary during our numerical examples, as the solution time of the auxiliary problems never exceeded reasonable limits.

The discretization technique for the advection-diffusion problem is a finite volume scheme with an upwind difference scheme, which is known to be numerically stable for large Peclet numbers. Unfortunately, this means that the adjoint problem corresponds to a downwind difference scheme, which is not so stable. It is observed that solving the adjoint problem requires up to \num{6} times the iterations before the convergence criteria is reached, compared to the forward problem for the advection-diffusion equation. This could be mitigated by explicitly constructing the adjoint operator using a upwind scheme, although this might introduce a small error due to the difference in the discretization schemes. 

Using a finite volume scheme for solving the advection-diffusion PDE means that the design field representation corresponds to one degree of freedom per design element. Hence no mapping is required between elements and nodes.

\section{Numerical examples}
\label{sec:results}
The introduced methodology is first demonstrated on a two dimensional cantilever beam example, which was also used by \cite{Langelaar2019}. The same cantilever beam is then optimized in three dimensions, showcasing how the methodology works on large scale. Finally the GE engine bracket \cite{bracketarticle} is optimized to illustrate how the methodology can be used in an industrial example. The three dimensional examples are implemented in a preexisting in-house unstructured topology optimization code \cite{Traff2020}. The optimization parameters for all three cases are given in \cref{tab:optparam}. Note that the homogeneous initial value of the design vector is also stated as an optimization parameter, as the choice of this parameter is important as the resulting initial design should neither be pure solid nor pure void. This is due to the sensitivities vanishing in SIMP at pure void and in the Heaviside projection if the agglomerated field ins $\rhoaggl>>1$.

\begin{table*}[htb]
    \centering
    \caption{Summary of used optimization parameters}
    \label{tab:optparam}
    \begin{tabular}{lrrr}
    \hline
         & 2D cantilever & 3D cantilever & GE Bracket \\\hline
        Projection sharpness $\beta$ & $8$ & $8$ & $10$\\
        Projection threshold $\eta$ & $0.5$ & $0.5$ & $0.5$\\
        SIMP penalization  $p$ & $3$, $5$ & $3$ & $3$ \\
        Filter radius $r_\mathit{min}$ & $0.03$ & $0.0085$ & \SI{0.7}{mm} \\
        Peclet number $\Pe$ & $10^4$& $10^4$ & $500$ \\
        Heat source factor  & $1$ & $100$ & $1$ \\
        $p$-mean penalization & $-3$ & $1$,$-4$,$-6$& $-4$ \\
        Young's module $E_\mathit{max}$ & $1$ & $1$  & \SI{113.8}{GPa}\\
        Young's module ratio $E_\mathit{min}/E_\mathit{max}$ & $10^{-9}$ & $10^{-4}\rightarrow10^{-7}$  & $10^{-3} \rightarrow 10^{-6}$\\
        Poisson's ratio $\nu$ & \num{0.3} & \num{0.3}  & \num{0.342}\\
        Initial design variable $\rhoinit$ & $0.005$, $0.02$ & $0.002$ & $0.004$ \\
        Volume fraction $V^*$ & $0.5$ & $0.15$ & $0.137135$
    \end{tabular}
\end{table*}

\subsection{Two dimensional cantilever examples}
\label{sec:2Dexamples}

\begin{figure}[htb]
    \centering
    \includegraphics[width=0.5\linewidth]{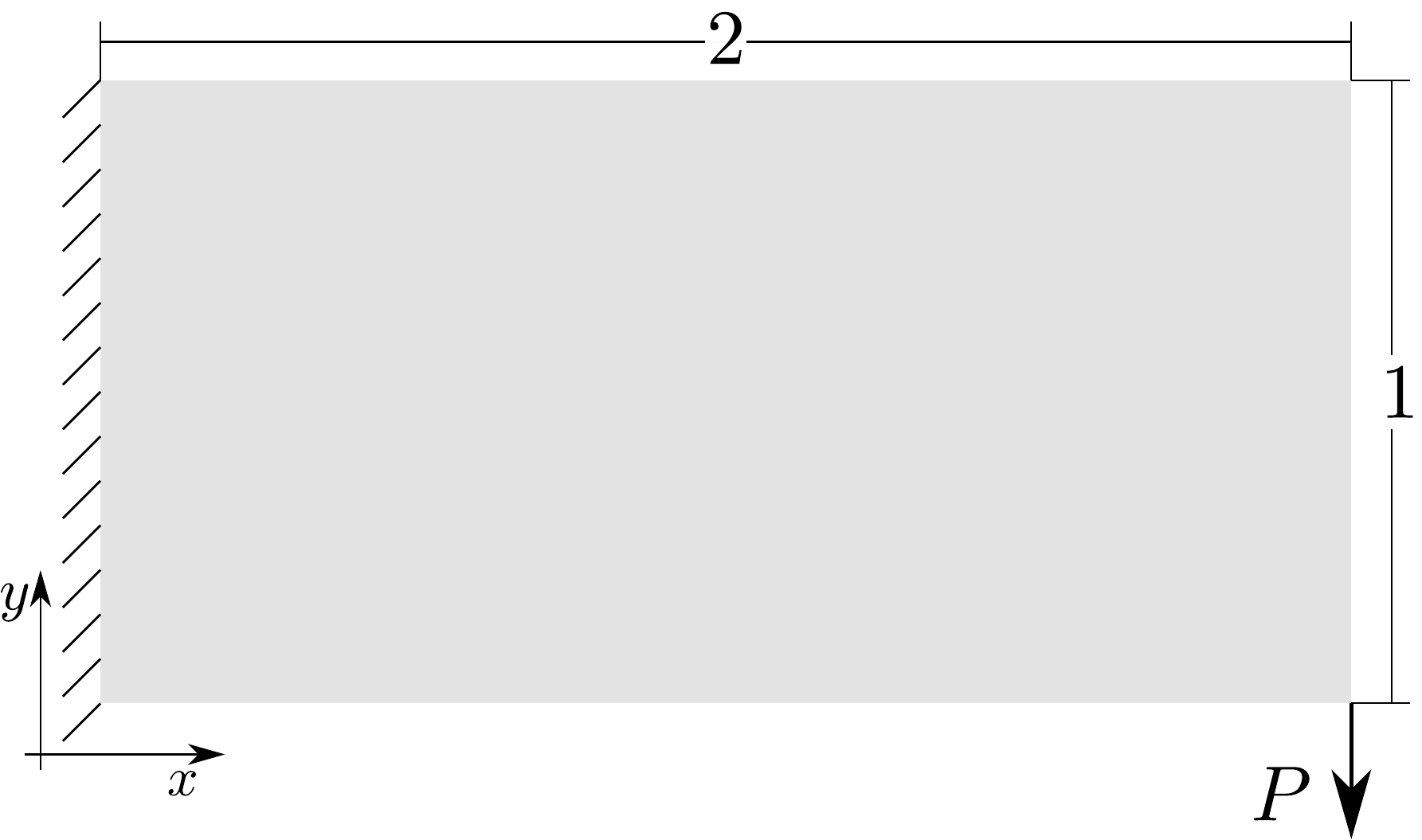}
    \caption{Problem setup of the 2D cantilever beam example. Material can be placed inside the design domain of dimensions $2\times 1$, depicted in gray. }
    \label{fig:2dproblem}
\end{figure}

In the following, the optimization results on a two dimensional cantilever beam are discussed. The design domain of dimensions $2\times 1$ is clamped on the left hand side and point loaded in the lower right corner, as shown in \cref{fig:2dproblem}.

The domain is discretized by $200\times100$ rectangular elements. In addition to the parameters seen in \cref{tab:optparam}, the milling directions are selected to a variety of combinations. The SIMP penalization is set to $p=5$ for the cases with one or two milling directions and to $p=3$ for other cases. The increased SIMP penalization is used in the cases with few tool directions to avoid designs with intermediate densities since these designs have a tendency to perform relatively well in terms of compliance, and therefore need to be penalized further. Furthermore, the initial design variable value is set to $\rhoinit=0.005$ for cases with a single milling direction and $\rhoinit=0.02$ if multiple directions are considered.

\begin{figure*}[htb]
    \centering
    \begin{subfigure}{0.3\textwidth}
        \includegraphics[trim=50 80 35 65,clip,width=\linewidth]{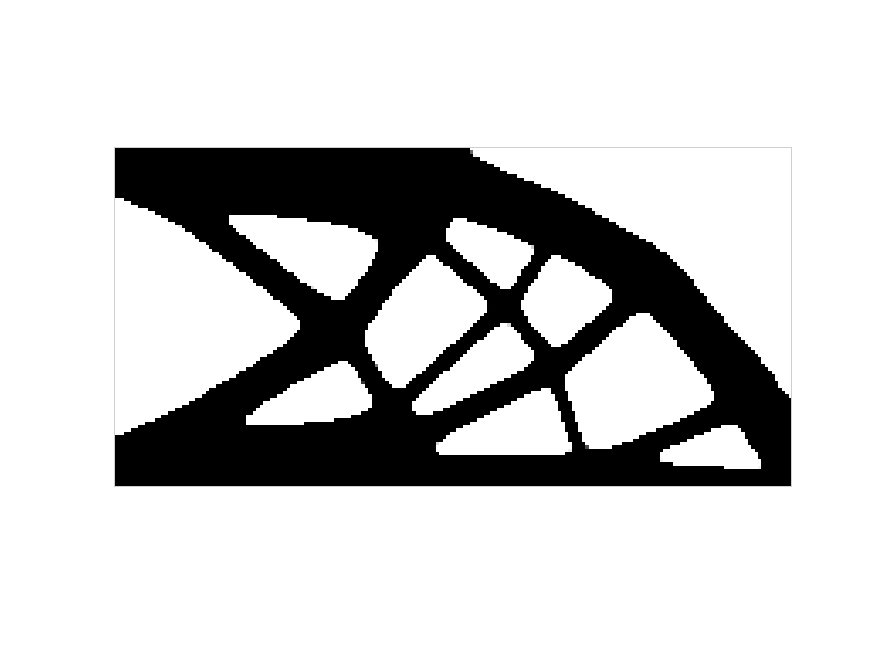}
        \caption{No milling, $C_\mathit{nominal}=\num{69.98}$}
        \label{fig:2Dbase}
    \end{subfigure}
    ~
    \begin{subfigure}{0.3\textwidth}
        \includegraphics[trim=50 80 35 65,clip,width=\linewidth]{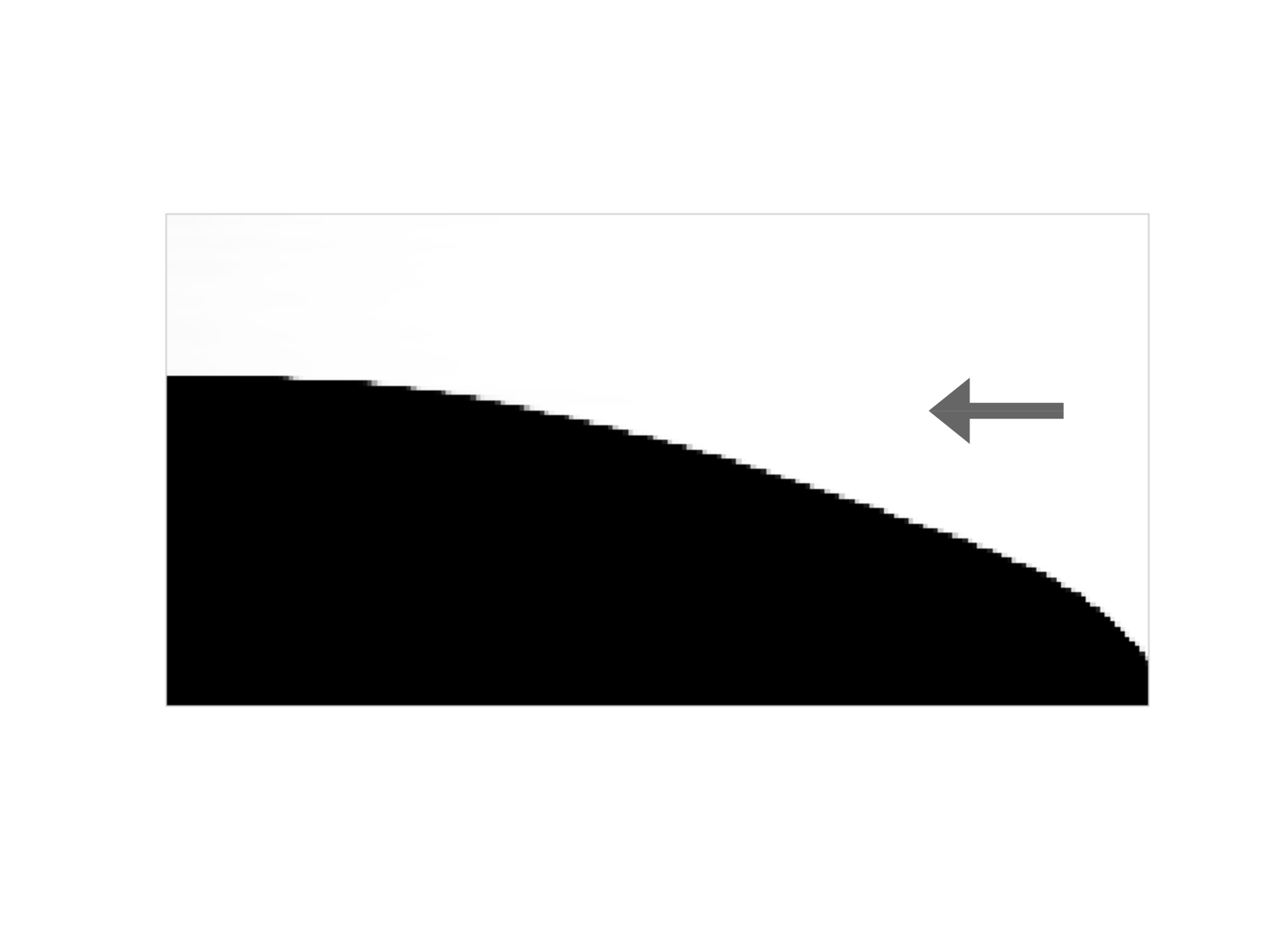}
        \caption{$\milldir=0$, $C=\num{179.68}$}
        \label{fig:2D1dir0}
    \end{subfigure}
    ~
    \begin{subfigure}{0.3\textwidth}
        \includegraphics[trim=50 80 35 65,clip,width=\linewidth]{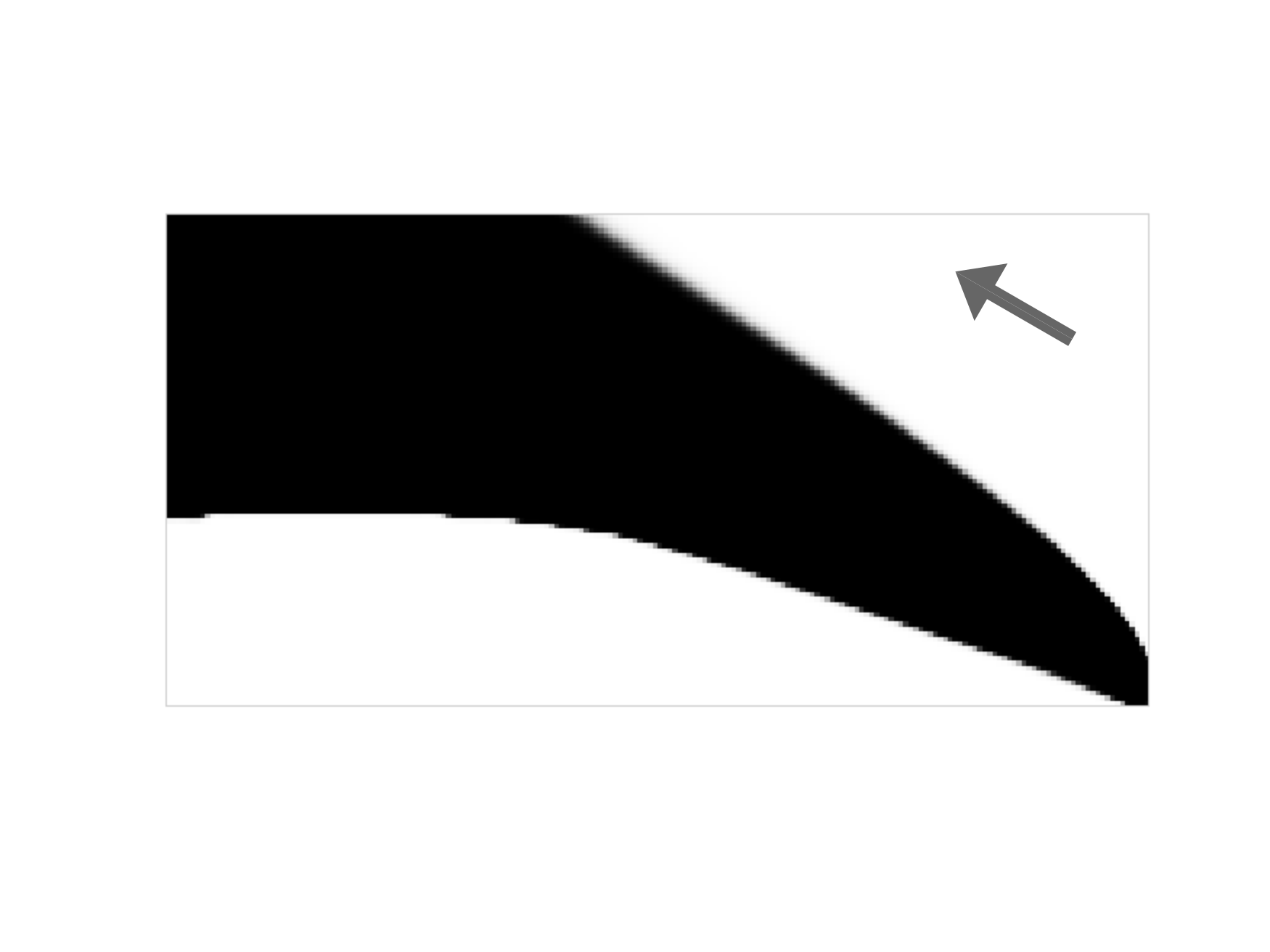}
        \caption{$\milldir=-30$, $C=\num{224.23}$}
        \label{fig:2D1dir-30}
    \end{subfigure}
    \\
    \begin{subfigure}{0.3\textwidth}
        \includegraphics[trim=50 80 35 65,clip,width=\linewidth]{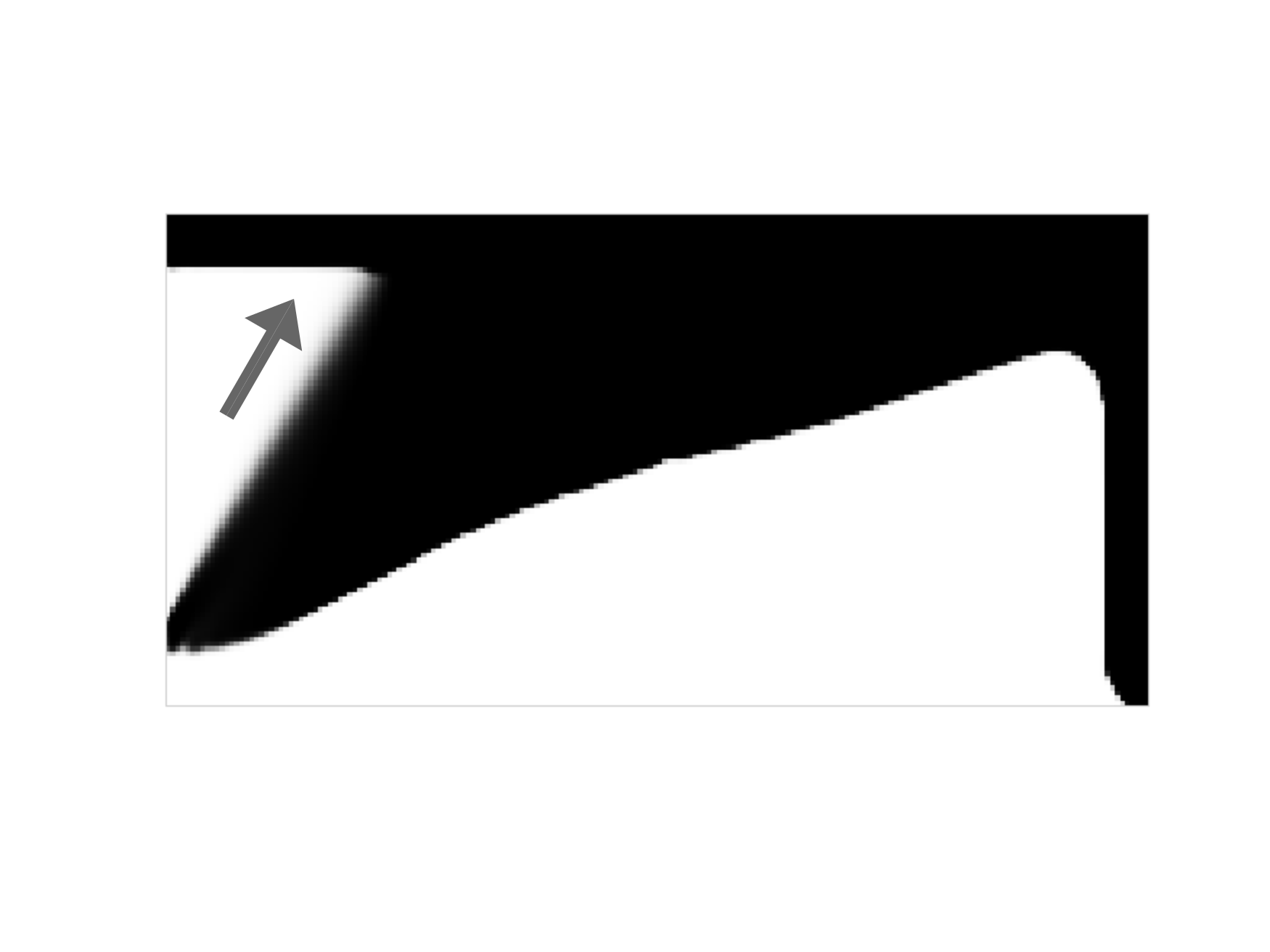}
        \caption{$\milldir=240$, $C=\num{211.04}$}
        \label{fig:2D1dir240}
    \end{subfigure}
    ~
    \begin{subfigure}{0.3\textwidth}
        \includegraphics[trim=50 80 35 65,clip,width=\linewidth]{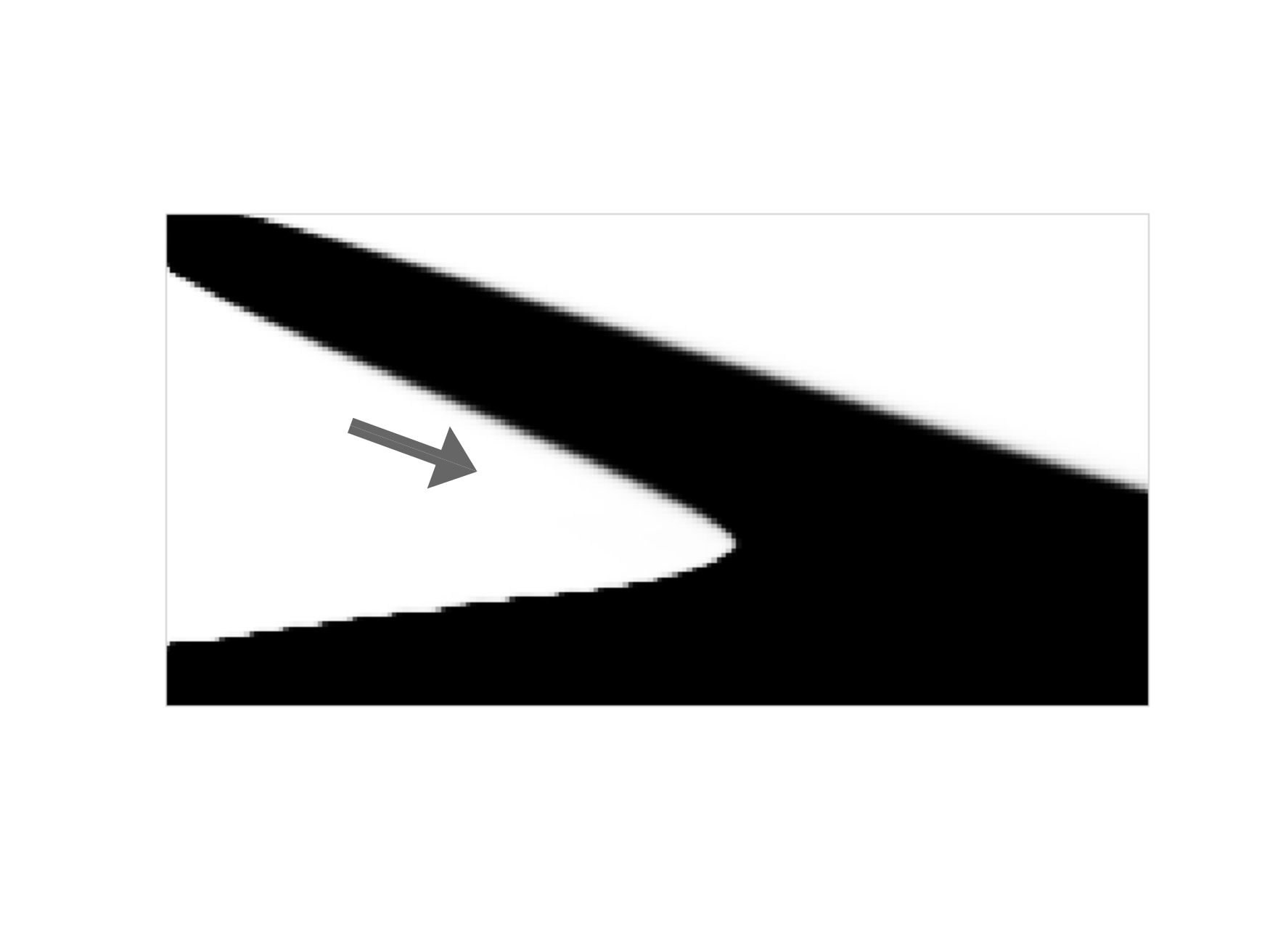}
        \caption{$\milldir=160$, $C=\num{87.49}$}
        \label{fig:2D1dir160}
    \end{subfigure}
    ~
    \begin{subfigure}{0.3\textwidth}
        \includegraphics[trim=50 80 35 65,clip,width=\linewidth]{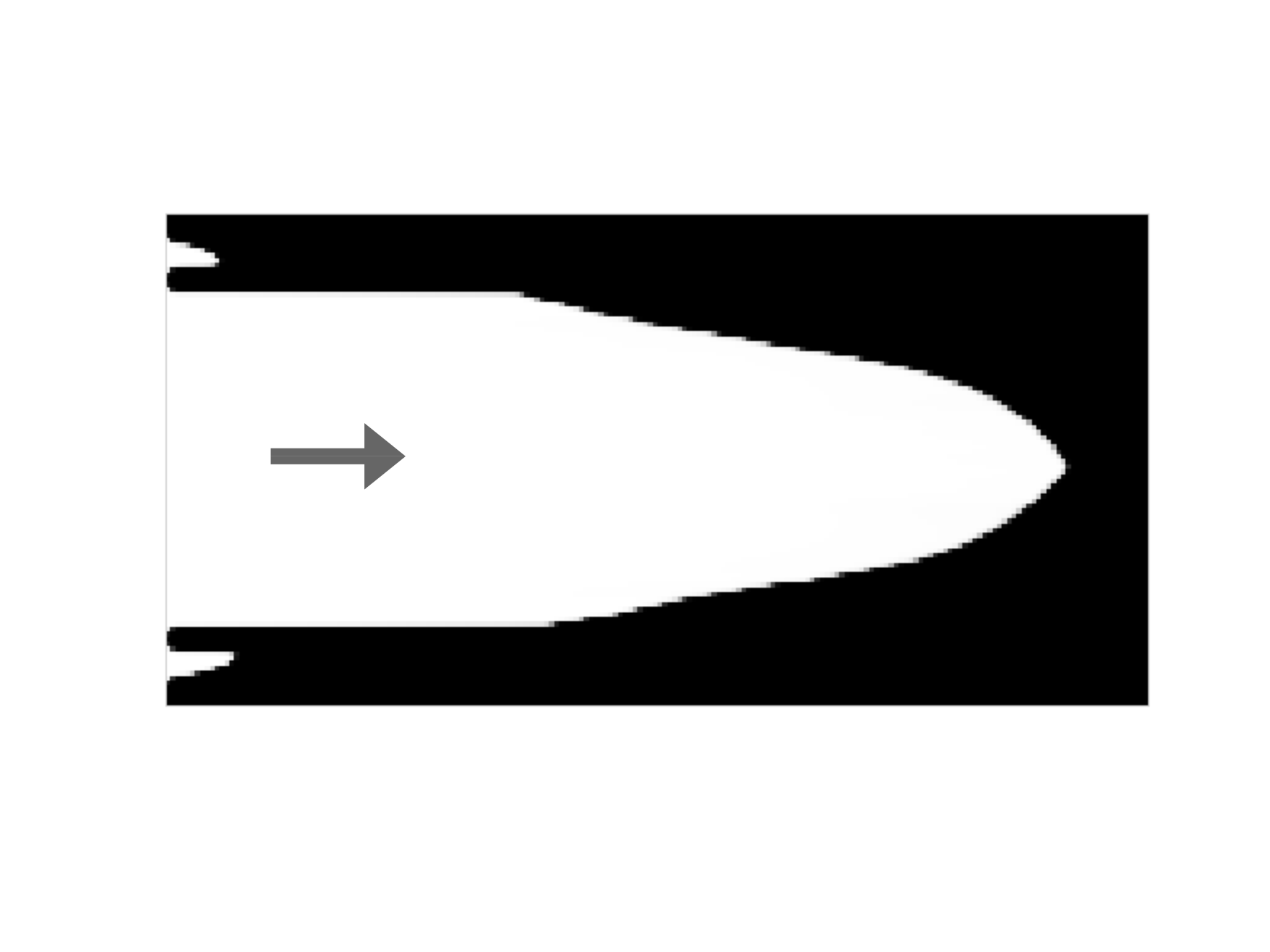}
        \caption{$\milldir=180$, $C=\num{287.93}$}
        \label{fig:2D1dir180}
    \end{subfigure}
    \caption{Reference design and single tool direction designs for the 2D case. The final projected variable, $\rhophys$ is shown.}
    \label{fig:2D1dir}
\end{figure*}

A reference example is optimized using the robust formulation \cite{Wang2011}. For the reference, the $\beta$ value is continuated, and the SIMP penalization power is set to $p=1$. The projection thresholds are set to $\eta_\mathit{dilated}=0.2$, $\eta_\mathit{nominal}=0.5$ and $\eta_\mathit{eroded}=0.8$, for the dilated, nominal and eroded fields, respectively. This choice of threshold values should ensure a minimum length scale of $0.9r_\mathit{min}=2.7$ elements. The volume fraction on the dilated field is updated regularly, such that the nominal volume fraction matches the desired one, shown in \cref{tab:optparam}. The obtained reference example is seen in \cref{fig:2Dbase}. The objective values and number of design iterations corresponding to the designs are seen in \cref{tab:2Dcompare}. It should be noted that this reference design is constrained by a minimal length-scale of the solid members, which the upcomming designs are not.

Results optimized with a single milling direction are shown in \cref{fig:2D1dir}. It is observed that all of the resulting designs are distinctly shaped by their corresponding tool direction. From \cref{tab:2Dcompare} it can be seen that all of the obtained structures have a higher compliance than the reference design. The best performing design \cref{fig:2D1dir160} has a compliance, which is $25\%$ higher than the reference design. From this, it is seen that constraining the design to be filtered from a single tool direction severely limits the design freedom of the optimization algorithm.

\begin{figure*}[htb]
    \centering
    \begin{subfigure}{0.3\textwidth}
        \includegraphics[trim=50 80 35 65,clip,width=\linewidth]{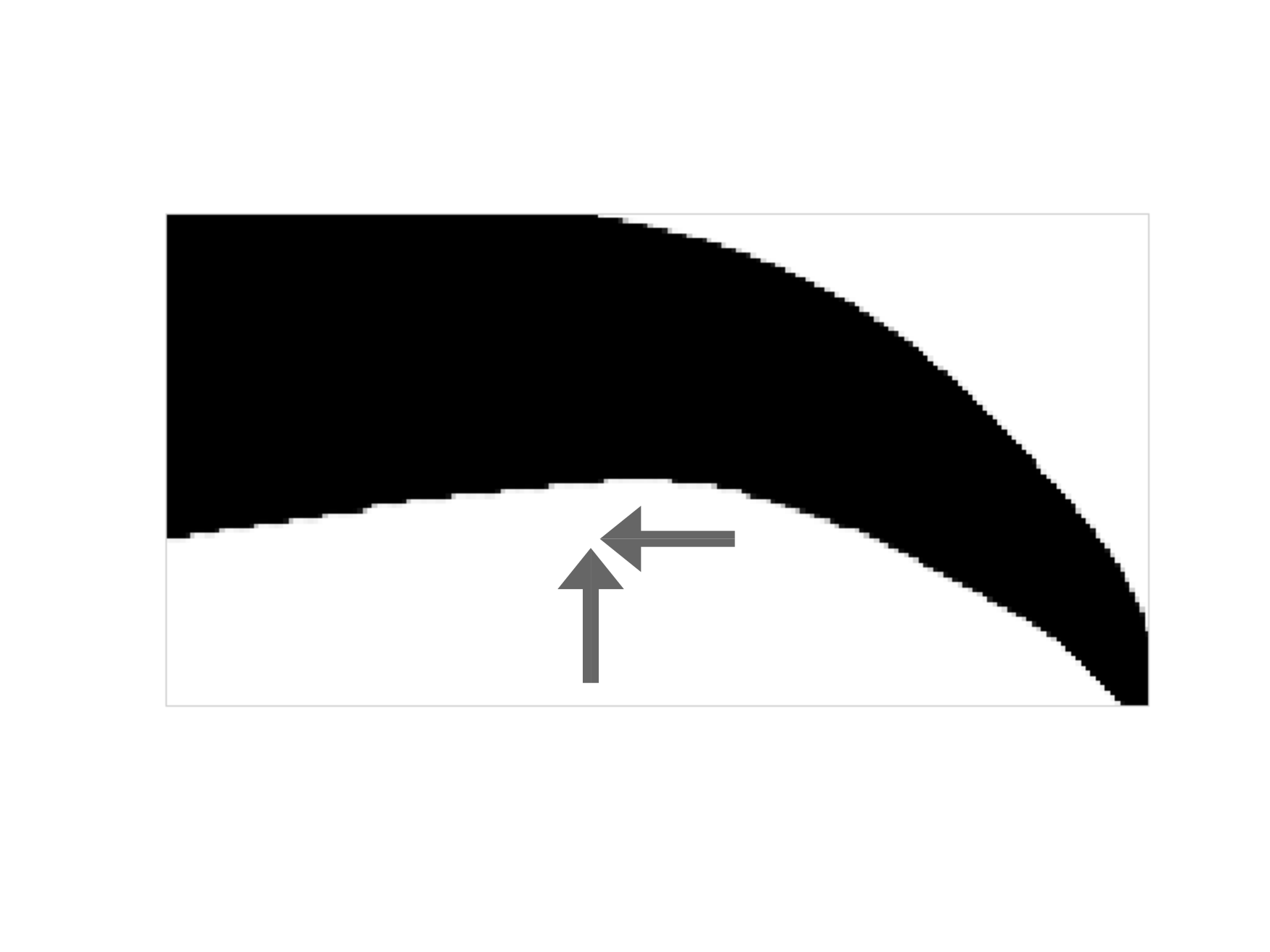}
        \caption{$\milldir=\left\{-90,\;0\right\}$, $C=\num{220.46}$}
        \label{fig:2D2dir}
    \end{subfigure}
    ~
    \begin{subfigure}{0.3\textwidth}
        \includegraphics[trim=50 80 35 65,clip,width=\linewidth]{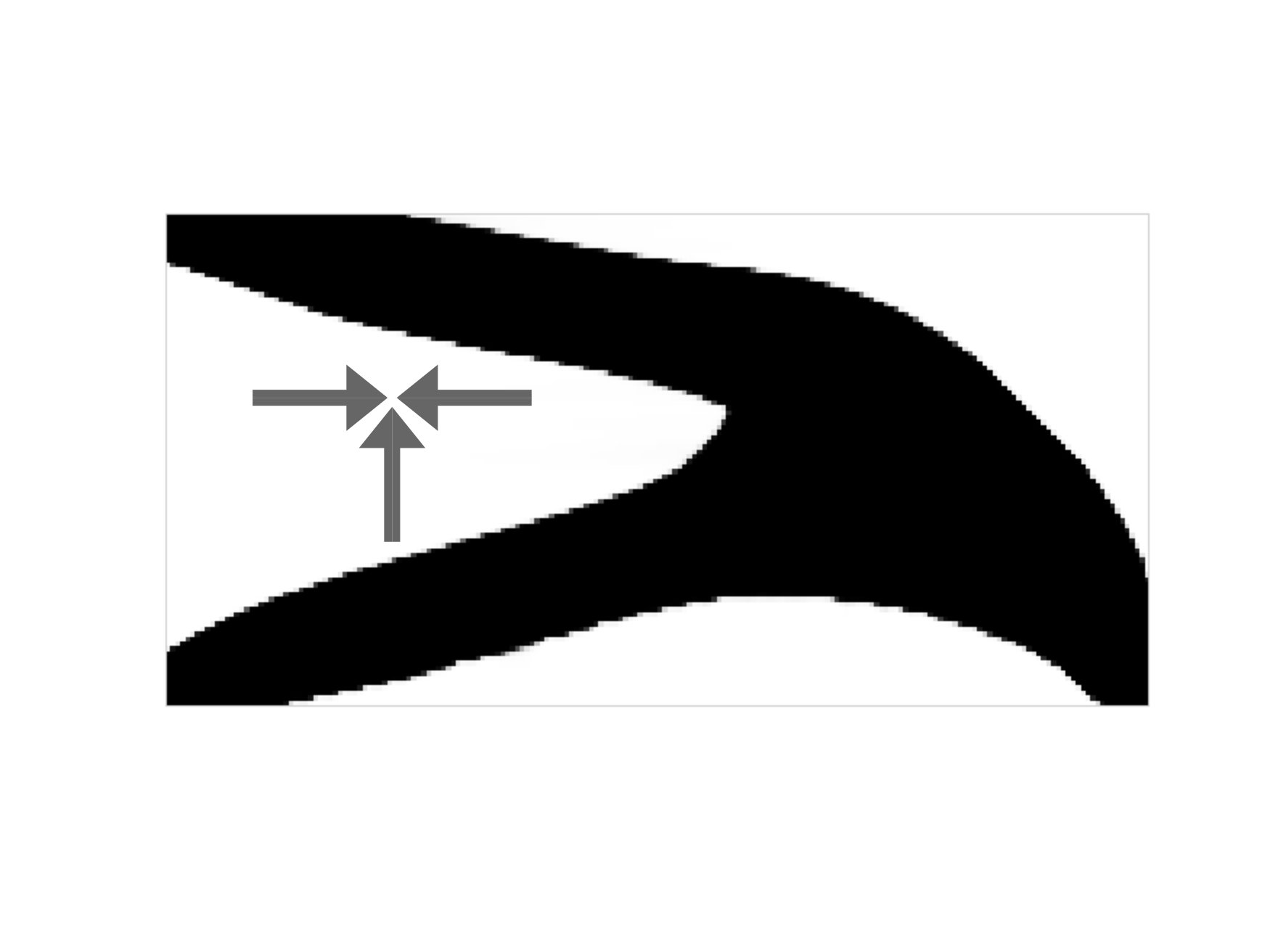}
        \caption{$\milldir=\left\{-90,\;0,\;180\right\}$, $C=\num{86.92}$}
        \label{fig:2D3dir}
    \end{subfigure}
    ~
    \begin{subfigure}{0.3\textwidth}
        \includegraphics[trim=50 80 35 65,clip,width=\linewidth]{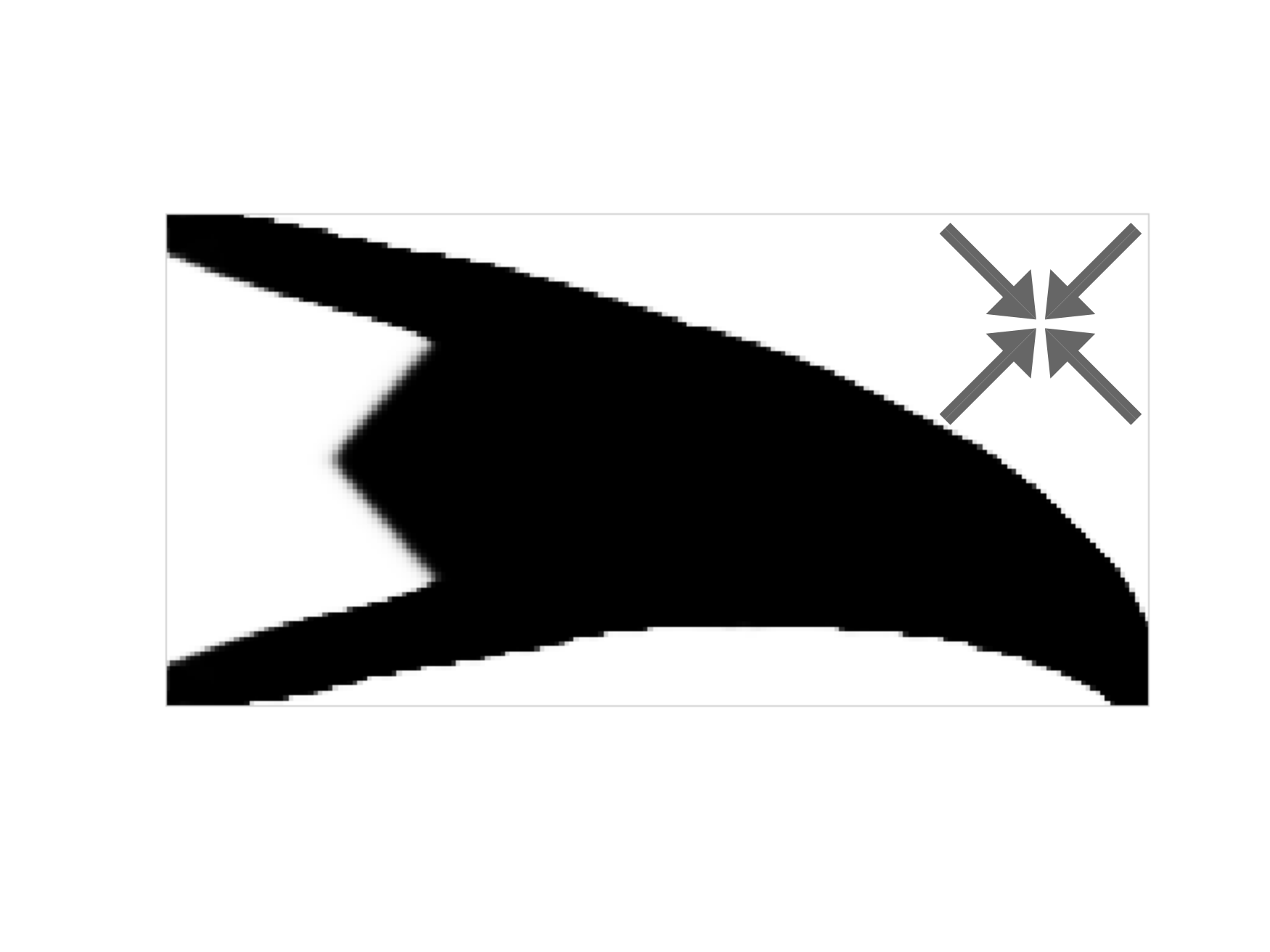}
        \caption{$\milldir=\left\{45,\;135,\;225,\;315\right\}$, $C=\num{115.14}$}
        \label{fig:2D4dir}
    \end{subfigure}
    \caption{Resulting designs when using multiple tool directions for the 2D case. The final projected variable, $\rhophys$ is shown.}
    \label{fig:2Dmultdir}
\end{figure*}

Designs optimized with multiple milling directions are shown in \cref{fig:2Dmultdir}. Having multiple milling directions impacts the obtained design, as material can be removed from multiple directions. This is most notably seen when comparing \cref{fig:2D1dir0,fig:2D2dir,fig:2D3dir}. In the first case, \cref{fig:2D1dir0}, material can only be removed from the right hand side, resulting in material being placed in the lower left triangle part of the design domain. In the second case, seen in \cref{fig:2D2dir}, material can also be removed from below, resulting in material being placed near the upper design domain boundary. In \cref{tab:2Dcompare}, it is seen that the first case, with a single direction performs better than the latter one, which indicates that a local minimum with an inferior performance has been obtained. The same local minimum was also observed by \citet{Langelaar2019}. The solutions find different local minima, due to the tool directions heavily affecting the initial density layout which is found using a homogeneous design variable distribution. This could potentially be avoided with an other starting guess.

With a third milling direction, as seen in \cref{fig:2D3dir}, the obtained design can become attached to both extremities of the supported side. In the performance comparison, \cref{tab:2Dcompare}, it is also seen that this design performs better than the ones presented in \cref{fig:2D1dir0,fig:2D2dir} and only $24\%$ worse than the reference design from \cref{fig:2Dbase}. 

The design obtained when optimizing with the four diagonal tool directions is seen in \cref{fig:2D4dir}. The design has some features similar to the one from \cref{fig:2D3dir}. However, the groove going into the structure from the support plane cannot be deeper with the selected tool orientations. The selected tool directions are also responsible for the triangular shape, seen inside the groove. This triangular shape does not bear any load, however it cannot be removed with the selected tool directions. The design performs $65\%$ worse than the one seen in \cref{fig:2D3dir}, which is most likely due to the high amount of non-load bearing material.

It is noted that the performance of the designs with $\milldir=160$ and $\milldir=\left\{-90,\;0,\;180\right\}$ have very similar compliances and are also similar in a qualitative manner. It is noted that the design with $\milldir=160$, seen in \cref{fig:2D1dir160}, is feasible with both sets of milling directions, this is not the case for the design obtained with $\milldir=\left\{-90,\;0,\;180\right\}$, seen in \cref{fig:2D3dir}. This might explain the slightly lower compliance of the design seen in \cref{fig:2D3dir}.

\begin{table}[htb]
    \centering
    \caption{Comparison of compliances of the two dimensional results.}
    \label{tab:2Dcompare}
    \begin{tabular}{lrrr}
    \hline
        Figure &   $\milldir$ &           $C$ &      $C/C_\mathit{ref}$ \\\hline
        \ref{fig:2Dbase} & &   $69.98$ &  $1.00$ \\
        \ref{fig:2D1dir0} & 0 &  $179.68$ & $2.57$ \\
        \ref{fig:2D1dir-30} &-30& $224.23$ & $3.20$ \\
        \ref{fig:2D1dir240} & 240&$211.04$ & $3.01$ \\
        \ref{fig:2D1dir160} & 160&$87.49$ &  $1.25$ \\
        \ref{fig:2D1dir180} & 180&$287.93$ & $4.11$ \\
        \ref{fig:2D2dir} &  $\left\{-90,\;0\right\}$ &  $220.46$ & $3.15$ \\
        \ref{fig:2D3dir} &  $\left\{-90,\;0,\;180\right\}$&  $86.92$ &  $1.24$ \\
        \ref{fig:2D4dir} &  $\left\{45,\;135,\;225,\;315\right\}$&  $115.14$ & $1.65$ 
    \end{tabular}
\end{table}

A large qualitative similarity is observed between the designs obtained by \citet{Langelaar2019} and the ones presented in \cref{fig:2D1dir,fig:2Dmultdir} - with the exception of the reference design (\cref{fig:2Dbase}). The differences in the reference designs are probably due to different formulations being used for obtaining the reference design by \cite{Langelaar2019} and the presented work.

\subsection{Three dimensional cantilever examples}
\label{sec:cantilever}

The three dimensional cantilever beam has a similar design domain as the two dimensional one from \Cref{fig:2dproblem}, where the out of plane direction is modeled with the thickness $1$. The applied load is a line load at the corresponding 3 dimensional position. The domain is discretized with a mesh containing $31.25$ million hexahedral elements. The applied filter radius corresponds to approximately $1.5$ elements. The optimization settings are shown in \Cref{tab:optparam}. The results are computed on 10 nodes at the DTU Sophia cluster, which has 2 AMD EPYC 7351 16-core processors and 125GB memory at every node. The solution time is between 55 and 450 seconds per design iteration, depending on the number of milling directions.

\begin{figure*}[htb]
    \centering
    \begin{subfigure}{0.33\textwidth}
        \includegraphics[width=\linewidth]{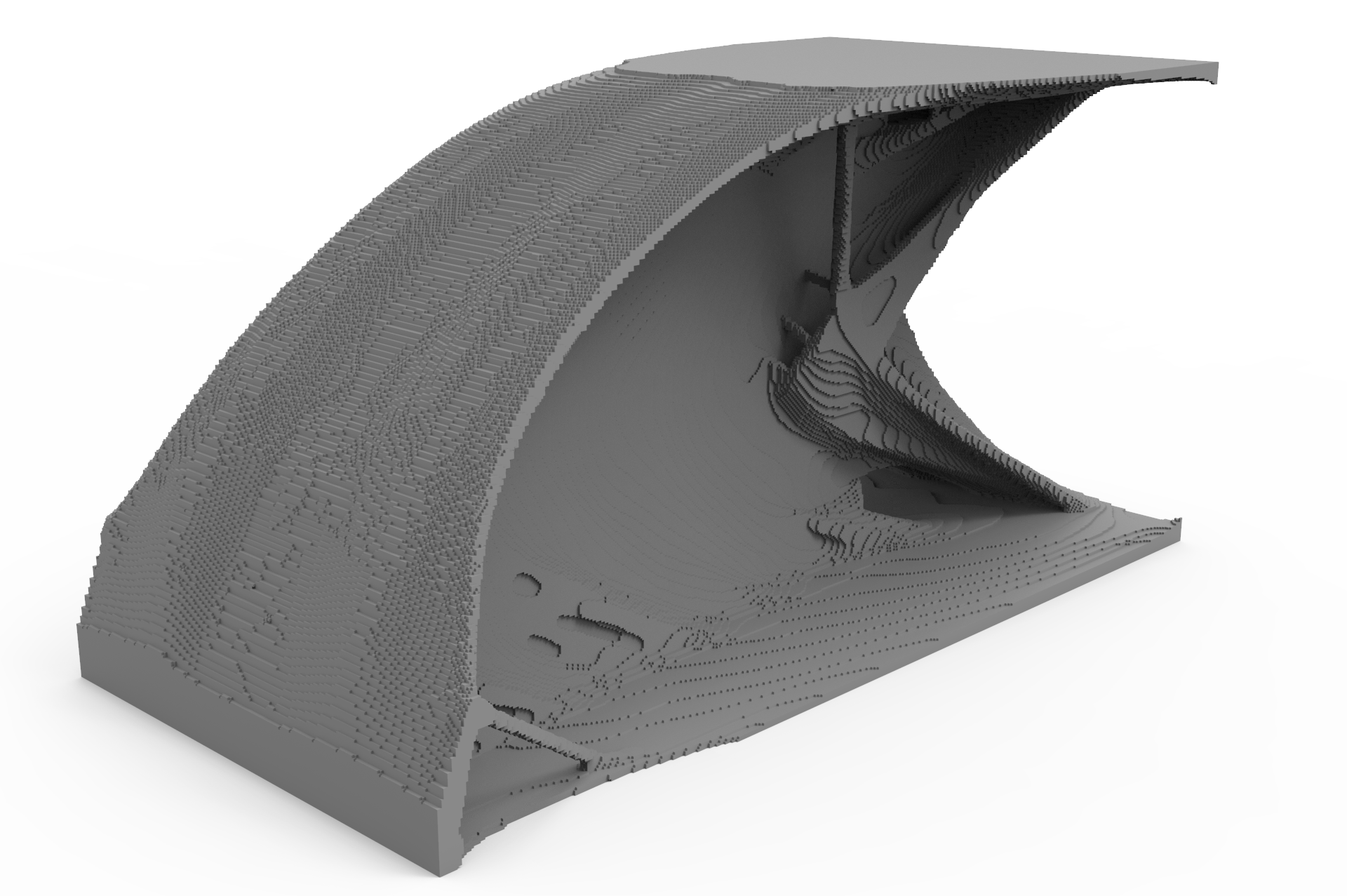}
    \end{subfigure}
    \begin{subfigure}{0.33\textwidth}
        \includegraphics[width=\linewidth]{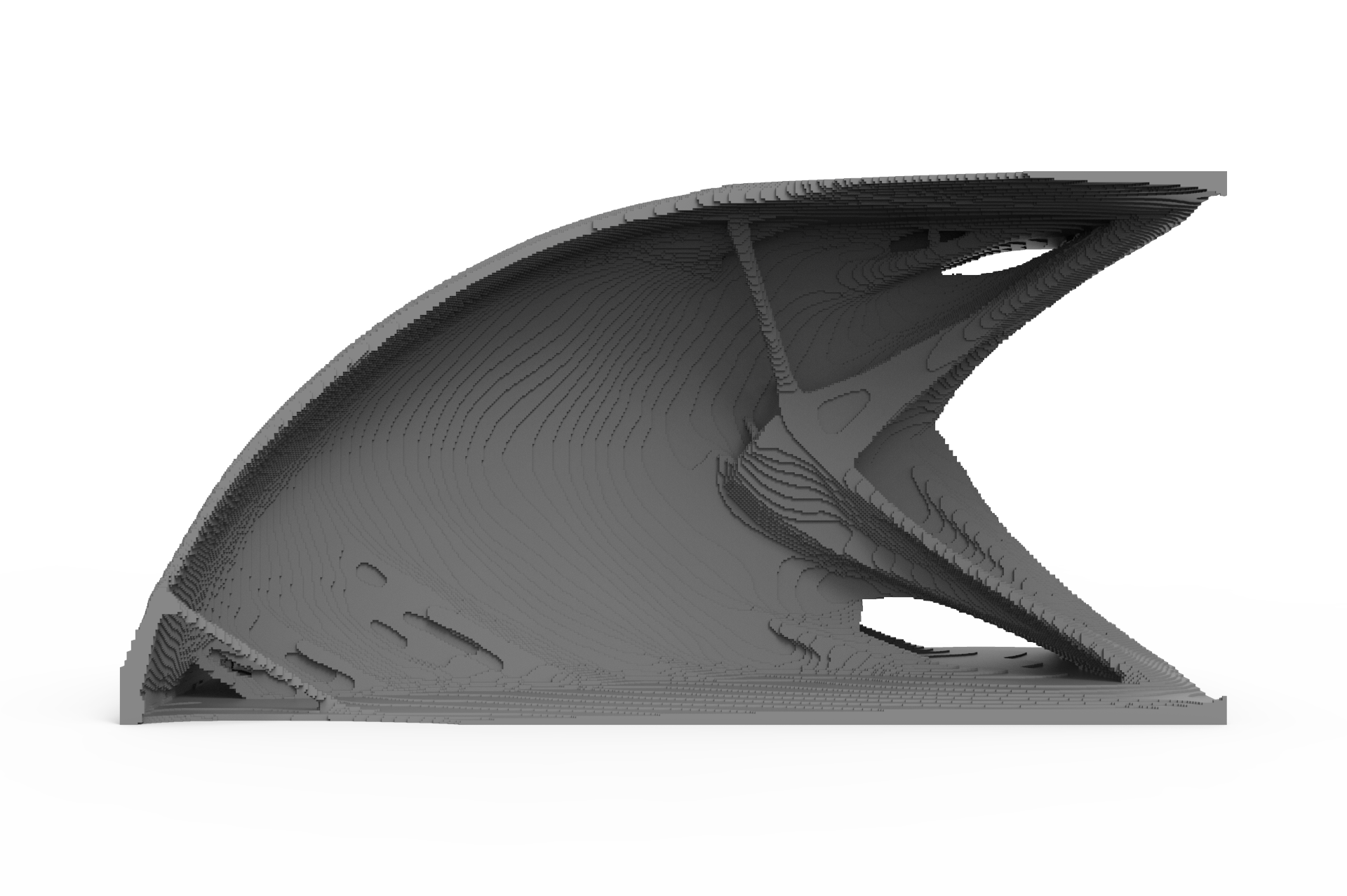}
    \end{subfigure}
    \begin{subfigure}{0.33\textwidth}
        \includegraphics[width=\linewidth]{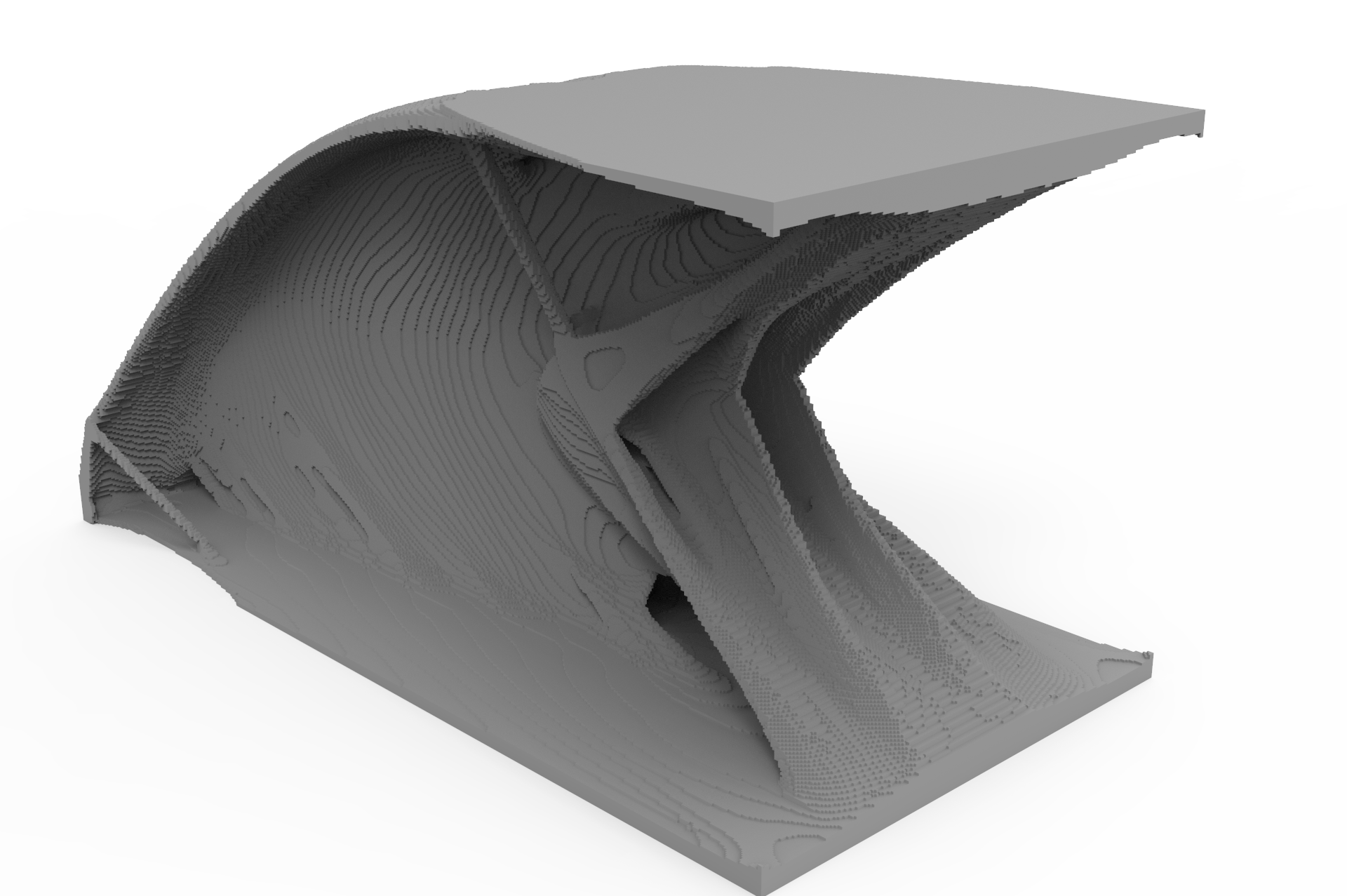}
    \end{subfigure}
    \caption{Reference cantilever beam obtained with no milling and a robust formulation (nominal design is shown). $C_\mathit{nominal}= \num{1.15e7}$. }
    \label{fig:cantilever_ref}
\end{figure*}

A benchmark cantilever beam without imposed manufacturability is optimized using the robust formulation \cite{Wang2011}. In this case, the $\beta$-value for the Heaviside projection sharpness is ramped up to a final value of $59.4$, and the optimization is process is run for $700$ iterations. The threshold values are set to $\eta=0.2$ in the dilated-, $\eta=0.5$ in the nominal- and $\eta=0.8$ in the eroded field - resulting in a minimum length scale of $0.00765$. The reference design obtained with the robust formulation is seen in \cref{fig:cantilever_ref}. The final robust objective, used as a reference is evaluated on the nominal field as a postprocessing step to the optimization.

In the early design iterations of optimizations considering machining, the structure is not connected to the support and load, due to a milling tool coming from the support plane - or loaded line. This renders the linear elasticity equation very difficult and slow to solve due to the ill conditioning. As a remedy, a continuation scheme is applied on the Youngs module contrast, which is reduced by a factor of 10 after 20, 40 and 60 iterations. The optimization is hence started with $E_\mathit{min} = \num{1e-4} E_\mathit{max}$ and reaches $E_\mathit{min} = \num{1e-7} E_\mathit{max}$ at iteration 60.

Several milling direction cases are considered. In the cases with a single milling direction, the $p$ norm power is set to $p=1$ as no element-wise minimum is required. In cases with less than 10 milling directions, it is set to $p=-4$ and to $p=-6$ for cases with more than 10 directions. The $p$ norm power is changed with the varying number of tool directions due to the changing characteristics of the $p$ norm. A lower value of $p$ results in a less accurate approximation of the $\min$ operator, but also results in a less non-linear operator, which is desired during the optimization.

\begin{figure*}[htb]
    \centering
    \begin{subfigure}{0.33\textwidth}
        \includegraphics[width=\linewidth]{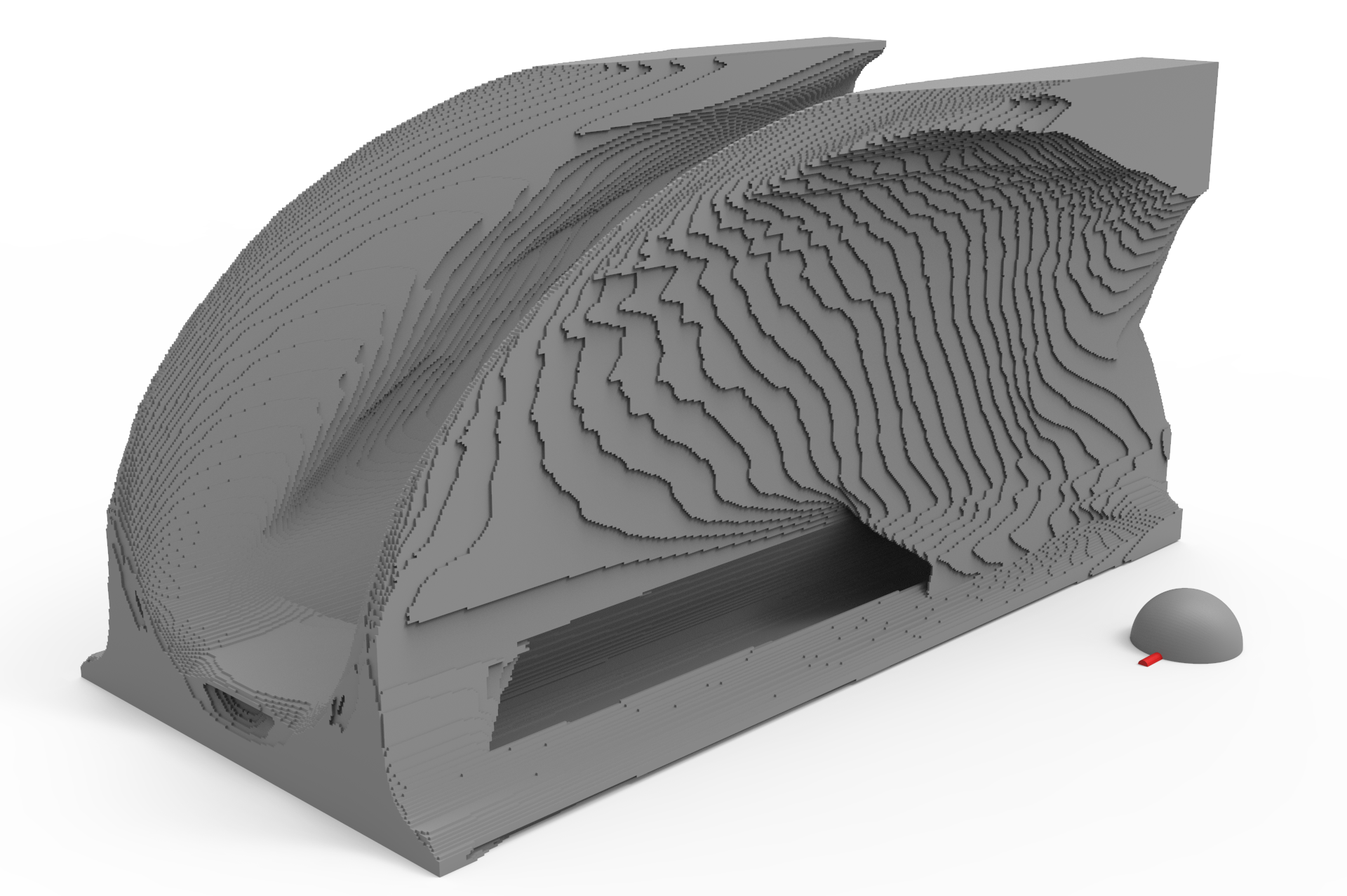}
    \end{subfigure}
    \begin{subfigure}{0.33\textwidth}
        \includegraphics[width=\linewidth]{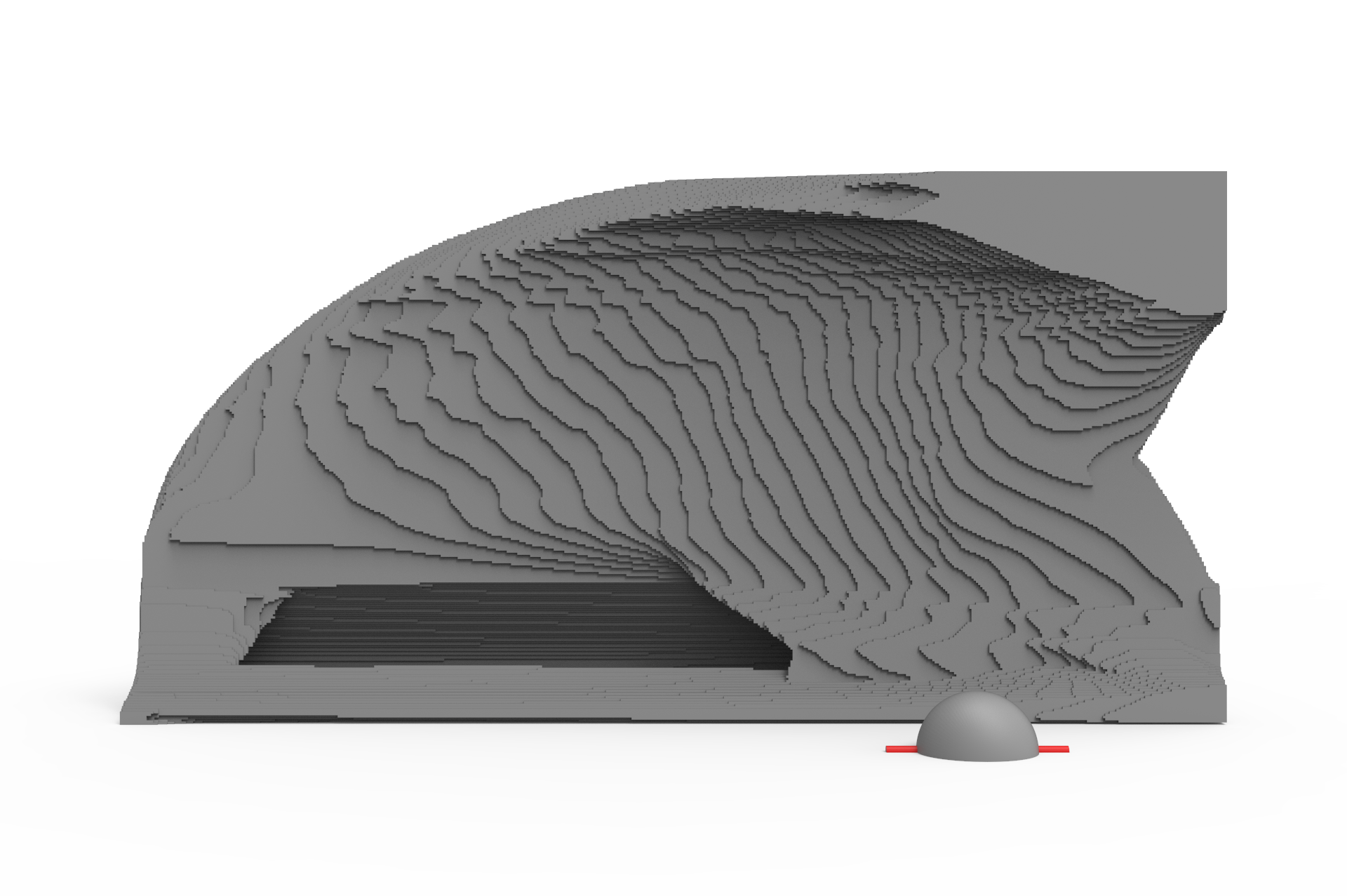}
    \end{subfigure}
    \begin{subfigure}{0.33\textwidth}
        \includegraphics[width=\linewidth]{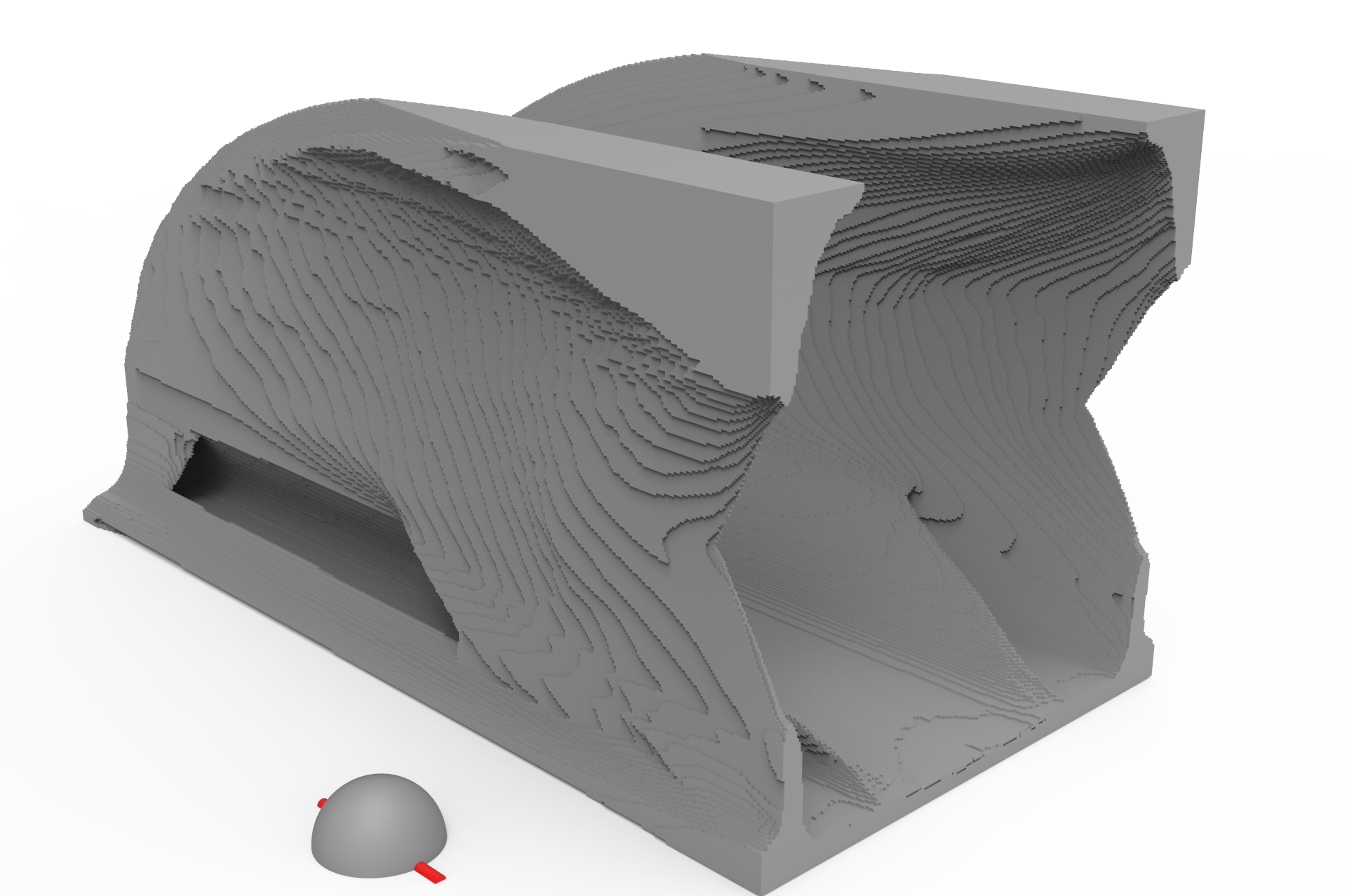}
    \end{subfigure}
    \caption{Cantilever beam with a single milling direction from the front and back. $C=\num{1.37e+07}$.}
    \label{fig:cantilever_x}
\end{figure*}

The cantilever beam is optimized using milling directions normal to the supported plane (in both directions). The obtained design is seen in \cref{fig:cantilever_x}. The obtained structure consists of two vertical plates that curve down towards the loaded line. A plate at the bottom of the domain connects the two vertical ones.

\begin{figure*}[htb]
    \centering
    \begin{subfigure}{0.33\textwidth}
        \includegraphics[width=\linewidth]{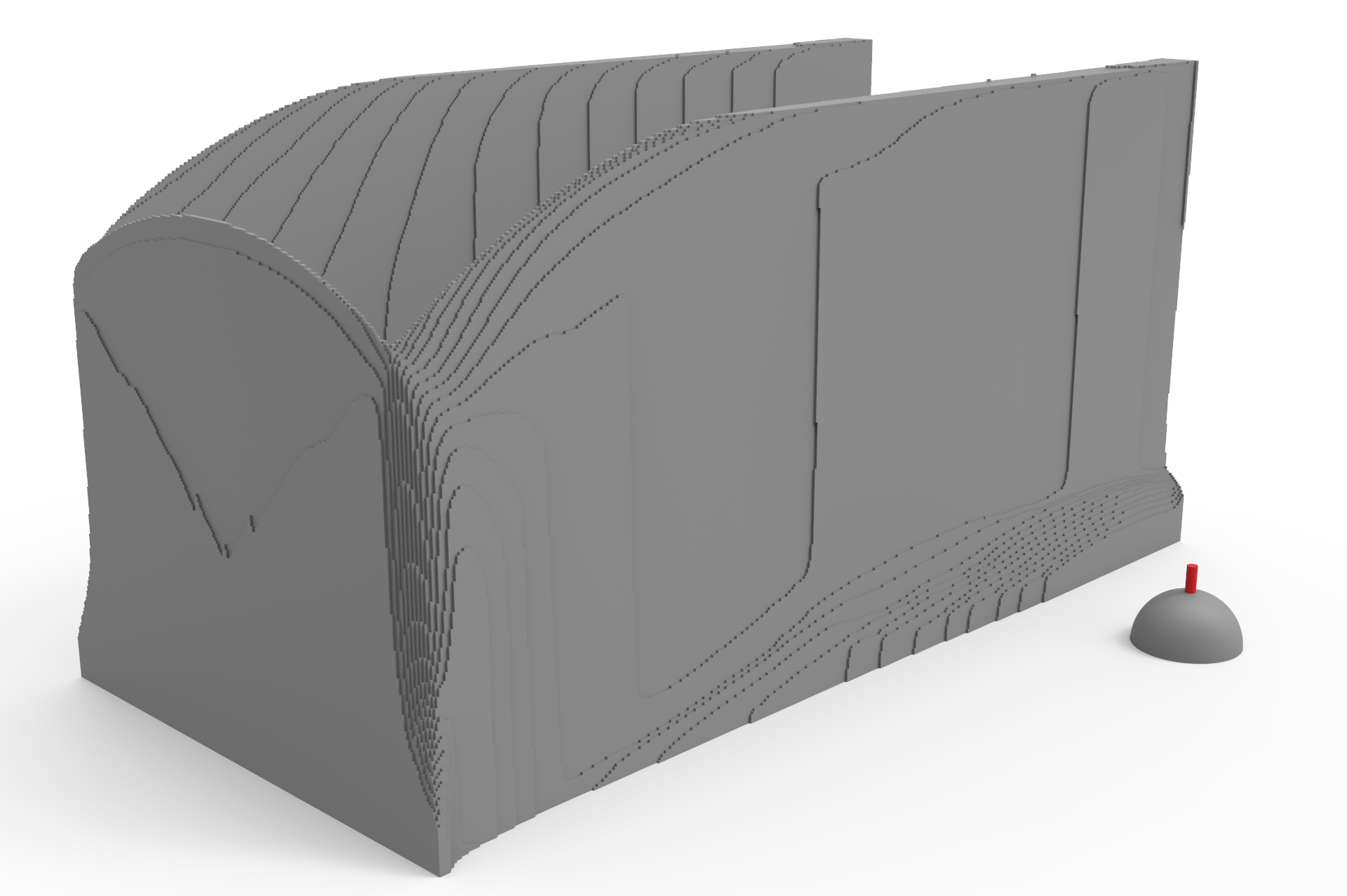}
    \end{subfigure}
    \begin{subfigure}{0.33\textwidth}
        \includegraphics[width=\linewidth]{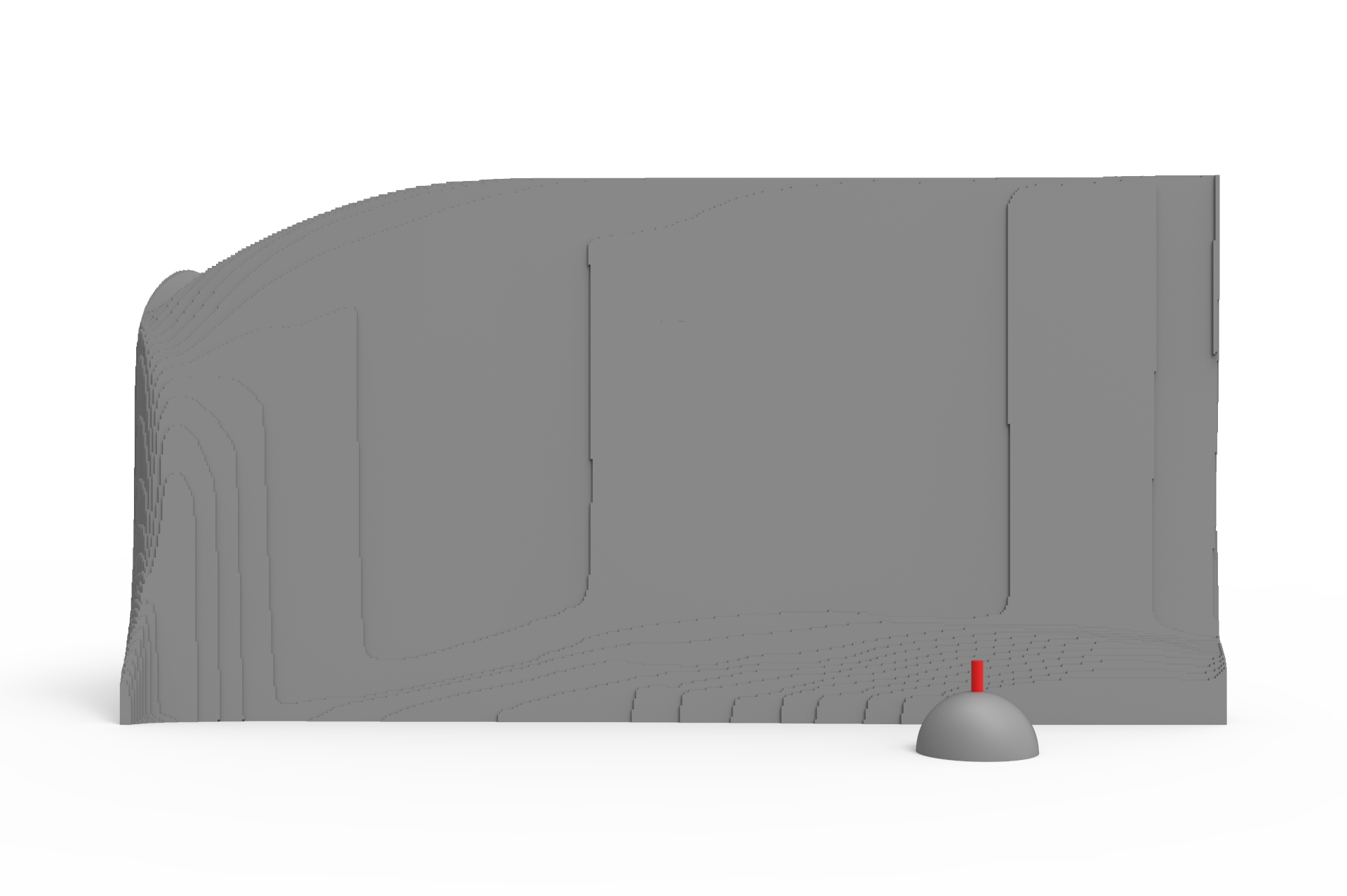}
    \end{subfigure}
    \begin{subfigure}{0.33\textwidth}
        \includegraphics[width=\linewidth]{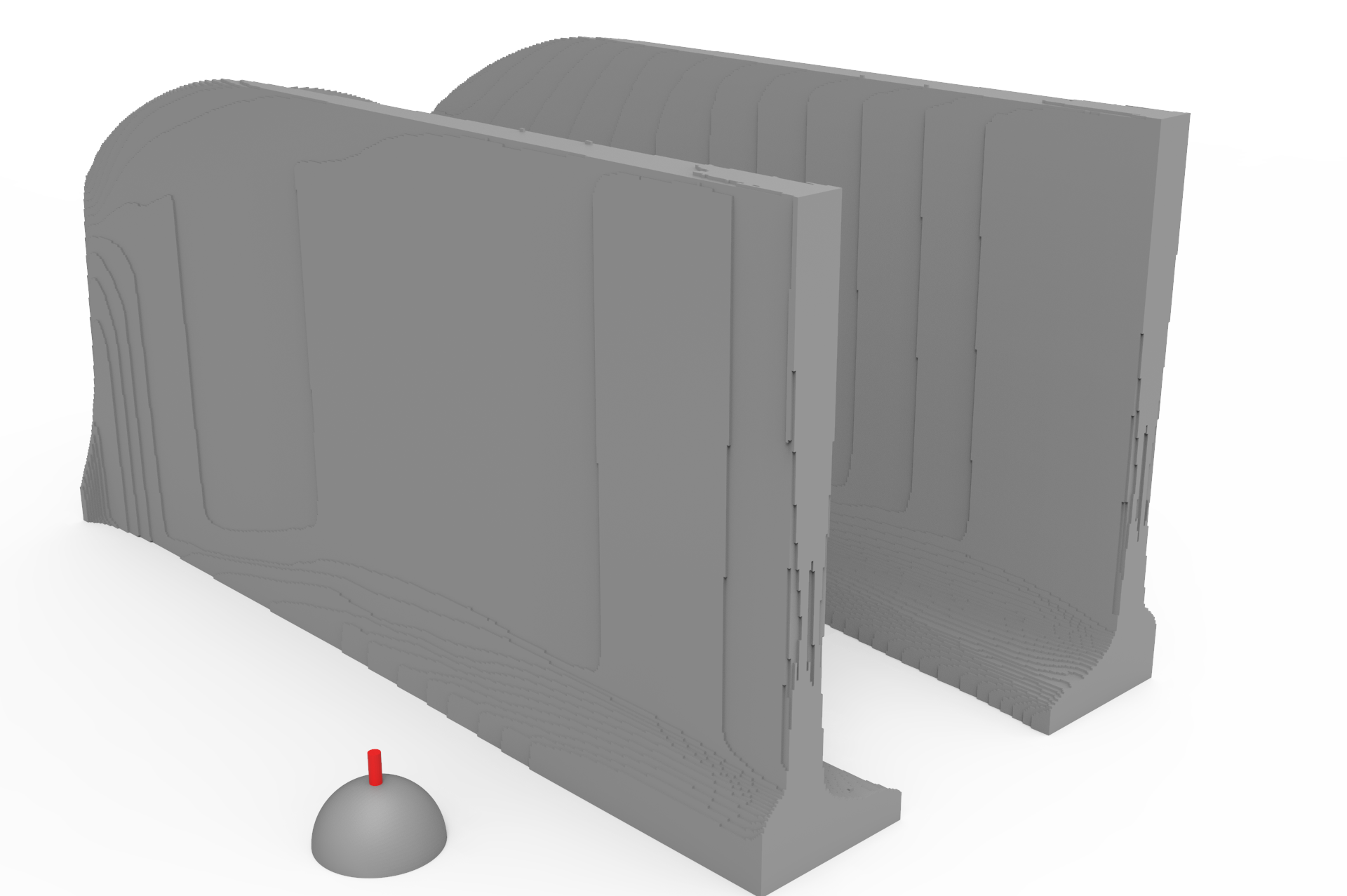}
    \end{subfigure}
    \caption{Cantilever beam with a single milling direction from the top. $C=\num{1.53e7}$. }
    \label{fig:cantilever_y}
\end{figure*}

In \cref{fig:cantilever_y}, the design obtained with a single milling direction from top is shown. The design consists of two vertical plates, however they are not connected at the bottom of the domain. A third vertical plate connects the two other ones along the loaded line. This third plate is higher than the plates to which it is connected. This could be to add some bending stiffness along the loaded line.

\begin{figure*}[htb]
    \centering
    \begin{subfigure}{0.33\textwidth}
        \includegraphics[width=\linewidth]{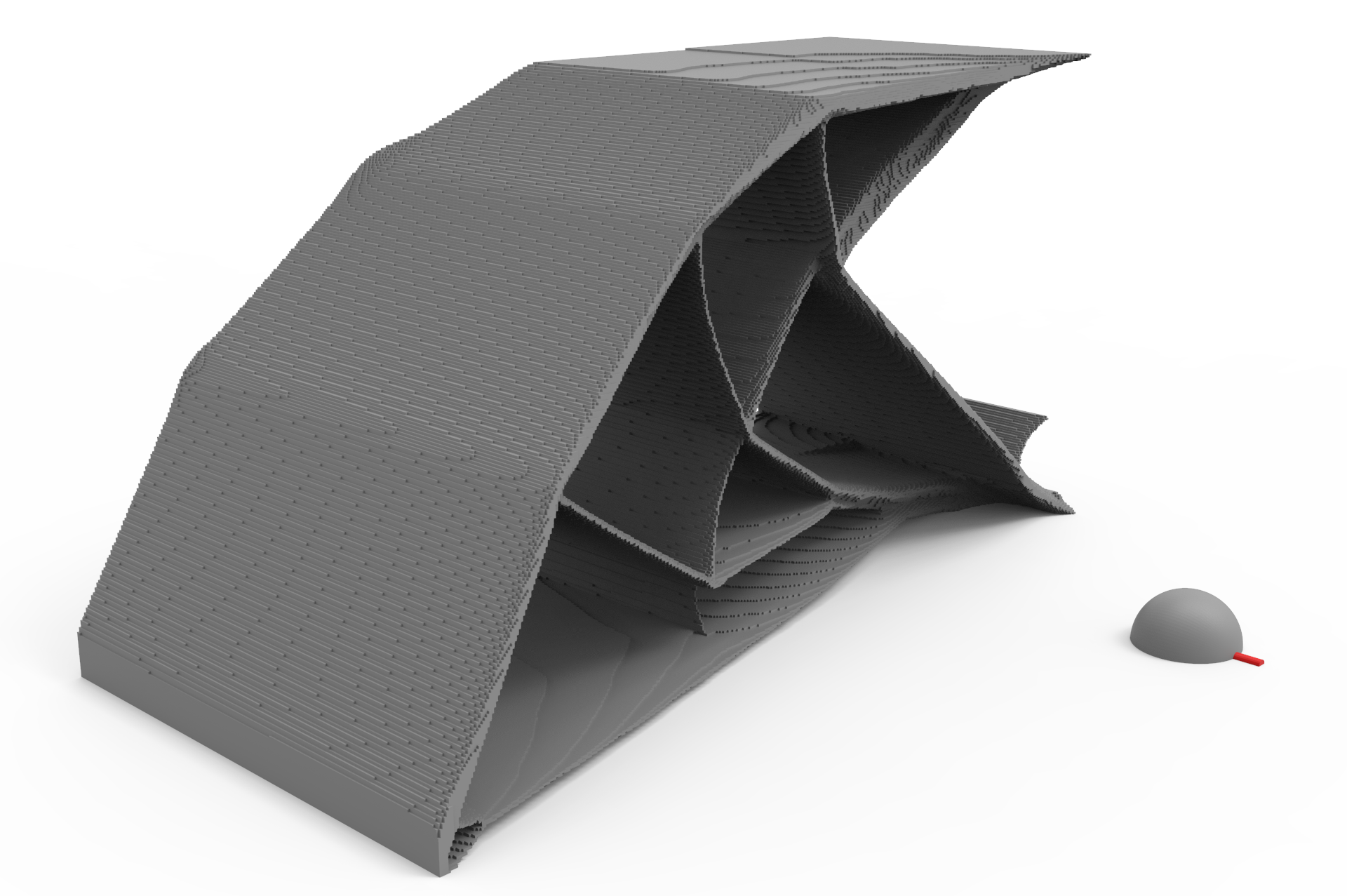}
    \end{subfigure}
    \begin{subfigure}{0.33\textwidth}
        \includegraphics[width=\linewidth]{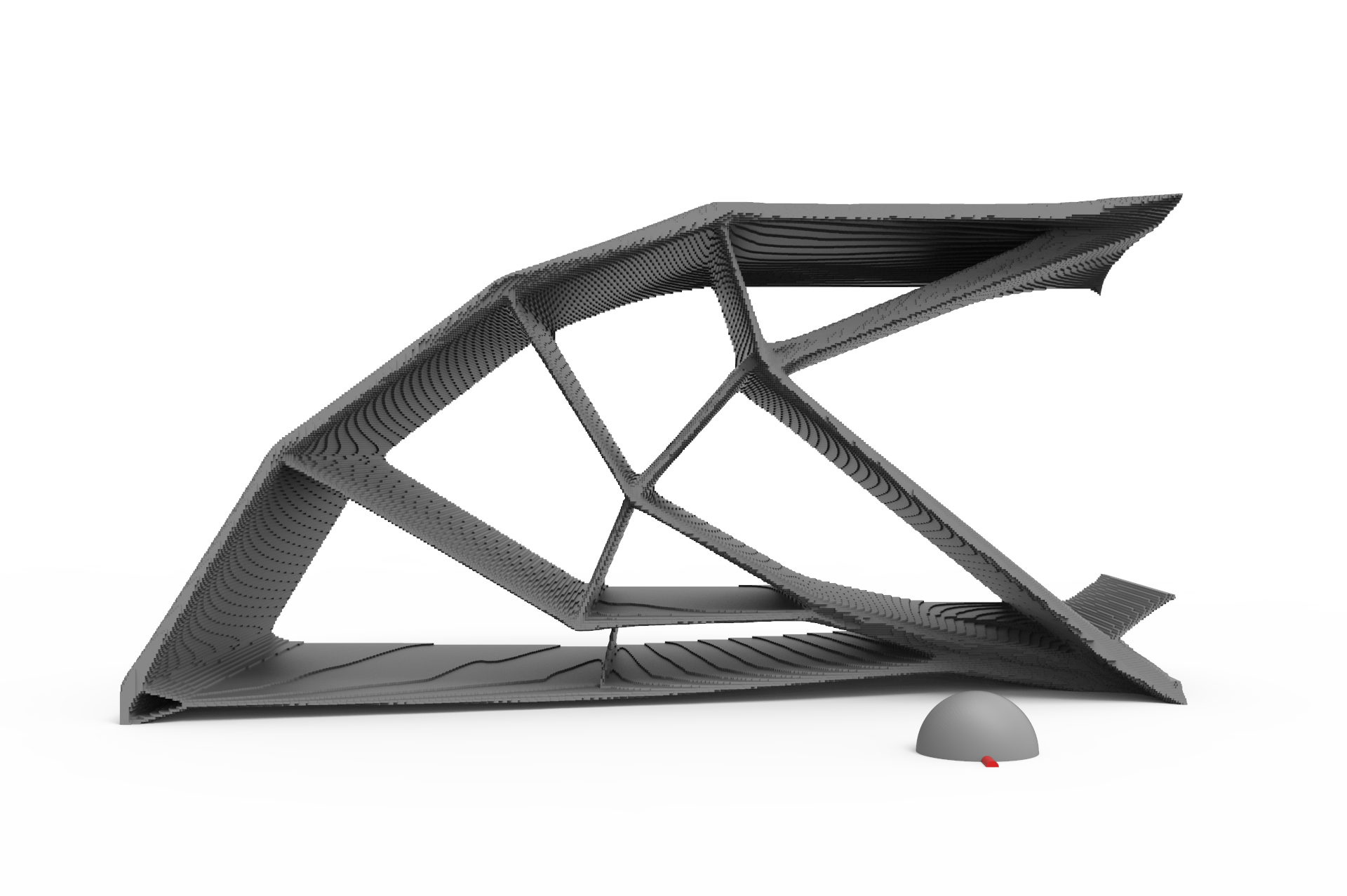}
    \end{subfigure}
    \begin{subfigure}{0.33\textwidth}
        \includegraphics[width=\linewidth]{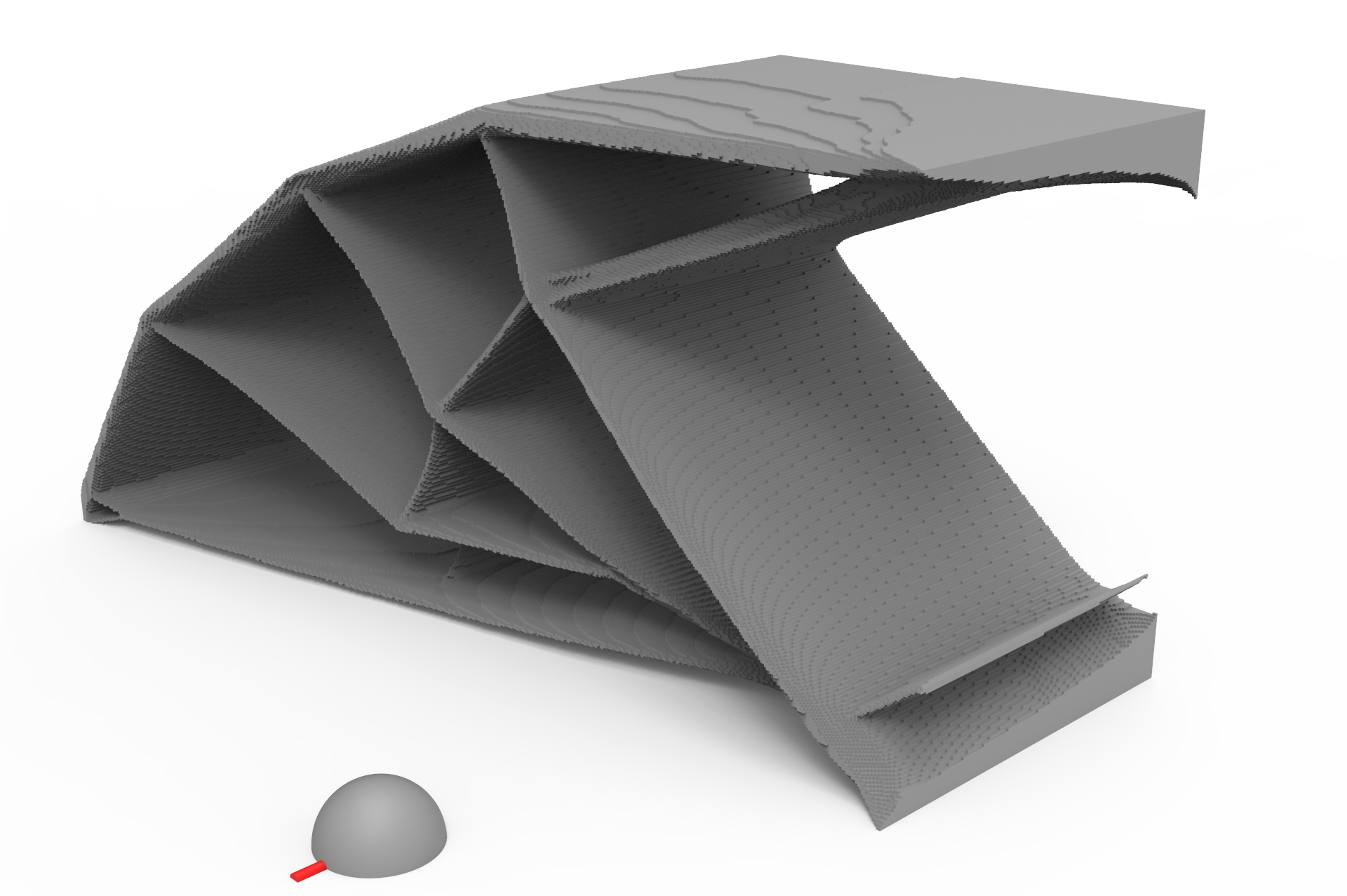}
    \end{subfigure}
    \caption{Cantilever beam with a single milling direction from the side. $C=\num{1.37e7}$. }
    \label{fig:cantilever_z}
\end{figure*}

A cantilever beam obtained with milling only from one side, orthogonal to both the load direction and the normal of the clamped surface, is shown in \cref{fig:cantilever_z}. The design cross section resembles the two dimensional cantilever beam design with no milling constraints seen in \cref{fig:2Dbase}. However, this three dimensional design is not a pure extruded version of the two dimensional design, as less structure is present in the side where the milling tool comes from. This probably reflects that a design, where the load can be transferred to a reduced part of the domain performs better.

\begin{figure*}[htb]
    \centering
    \begin{subfigure}{0.33\textwidth}
        \includegraphics[width=\linewidth]{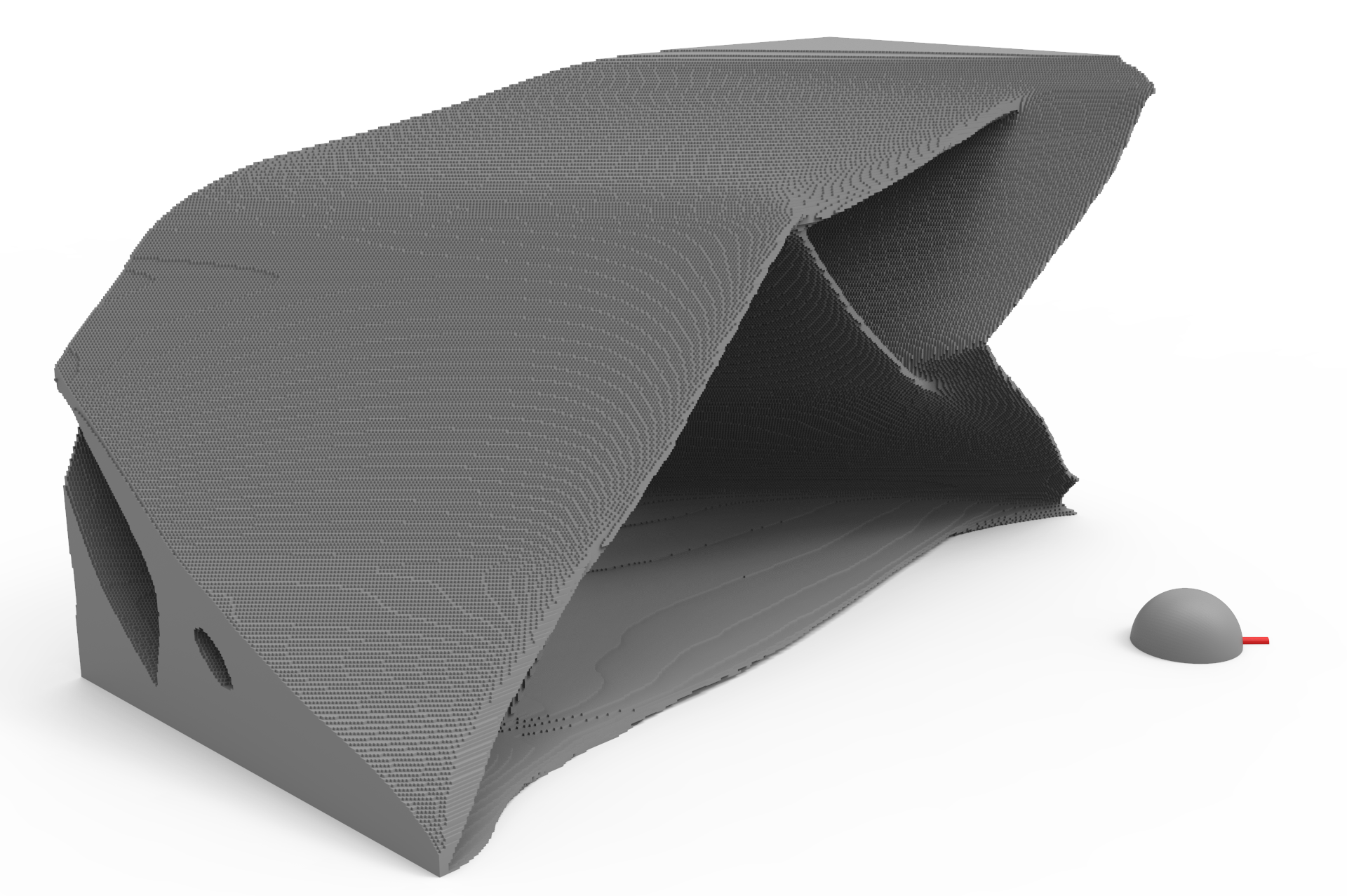}
    \end{subfigure}
    \begin{subfigure}{0.33\textwidth}
        \includegraphics[width=\linewidth]{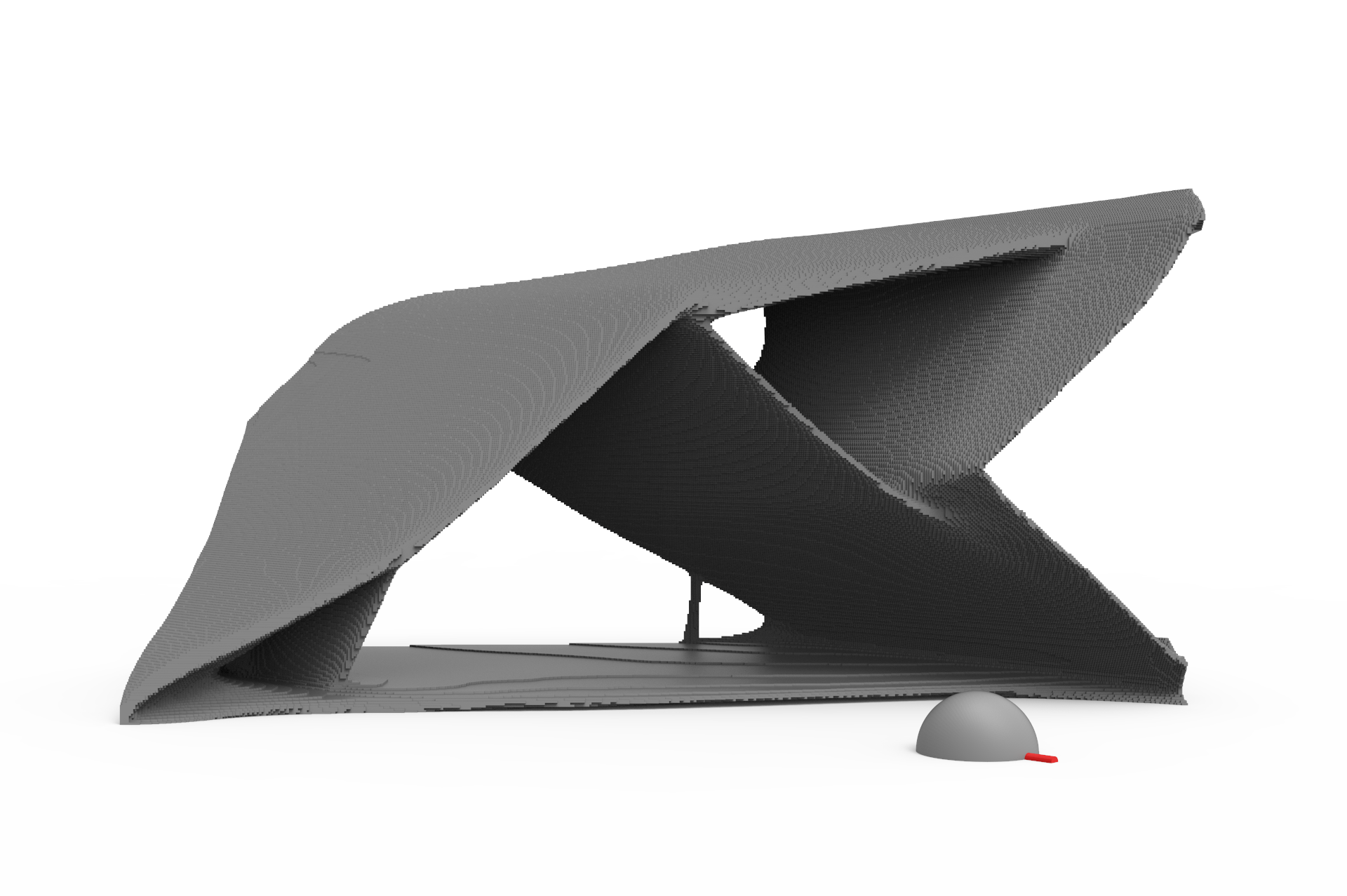}
    \end{subfigure}
    \begin{subfigure}{0.33\textwidth}
        \includegraphics[width=\linewidth]{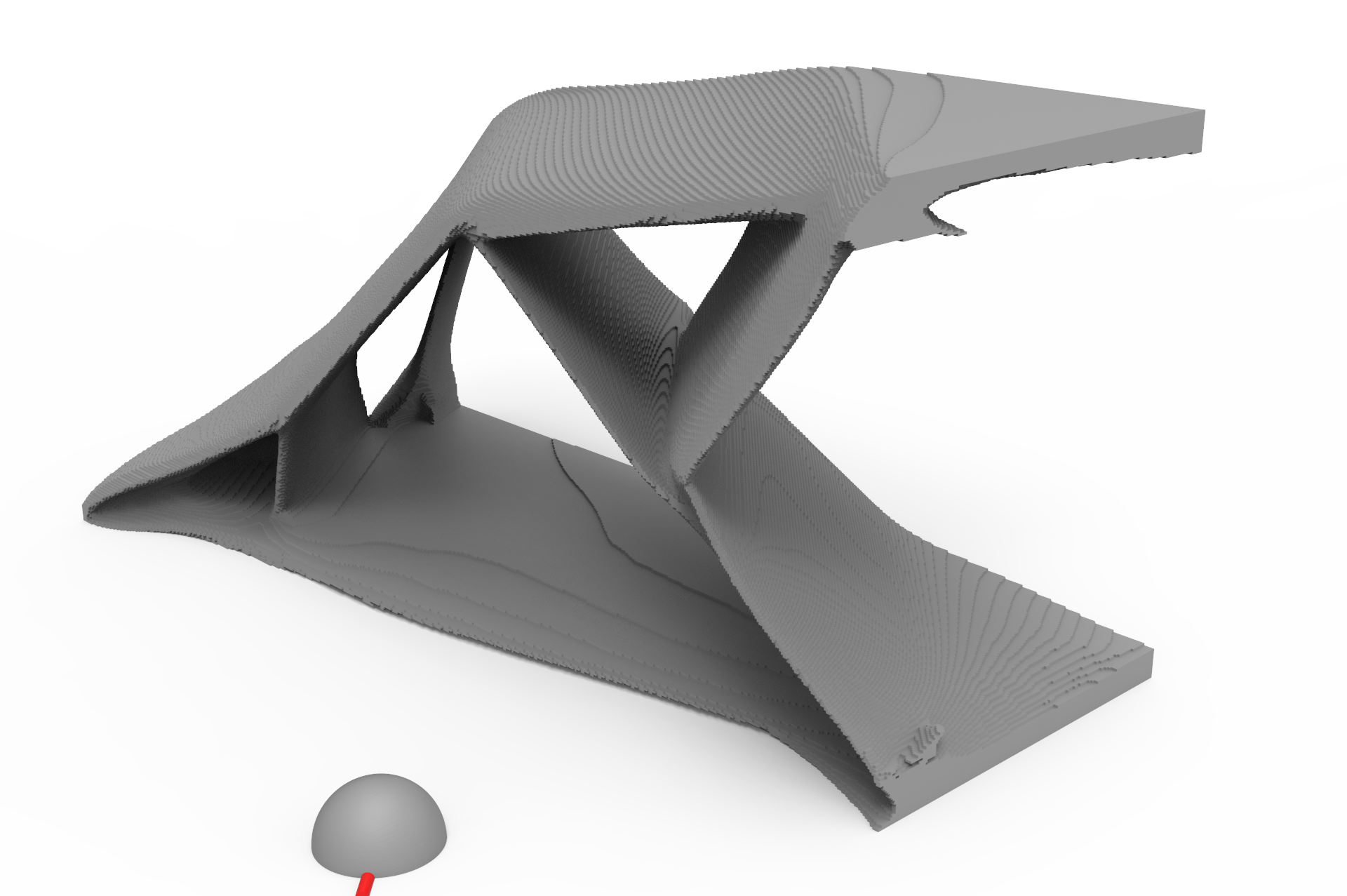}
    \end{subfigure}
    \caption{Cantilever beam with a single milling direction with a $45$ degree angle in the plane. $C=\num{1.62e7}$. }
    \label{fig:cantilever_45}
\end{figure*}

A beam optimized with a similar direction, which has been rotated 45 degrees towards the supporting plane, is seen in \cref{fig:cantilever_45}. The resulting cross section of the structure is seen to resemble the one from \cref{fig:cantilever_z}. However, the skewed milling direction affects the design drastically, both qualitatively and in terms of compliance value. It can be observed that the angle forces material to be placed at locations that are unfavorable, as none of the other designs make use of them (most notably the upper left side over the loaded line).

\begin{figure*}[htb]
    \centering
    \begin{subfigure}{0.33\textwidth}
        \includegraphics[width=\linewidth]{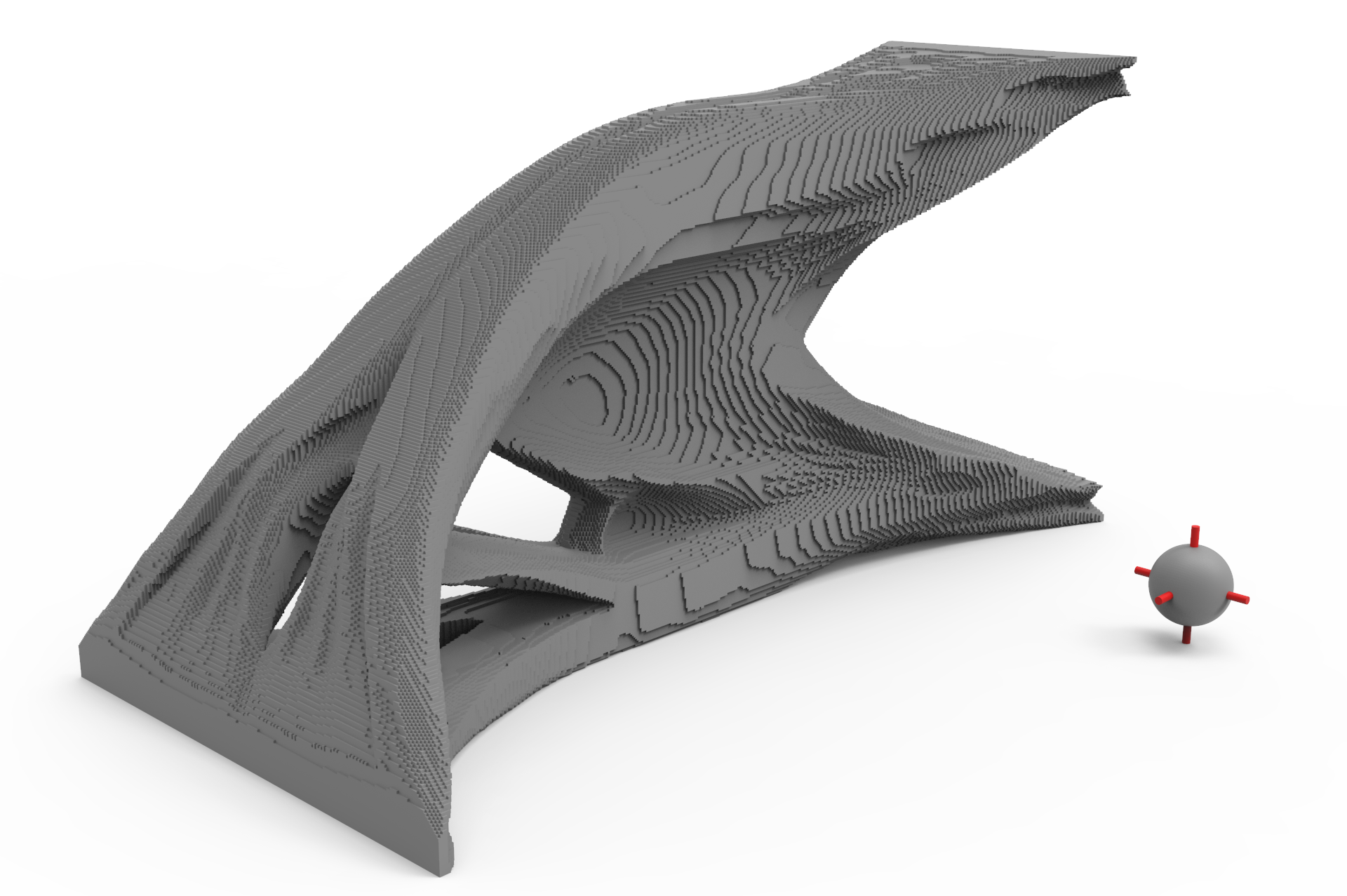}
    \end{subfigure}
    \begin{subfigure}{0.33\textwidth}
        \includegraphics[width=\linewidth]{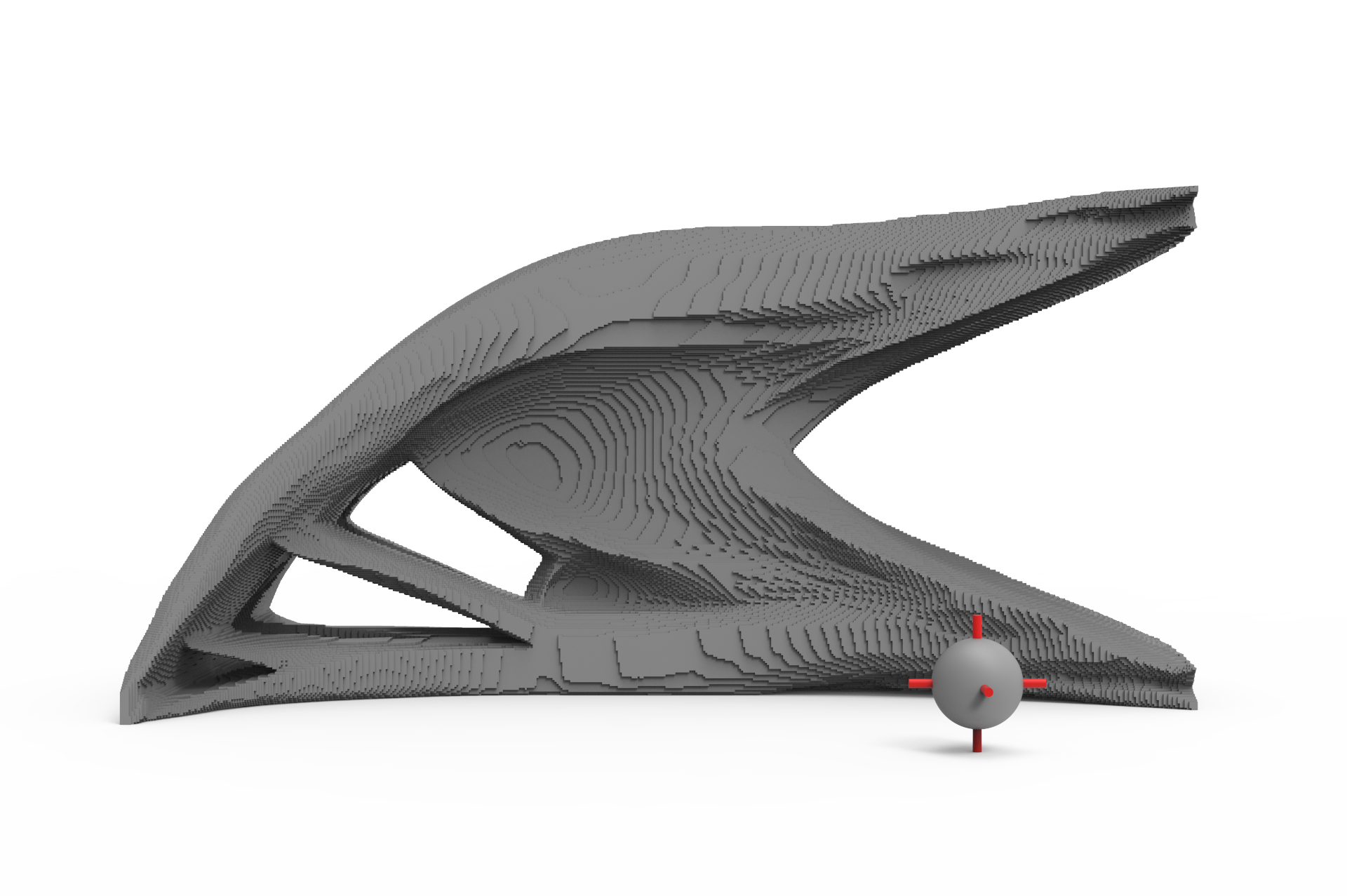}
    \end{subfigure}
    \begin{subfigure}{0.33\textwidth}
        \includegraphics[width=\linewidth]{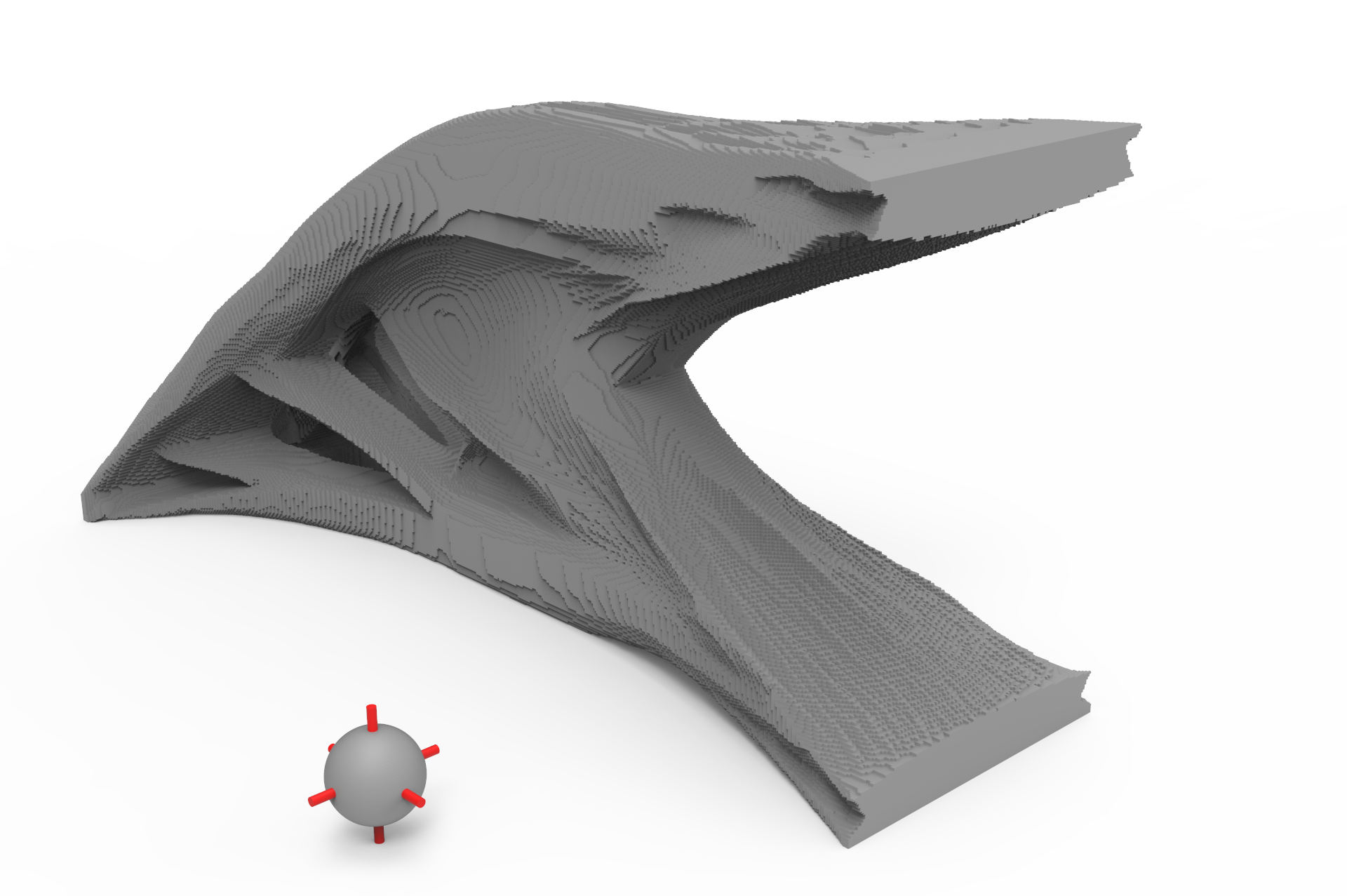}
    \end{subfigure}
    \caption{Cantilever beam with six milling directions, normal to the bounding box surfaces. $C=\num{1.33e7}$.}
    \label{fig:cantilever_6}
\end{figure*}

The three previously shown cases with directions normal to the domain boundaries are combined to a single case, with six milling directions, where the milling directions follow the cartesian coordinate system, in positive and negative directions. The design obtained with six milling directions is shown in \cref{fig:cantilever_6}. The obtained design resembles the one obtained with no milling constraint, \cref{fig:cantilever_ref}. However, the hollow interior of the side walls from the reference is infeasible with the milling filter, and has hence been replaced by holes in the structure connecting the top and bottom. 

\begin{figure*}[htb]
    \centering
    \begin{subfigure}{0.33\textwidth}
        \includegraphics[width=\linewidth]{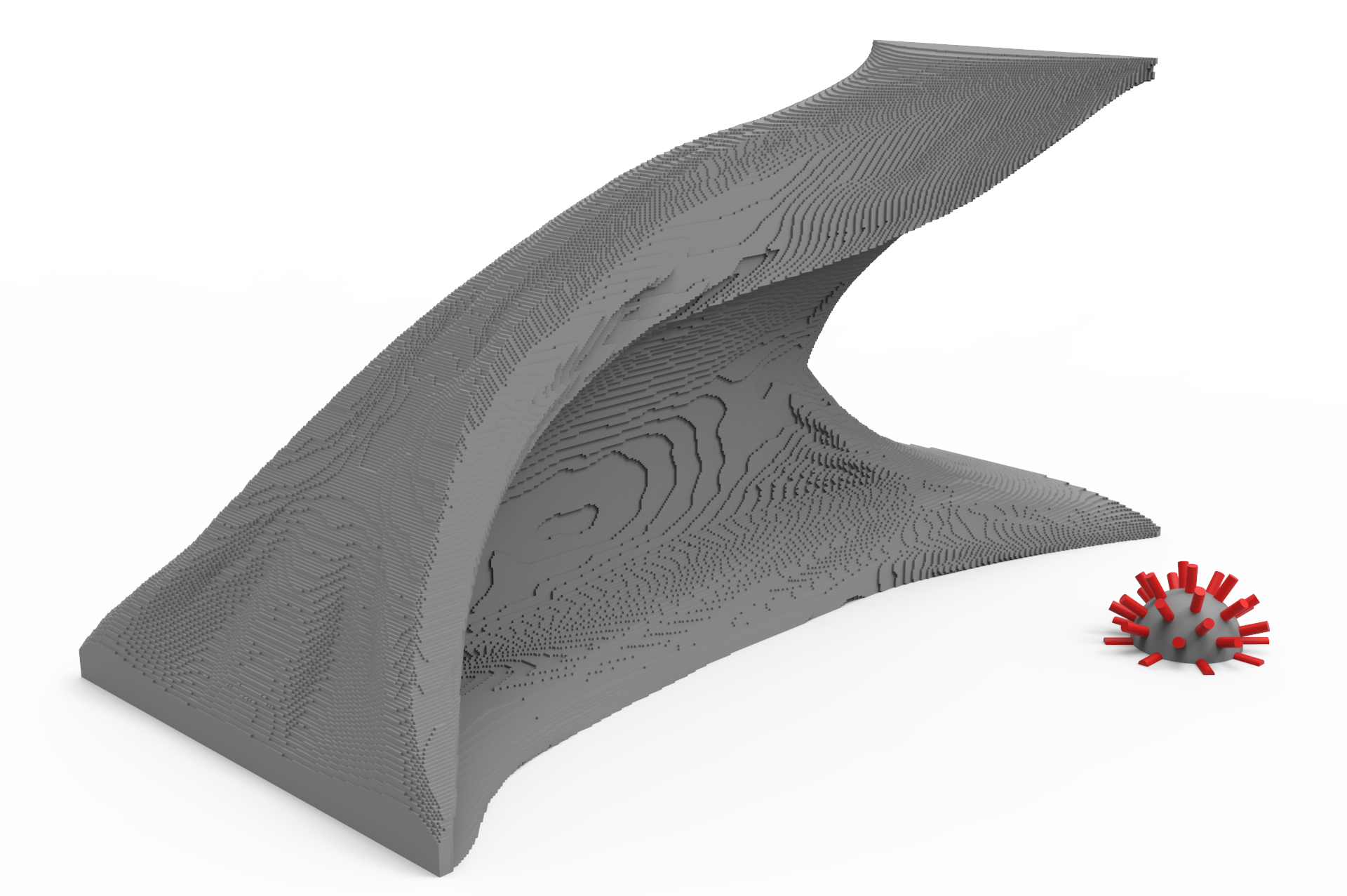}
    \end{subfigure}
    \begin{subfigure}{0.33\textwidth}
        \includegraphics[width=\linewidth]{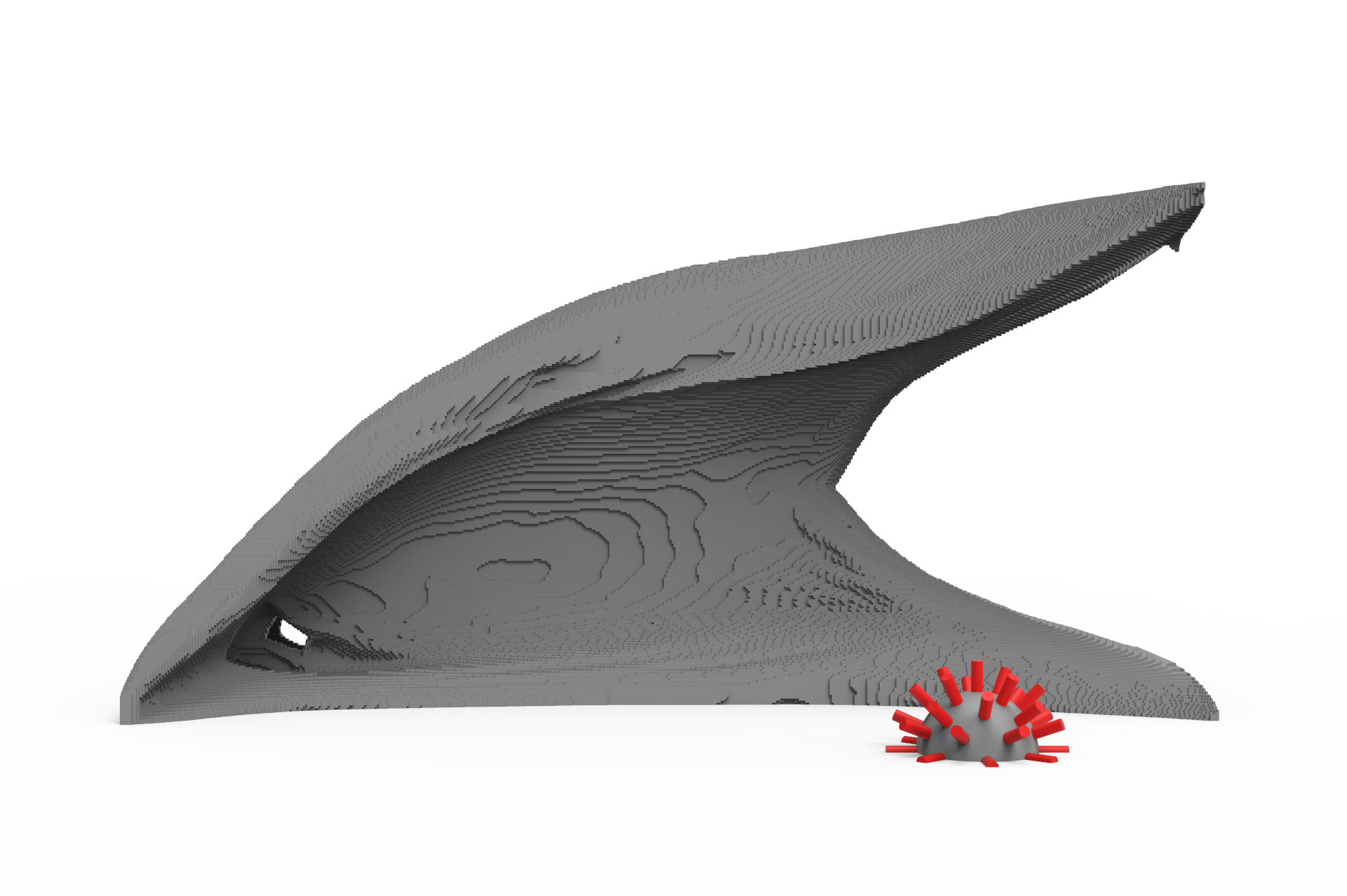}
    \end{subfigure}
    \begin{subfigure}{0.33\textwidth}
        \includegraphics[width=\linewidth]{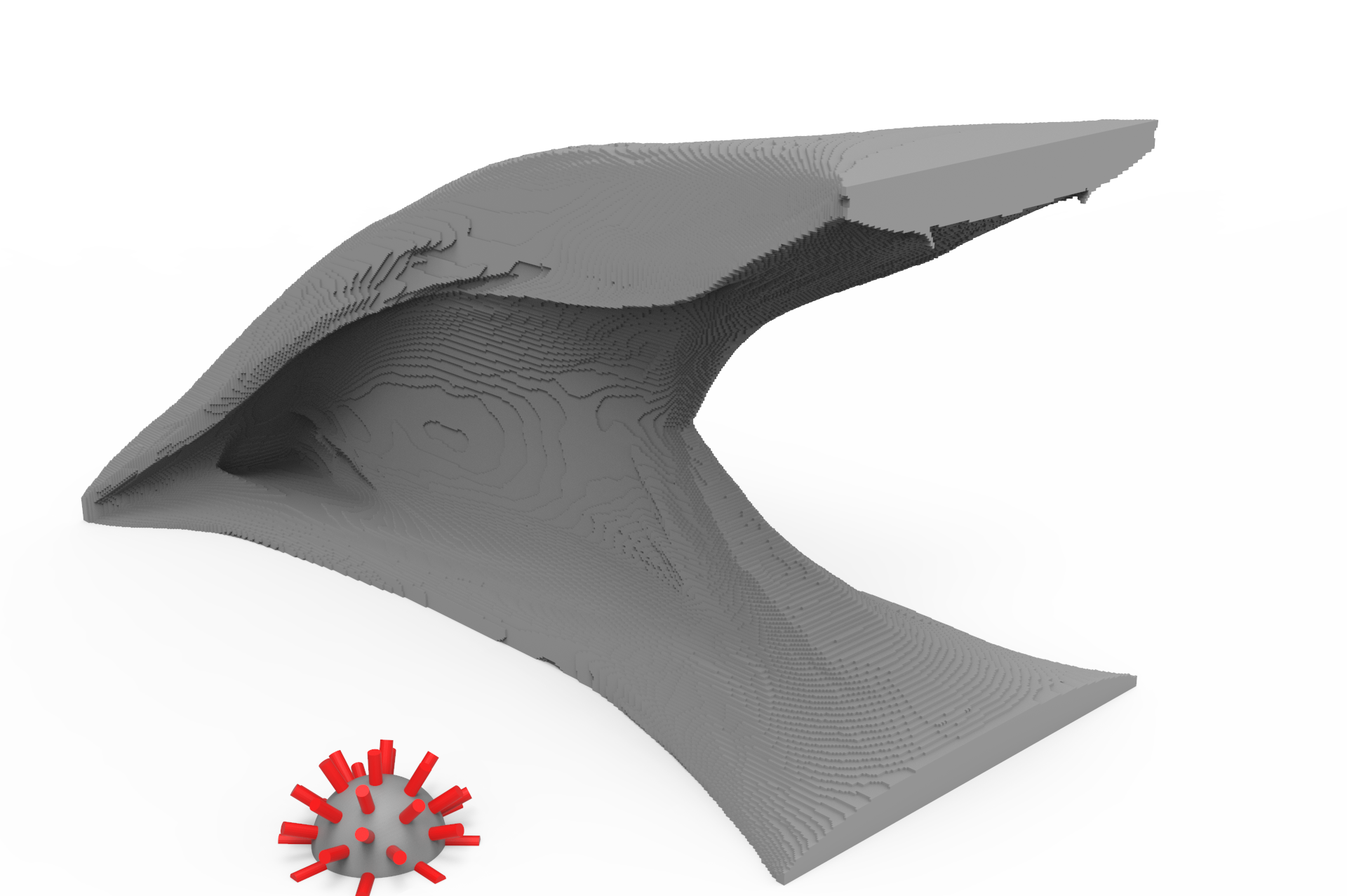}
    \end{subfigure}
    \caption{Cantilever beam with 29 milling directions, coming from direction from the northern hemisphere. $C=\num{1.32e7}$.}
    \label{fig:cantilever_29}
\end{figure*}

A cantilever beam is also optimized using a set of 29 milling directions, all coming from the northern hemisphere, as described by \cite{Langelaar2019}. The design obtained with this combination of milling directions is seen in \cref{fig:cantilever_29}. The design is very similar to the one using 6 directions seen in \cref{fig:cantilever_6}. However, most of the holes through the structure in the middle have been filled, and the thickness of that structure is varied instead. Some of the overhang near the only remaining hole next to the loaded line are only realizable due to some of the milling directions with the skewed angles.

The performance of the obtained designs is compared in \cref{tab:cantcompare}. As expected, all designs obtained with the machinability constraints perform worse than the benchmark using the robust formulation. Furthermore, it is also observed that adding more milling directions improves the performance of the design, as this also increases the design freedom of the optimization process.

\begin{table}[htb]
    \centering
    \caption{Comparison of performance of the 3D cantliver beam designs, obtained with different milling directions and comparison with reference design obtained without milling.}
    \label{tab:cantcompare}
    \begin{tabular}{lrrr}
        \hline
        Figure &  Num. tools & $C$ & $C/C_\mathit{ref}$\\\hline
        \ref{fig:cantilever_ref} & 0 & \num{1.15e7} & 1.0\\
        \ref{fig:cantilever_x} & 2 & \num{1.37e+07} & $1.192$\\
        \ref{fig:cantilever_y} & 1 & \num{1.53e7} & $1.331$\\
        \ref{fig:cantilever_z} & 1 & \num{1.37e7} & $1.194$\\
        \ref{fig:cantilever_45} & 1 & \num{1.62e7} & $1.408$\\
        \ref{fig:cantilever_6} & 6 & \num{1.33e7} & $1.159$\\
        \ref{fig:cantilever_29} & 29 & \num{1.32e7} & $1.148$
    \end{tabular}
\end{table}

As in the two dimensional results, discussed in \cref{sec:2Dexamples}, it is again seen that the chosen milling directions have a large influence influence on the performance of the obtained designs. If one wishes to only consider only a single milling direction, the direction should be chosen carefully, as it can have a large influence on the performance of the optimized design. A surprising discrepancy in performance is observed between the designs obtained with the milling directions normal to the supported plane from \cref{fig:cantilever_x} and the one obtained with the milling direction from above from \cref{fig:cantilever_y}. The design with the milling direction from above performs $\approx 11\%$ worse, even though the obtained designs are somewhat similar, qualitatively. 

\subsection{Three dimensional GE bracket examples}
\label{sec:bracket}
\begin{figure}[htb]
    \centering
    \includegraphics[width=0.4\textwidth]{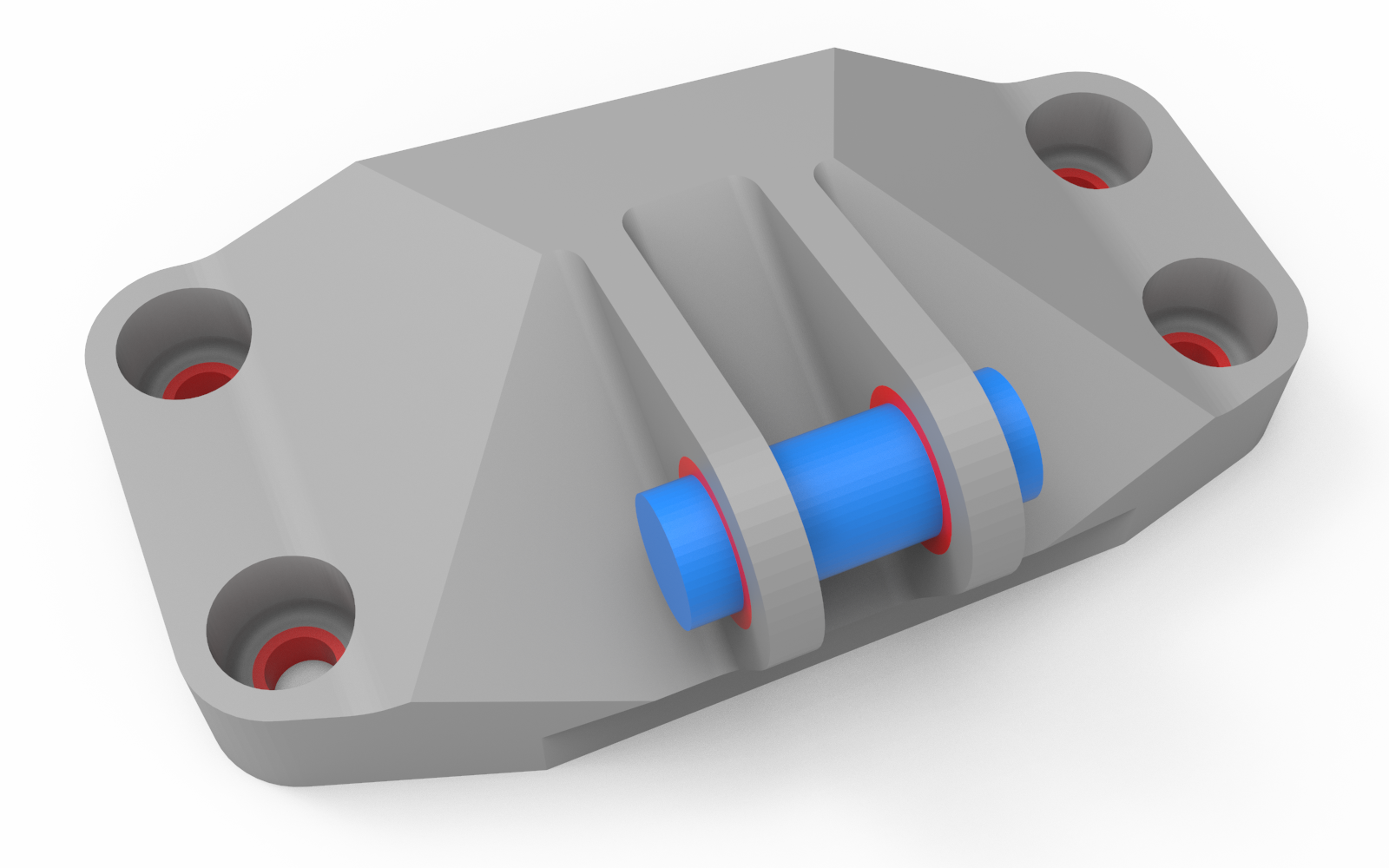}
    \caption{Geometry of GE Jet Engine Bracket example. The pin with increased stiffness is shown in blue, while the passive rings which are kept solid are shown in red.}
    \label{fig:bracketGeometry}
\end{figure}

To demonstrate the real-world possibilities of the milling filter, the GE jet engine bracket \cite{bracketwebpage} is used as an industrial example, which is shown in \cref{fig:bracketGeometry}. The original bracket geometry, material properties, and load cases are adapted from \cite{bracketwebpage,bracketarticle}, with slight simplifications. The pinned boundary interfaces are modeled by clamping the inner surface of the bolt interface. The rigid pin is included in the model using 10 times the Young's modulus of the used material. All six interfaces in the bracket model are assigned a passive solid ring, in order to ensure that an interface exists. The compliance of the structure is minimized subject to a constraint of volume fraction $V^*=0.137$ in the design domain, corresponding to a total bracket weight of \SI{300}{g} for the design and the passive solid rings.

The minimum Young's modulus $E_\mathit{min}$ is updated using a simple continuation scheme, as also described in \cref{sec:cantilever}. The optimization begins with $E_\mathit{min} = \num{1e-3} E_\mathit{max}$ which is decreased by an order of magnitude every 15 iterations until reaching $E_\mathit{min} = \num{1e-6} E_\mathit{max}$ at iteration 45.

The large size of the computational domain ($\num{160}\times\num{110}\times\num{90}$mm) allows to use a lower Peclet number than in the previous examples. This is due to the Peclet number being a function of the characteristic domain length. 

It should be noted that the bracket design domain is non-convex due to the two flanges which support the bolt. This non-convexity has some important implications for the solution of the advection-diffusion equation on the bracket domain. The boundaries will not transport information across the gaps in the design domain, treating every domain surface as an outer surface from where a tool might remove material. This will have an effect, as placing material in one flange, will not force the formulation to place material in the other flange due to the milling filter. As the flanges represent a small part of the domain, this error is limited, although present.

As discussed in \cref{sec:shadow}, the surfaces which separate the passive rings from the design domain are modeled using a Dirichlet boundary condition in the advection-diffusion equation in order to ensure that the passive rings also introduce material downstream into the design domain. Futhermore, it should be noted that while the stiffened pin, shown in blue in \cref{fig:bracketGeometry}, is included in the model, it is omitted from the visualization.

The bracket examples are solved using a mesh of 64 million hexahedral elements. The results are computed on 20 nodes at the DTU Sophia cluster. The computations take between 17 and 31 hours, where the milling filters and their adjoint problems account for up to 40\% of the runtime during a design iteration in the worst case with five milling directions. While this percentage seems like a high cost of filtering, it should be noted that the total time of a design iteration at this point is 6 minutes for a design iteration, including solving four linear elastic problems with 193M dof, five advection-diffusion equations, and ten adjoint advection-diffusion problems.

\begin{figure*}[htb]
    \centering
    \begin{subfigure}[t]{0.33\textwidth}
        \includegraphics[width=\linewidth]{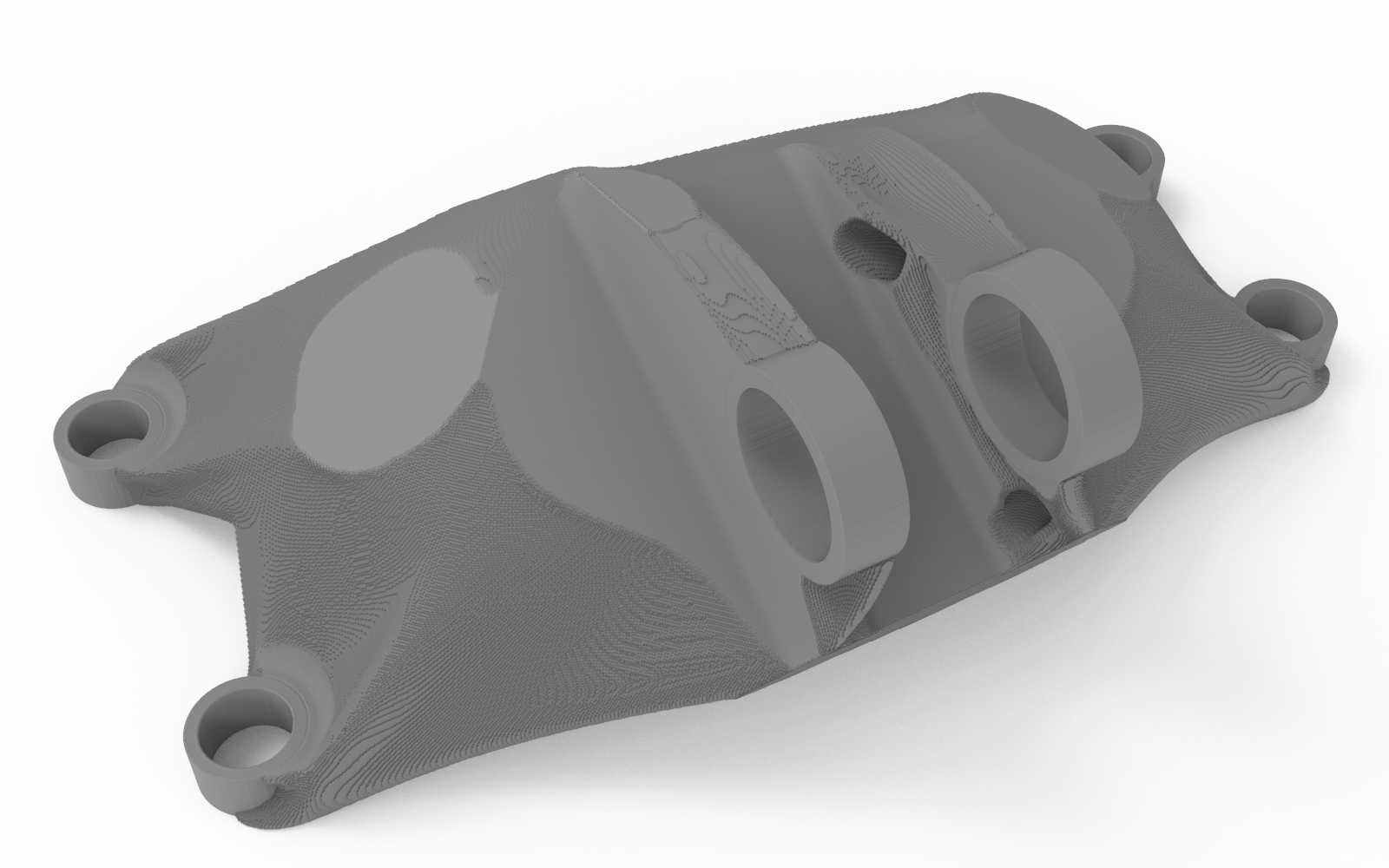}
    \end{subfigure}
    \begin{subfigure}[t]{0.33\textwidth}
        \includegraphics[width=\linewidth]{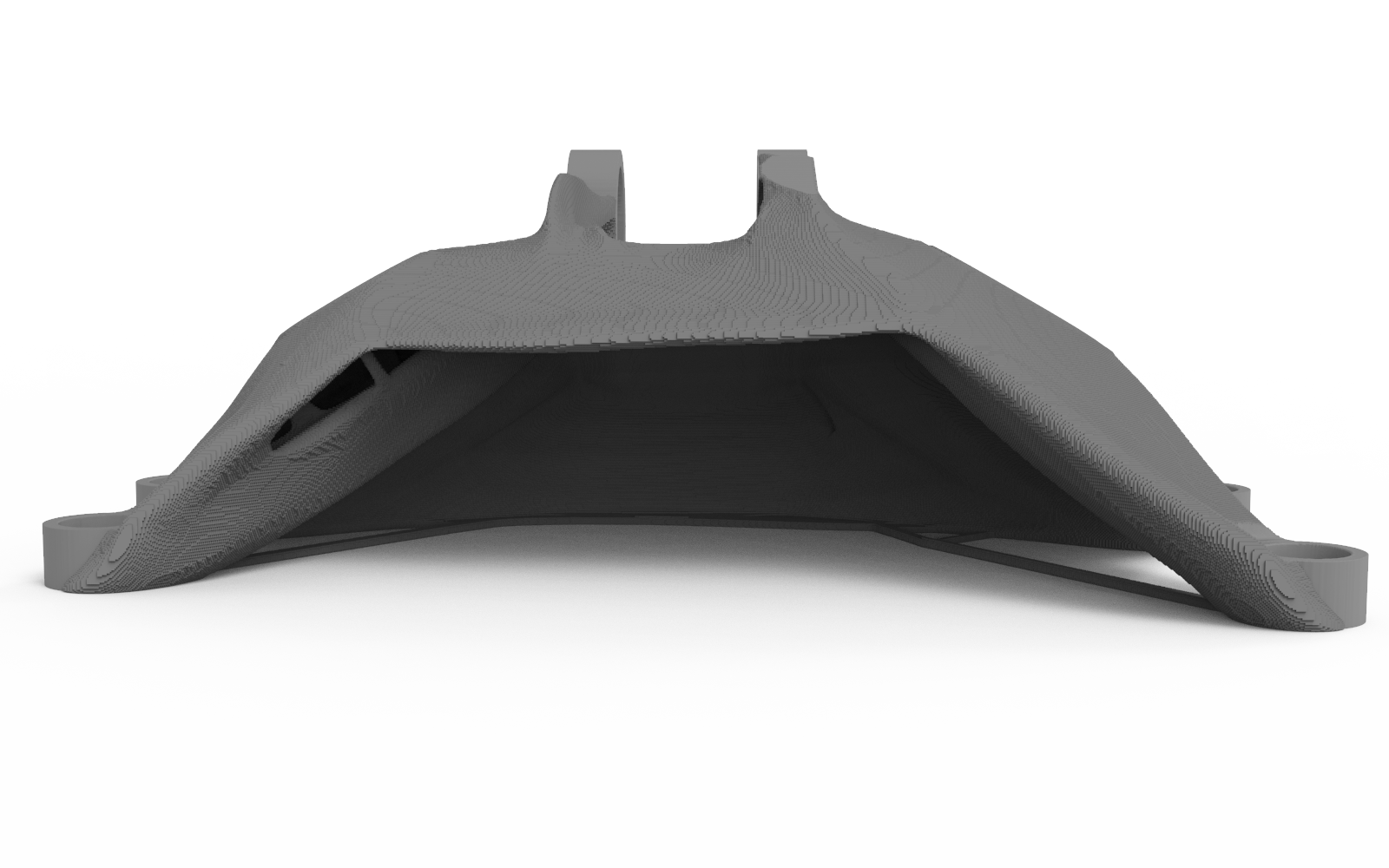}
    \end{subfigure}
    \begin{subfigure}[t]{0.33\textwidth}
        \includegraphics[width=\linewidth]{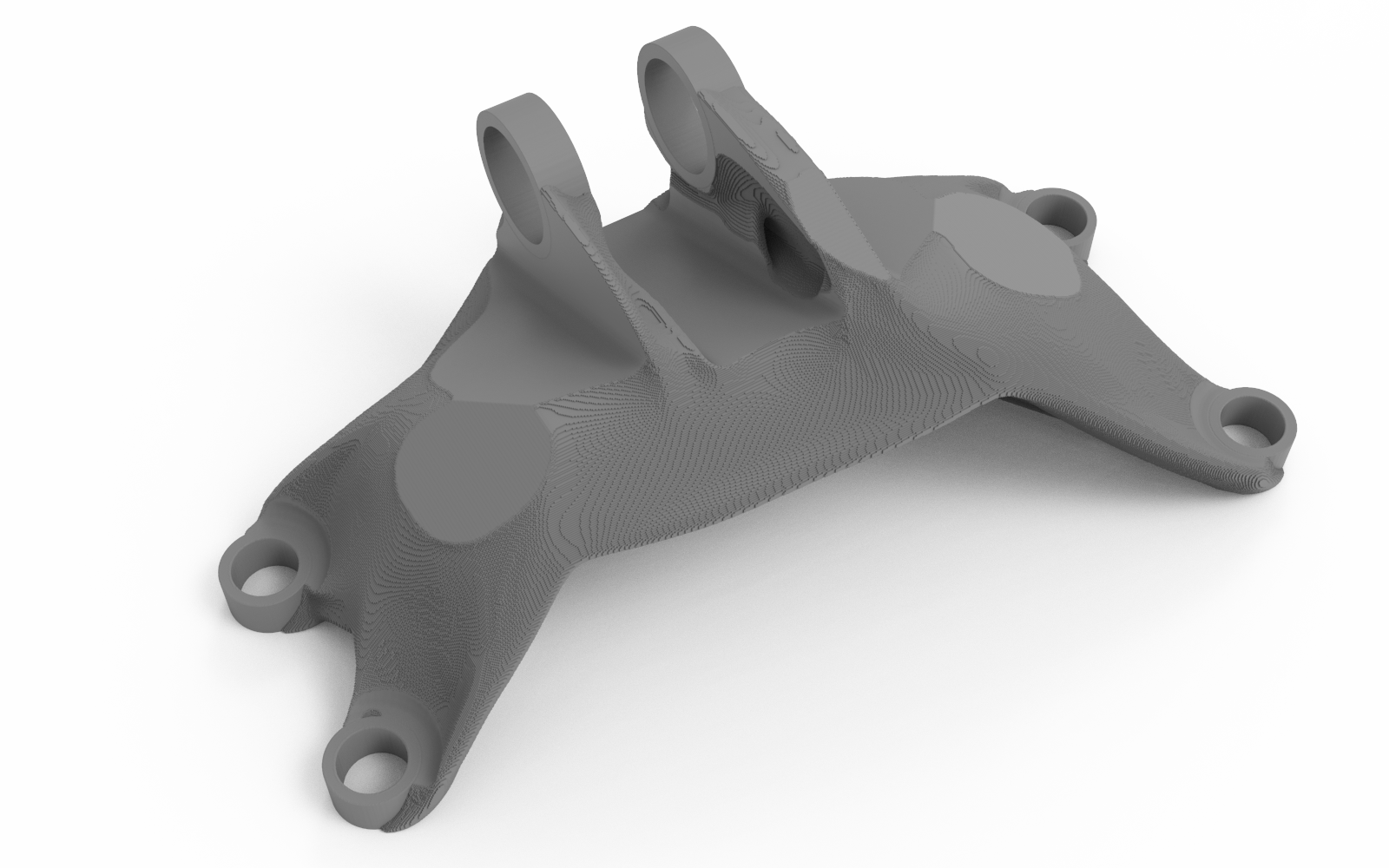}
    \end{subfigure}
    \caption{Reference design of the bracket example with Robust formulation and consistent boundary condition for PDE-filter. $C=\SI{4.21e7}{J}$.}
    \label{fig:bracketRef}
\end{figure*}

\begin{figure*}[htb]
    \centering
    \begin{subfigure}[t]{0.33\textwidth}
        \includegraphics[width=\linewidth]{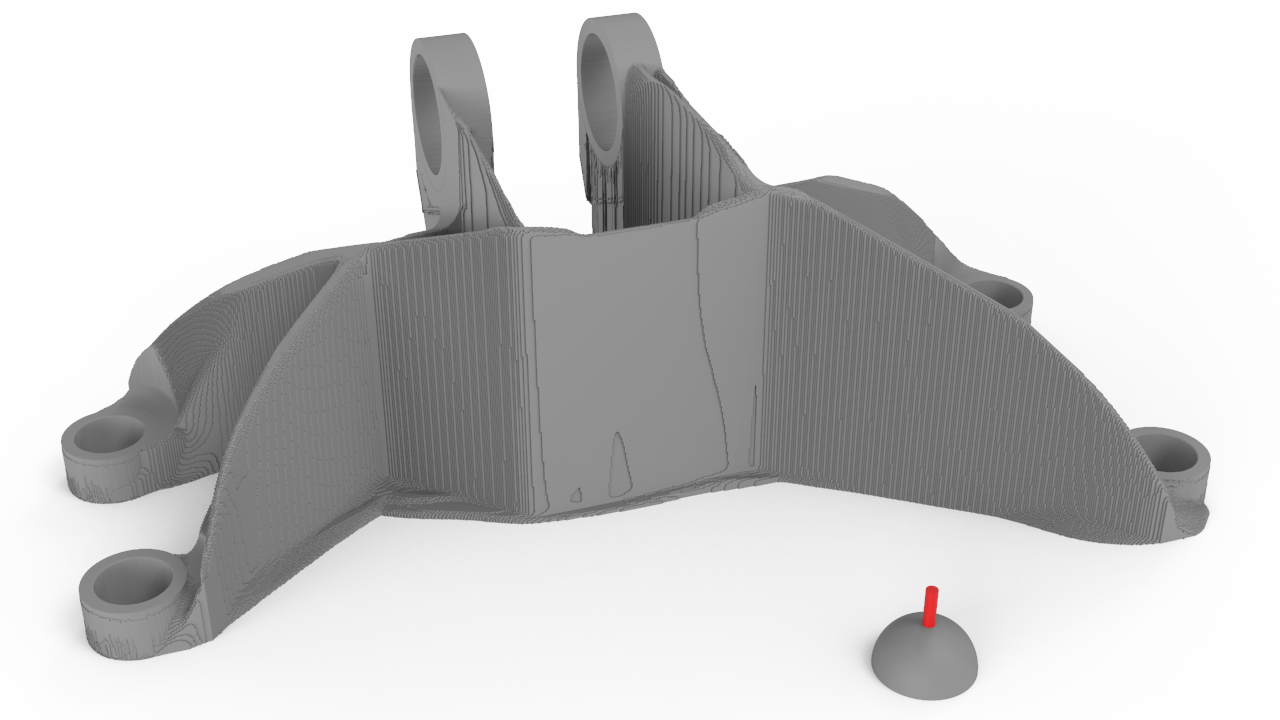}
    \end{subfigure}
    \begin{subfigure}[t]{0.33\textwidth}
        \includegraphics[width=\linewidth]{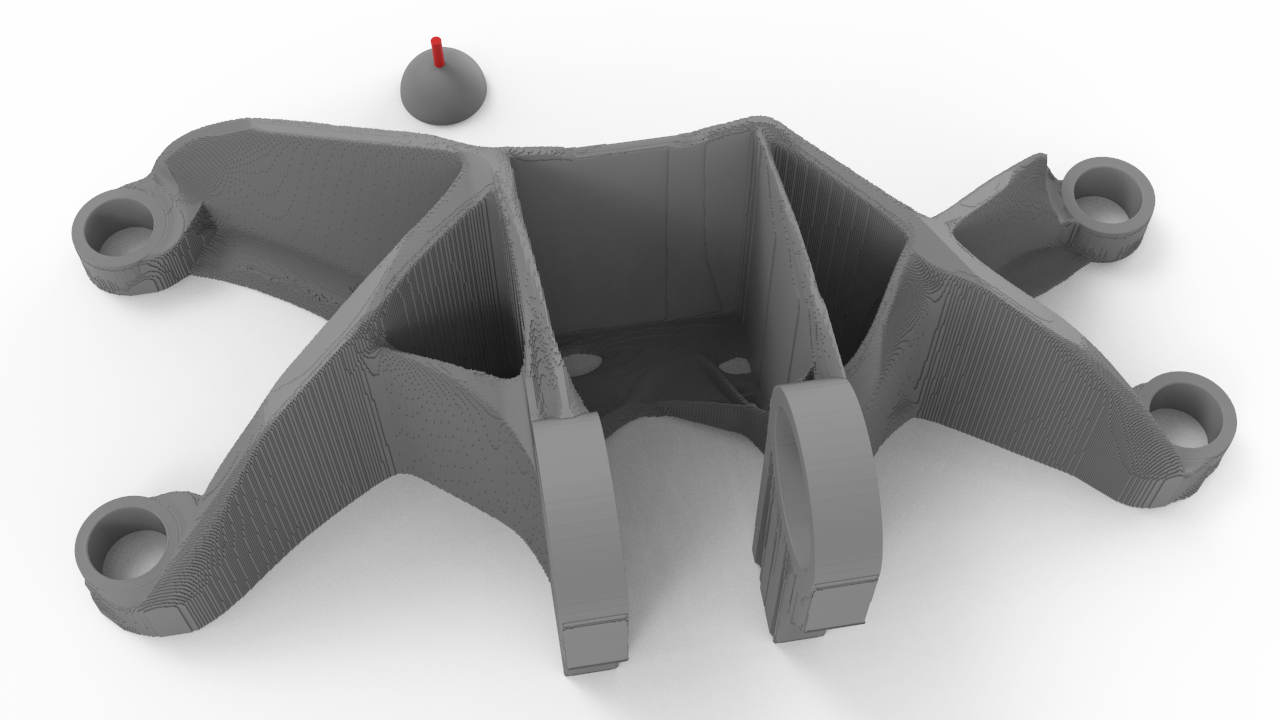}
    \end{subfigure}
    \begin{subfigure}[t]{0.33\textwidth}
        \includegraphics[width=\linewidth]{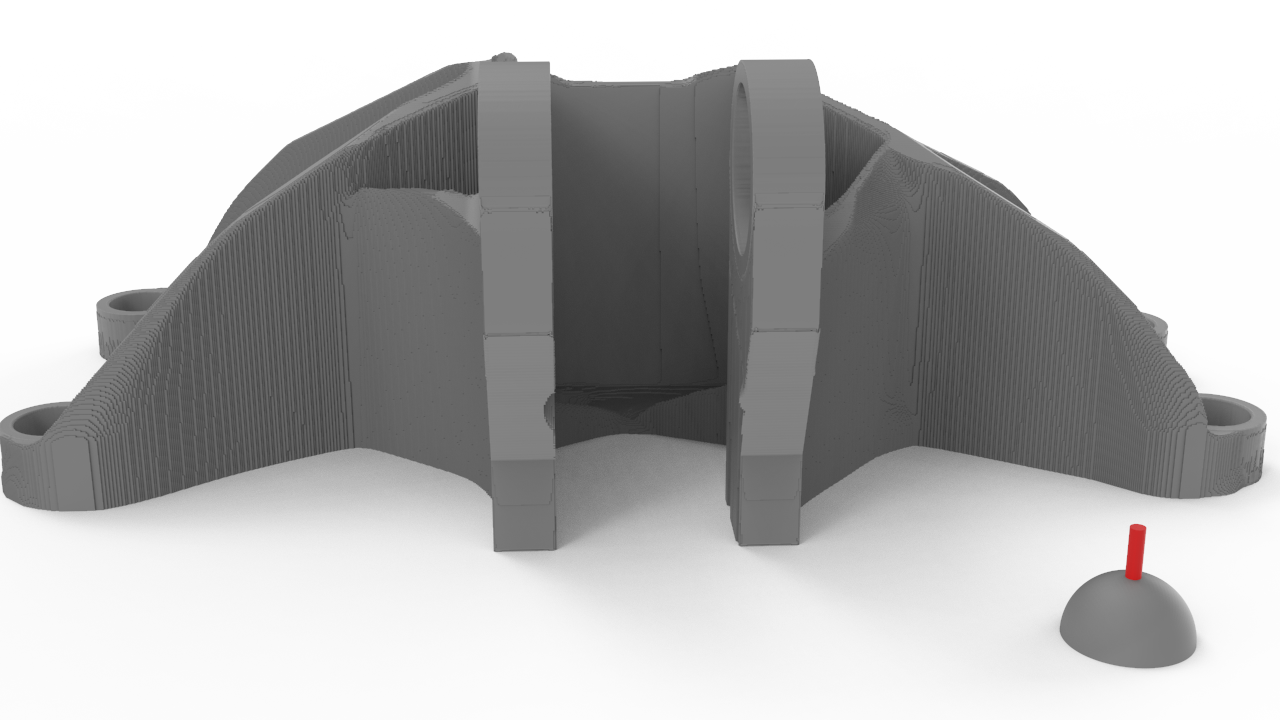}
    \end{subfigure}
    \caption{Design of the bracket example with one milling direction in the z-axis. $C=\SI{9.28e7}{J}$.}
    \label{fig:bracketOne}
\end{figure*}

A reference example is included for the bracket in order to compare the resulting structures and compliance values. The reference example is computed using the robust formulation \cite{Wang2011} with $\beta$-continuation, using a PDE-based filter with consistent boundary conditions \cite{Wallin2020}, which emulate a padded domain. The robust formulation is used due to the included heavy-side filter, which results in discrete 0-1 designs, like the proposed milling filter. This allows for a more accurate comparison of compliance values, as neither method is forced to include intermediate densities in the final design. Though, it should be noted that the robust formulation imposes a length-scale on the final design, which is not present in the milling filter. The resulting reference structure is shown in \cref{fig:bracketRef}.

\begin{figure*}[htb]
    \centering
    \begin{subfigure}[t]{0.33\textwidth}
        \includegraphics[width=\linewidth]{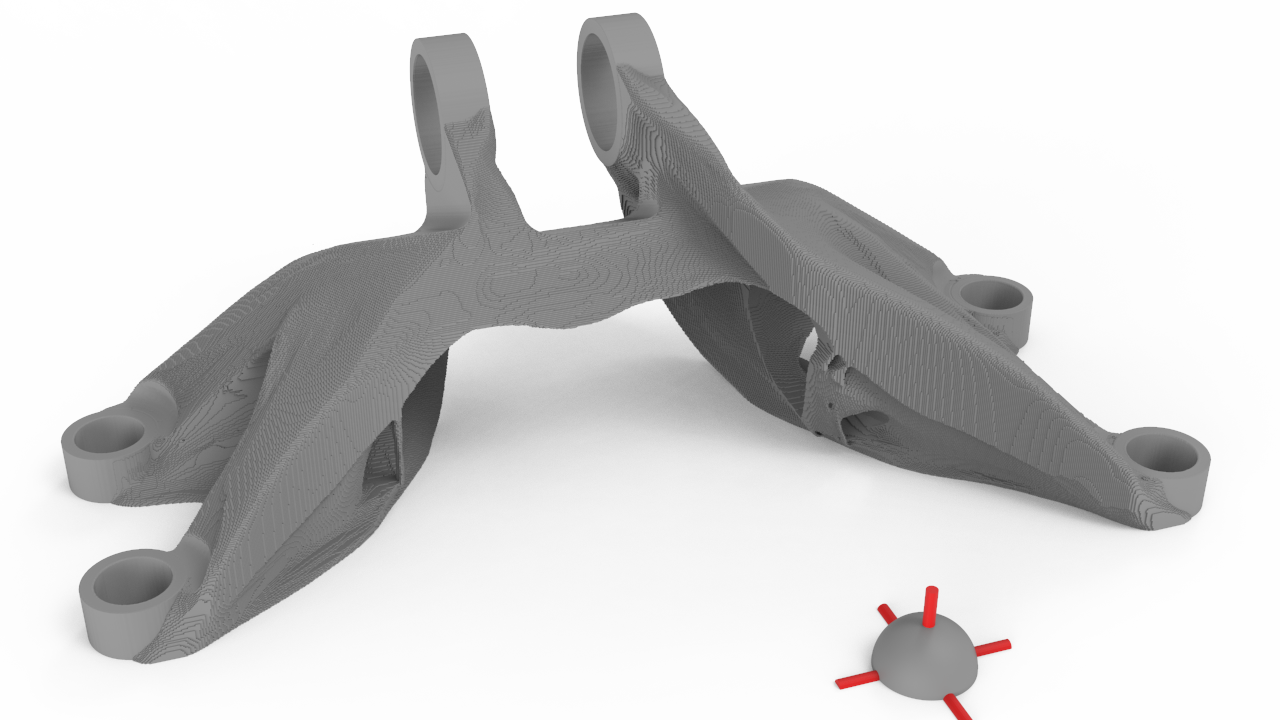}
    \end{subfigure}
    \begin{subfigure}[t]{0.33\textwidth}
        \includegraphics[width=\linewidth]{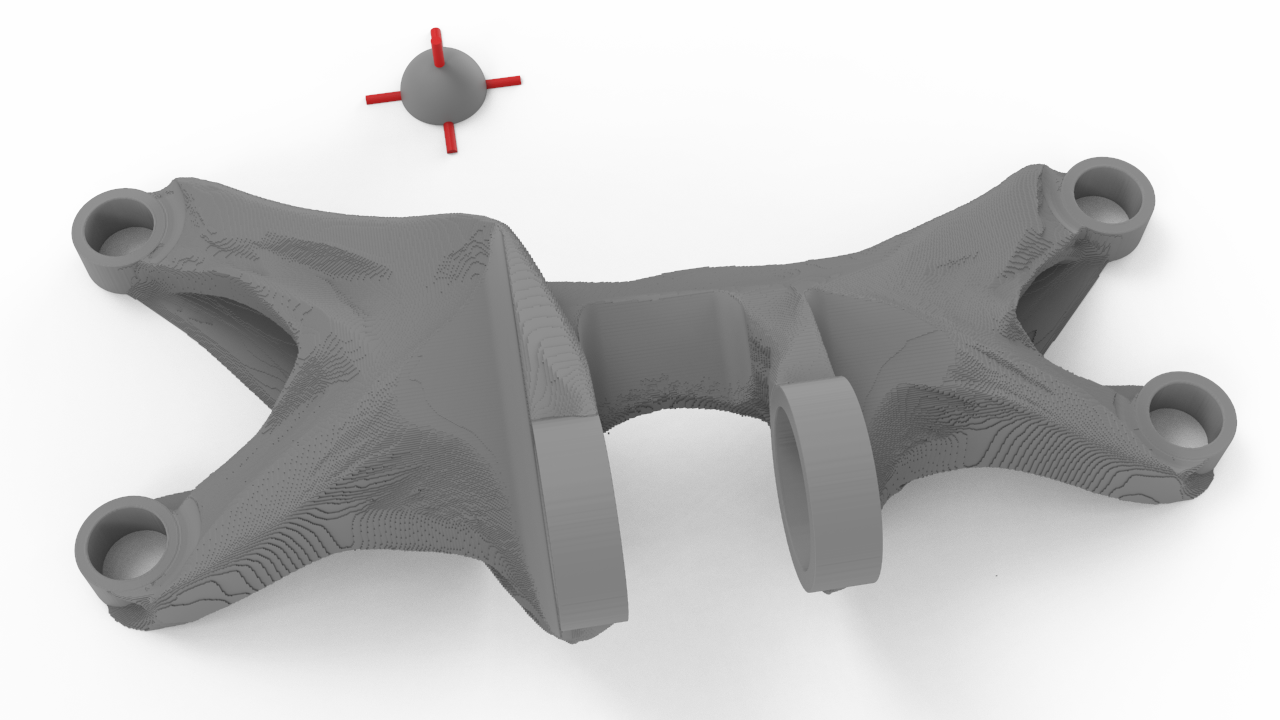}
    \end{subfigure}
    \begin{subfigure}[t]{0.33\textwidth}
        \includegraphics[width=\linewidth]{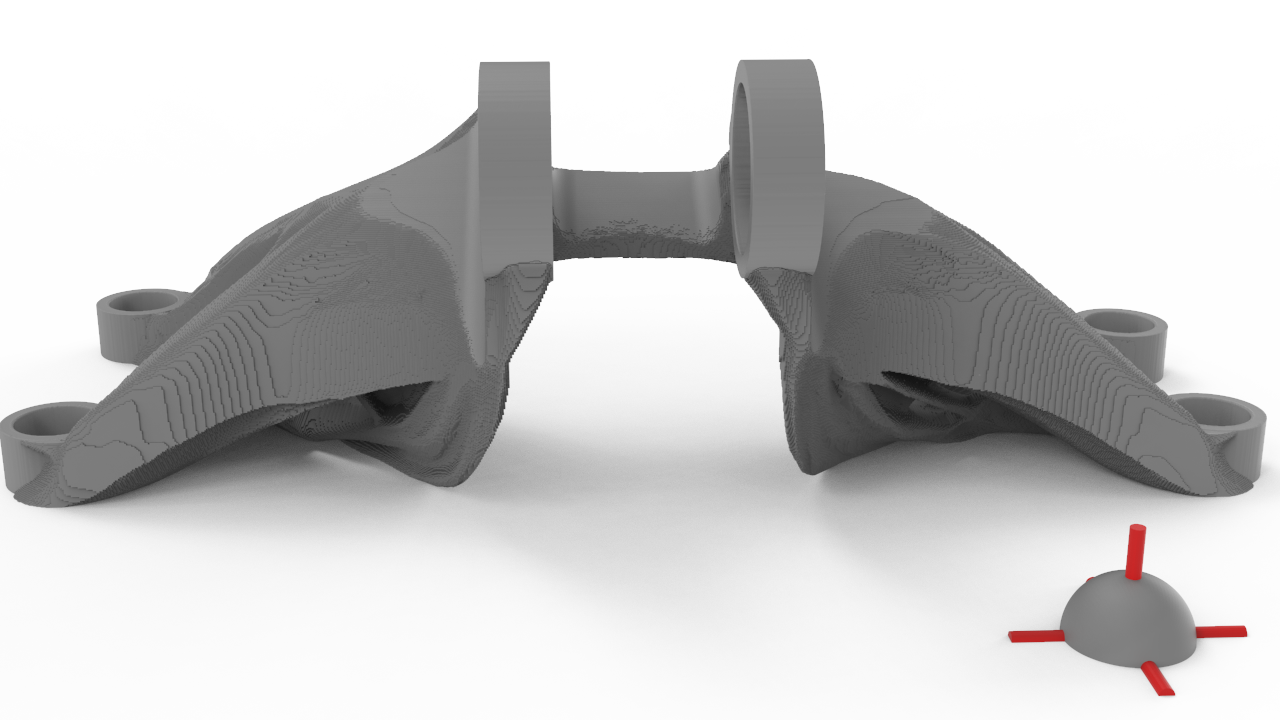}
    \end{subfigure}
    \caption{Design with five milling directions. $C=\SI{4.89e7}{J}$.}
    \label{fig:bracketFive}
\end{figure*}

\begin{figure*}[htb]
    \centering
    \begin{subfigure}[t]{0.33\textwidth}
        \includegraphics[width=\linewidth]{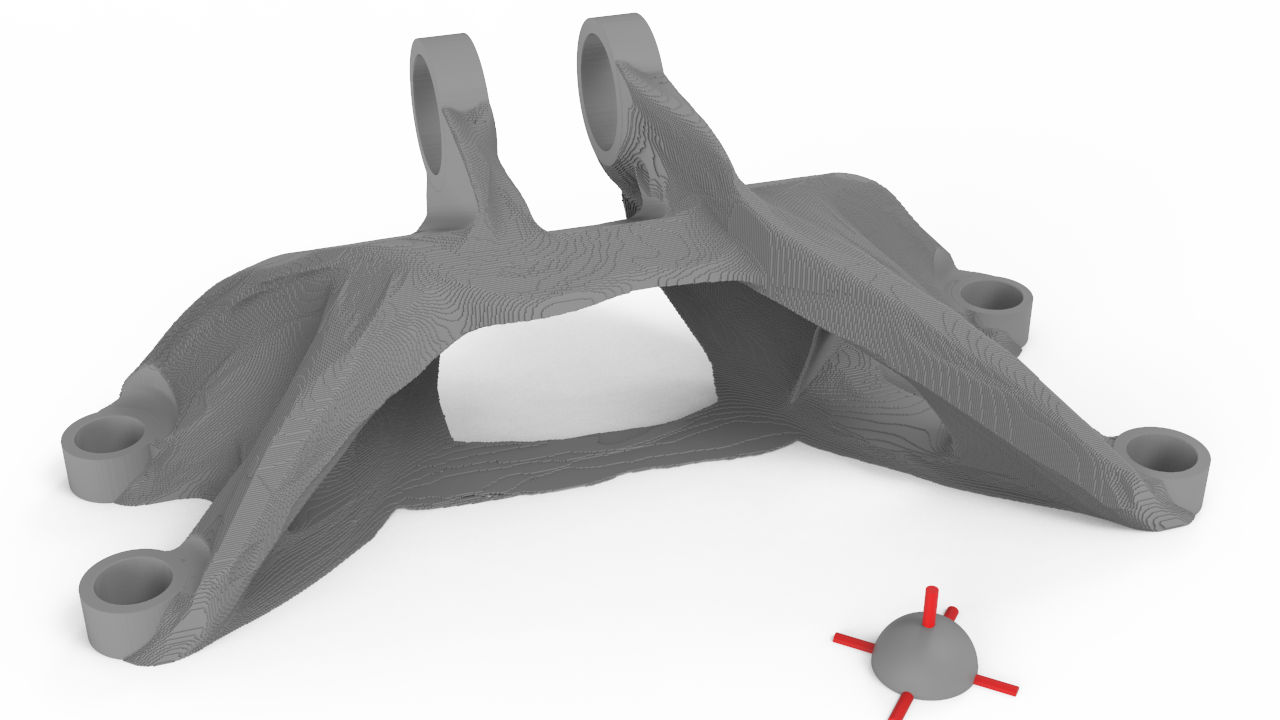}
    \end{subfigure}
    \begin{subfigure}[t]{0.33\textwidth}
        \includegraphics[width=\linewidth]{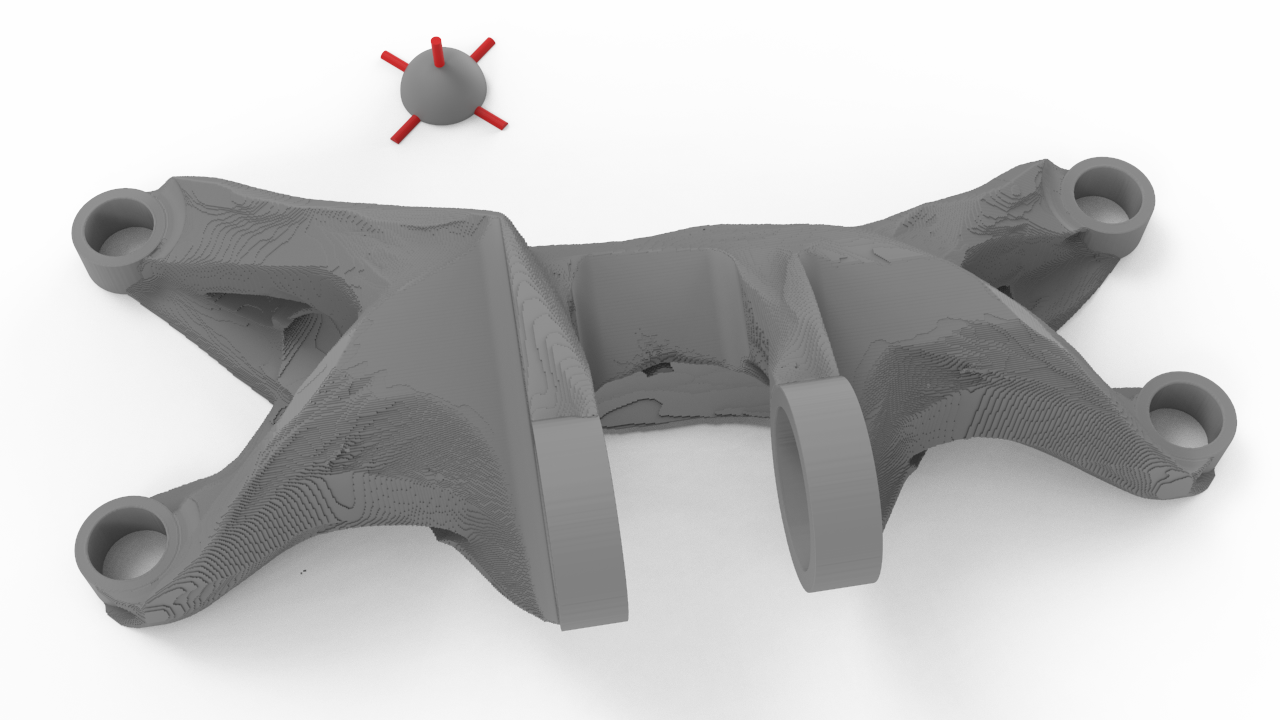}
    \end{subfigure}
    \begin{subfigure}[t]{0.33\textwidth}
        \includegraphics[width=\linewidth]{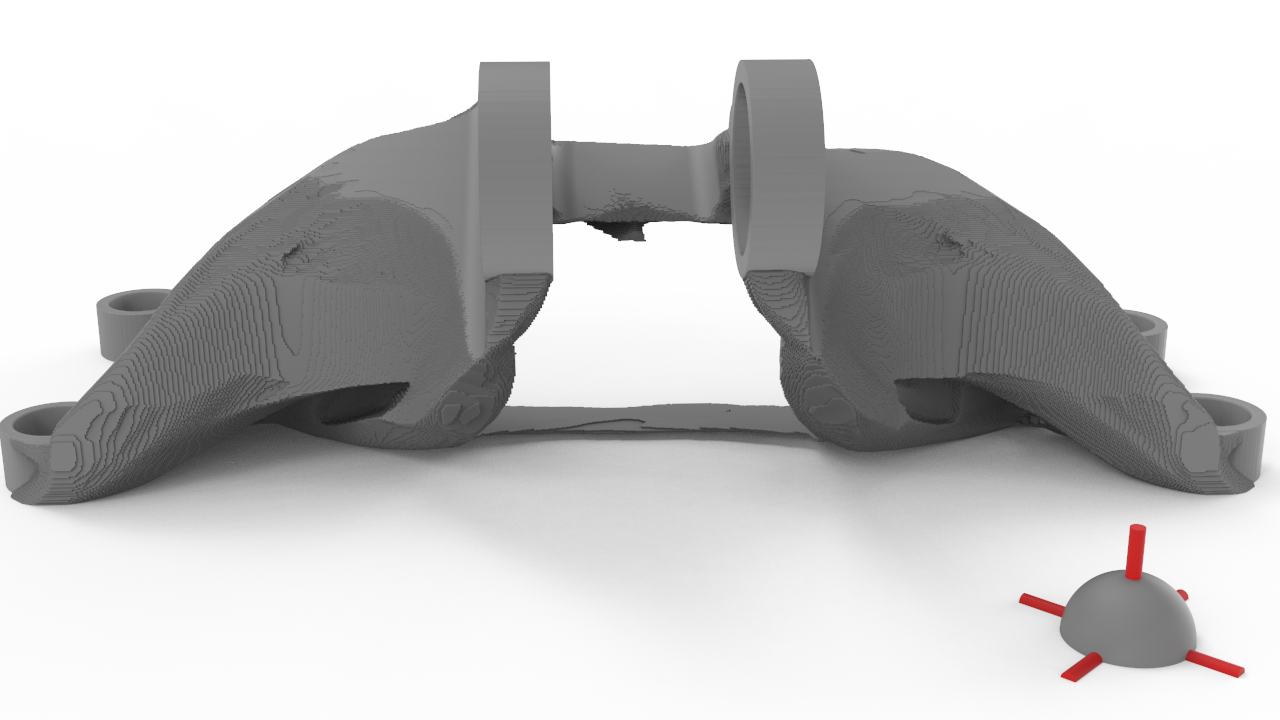}
    \end{subfigure}
    \caption{Design with five milling directions. $C=\SI{4.91e7}{J}$.}
    \label{fig:bracketFiveAlt}
\end{figure*}

\begin{figure*}[htb]
    \centering
    \begin{subfigure}[t]{0.33\textwidth}
        \includegraphics[width=\linewidth]{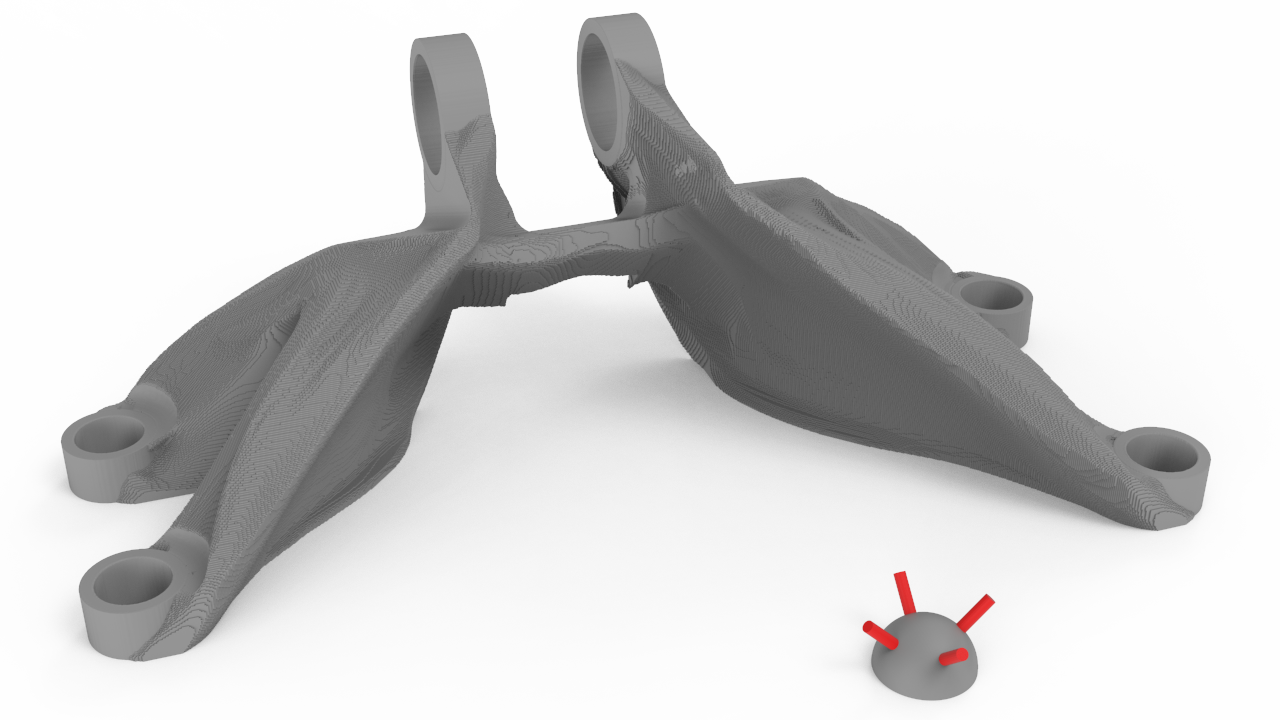}
    \end{subfigure}
    \begin{subfigure}[t]{0.33\textwidth}
        \includegraphics[width=\linewidth]{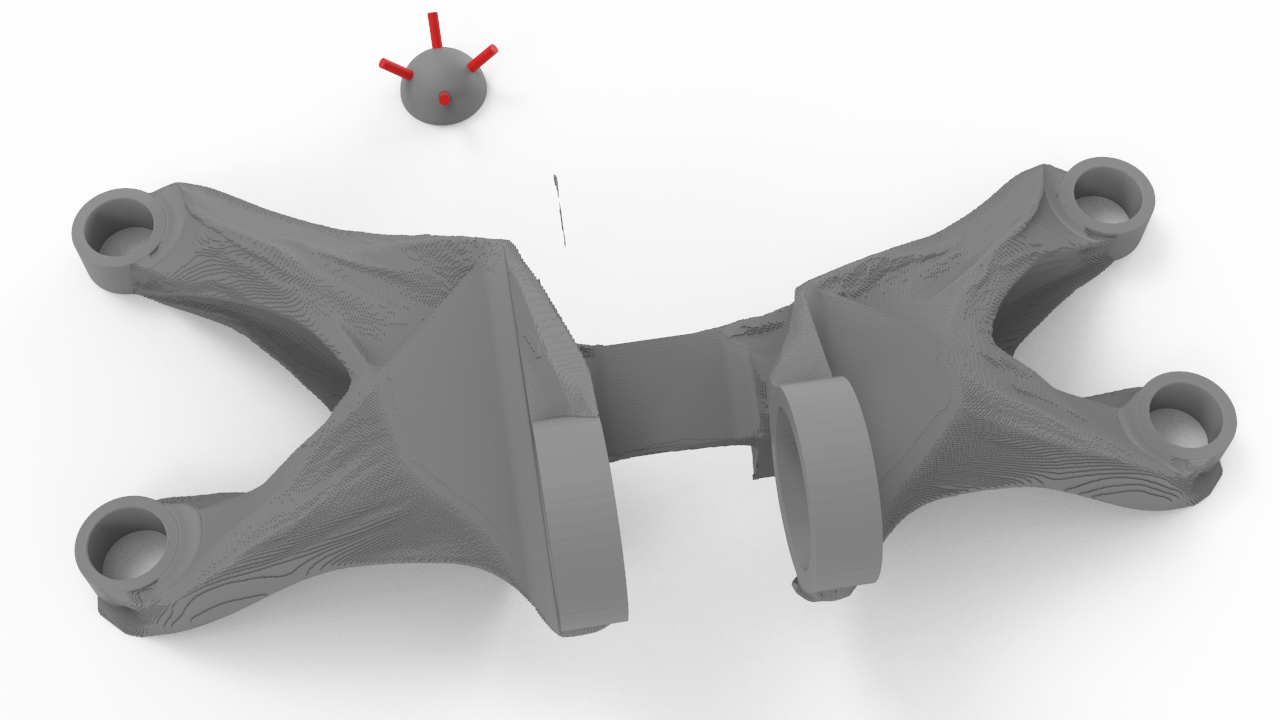}
    \end{subfigure}
    \begin{subfigure}[t]{0.33\textwidth}
        \includegraphics[width=\linewidth]{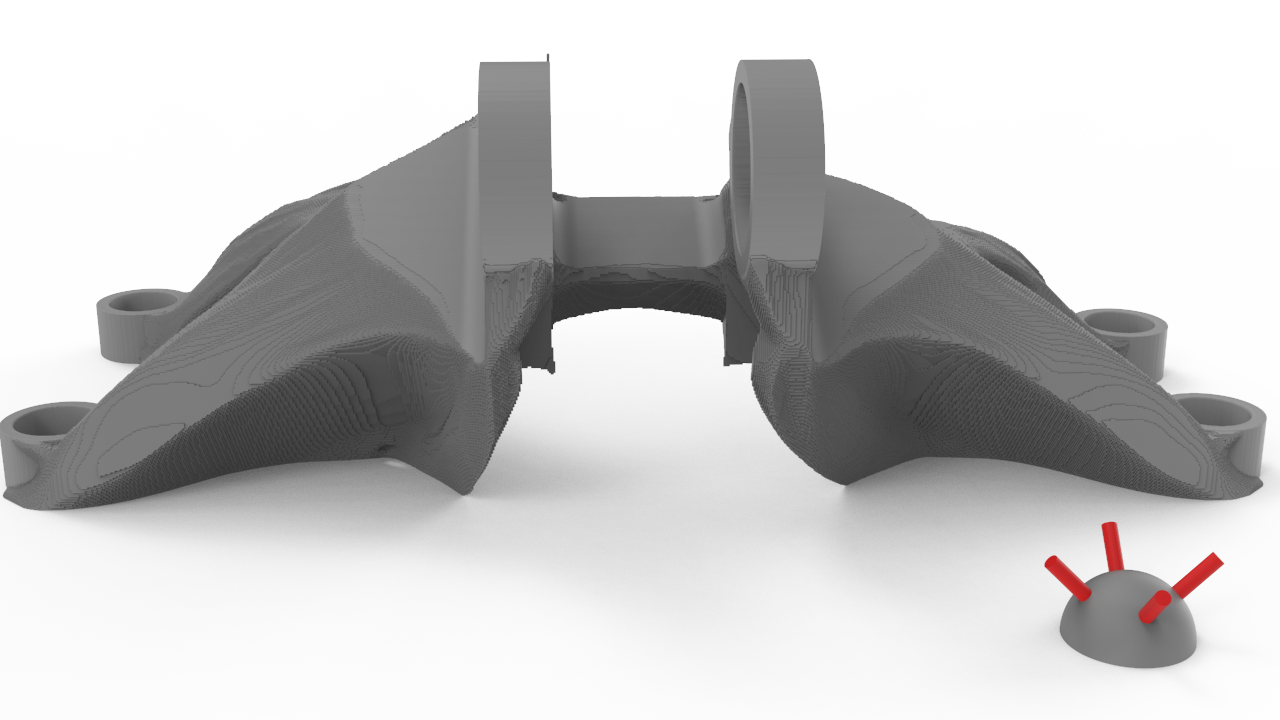}
    \end{subfigure}
    \caption{Design with four milling directions at an 45 degree angle. $C=\SI{5.12e7}{J}$.}
    \label{fig:bracketFour}
\end{figure*}

\begin{figure*}[htb]
    \centering
    \begin{subfigure}[t]{0.33\textwidth}
        \includegraphics[width=\linewidth]{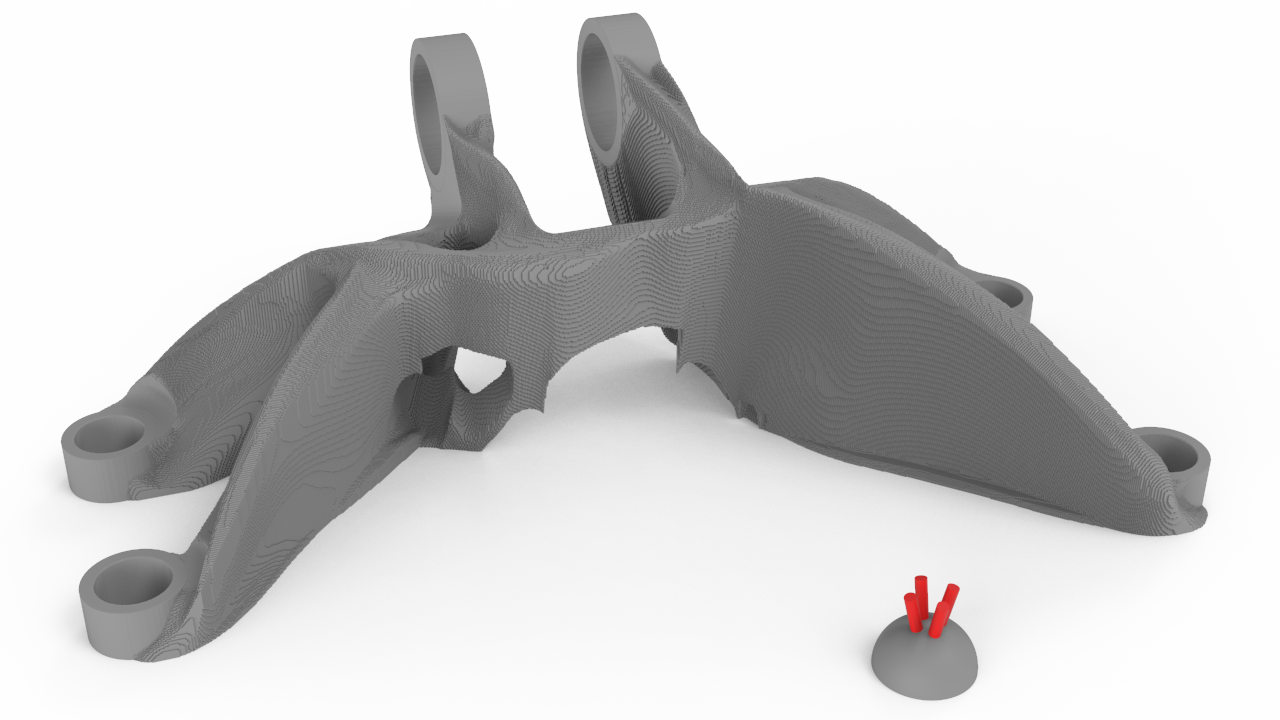}
    \end{subfigure}
    \begin{subfigure}[t]{0.33\textwidth}
        \includegraphics[width=\linewidth]{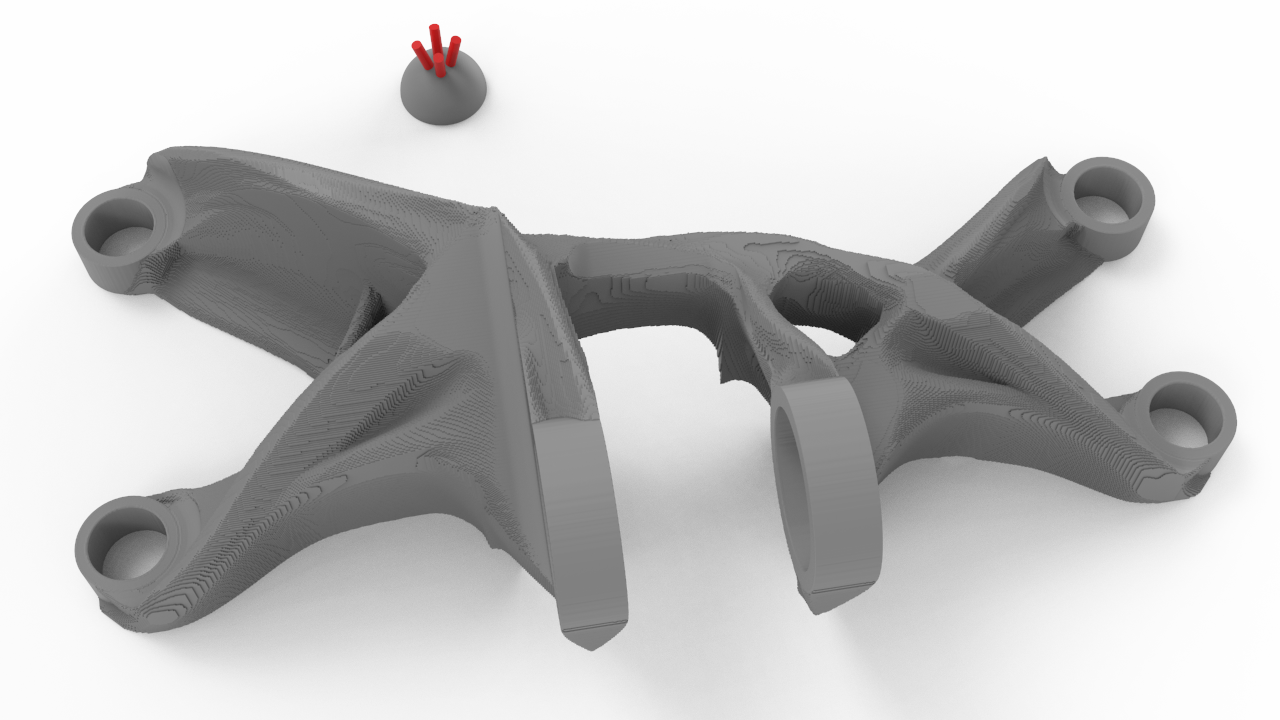}
    \end{subfigure}
    \begin{subfigure}[t]{0.33\textwidth}
        \includegraphics[width=\linewidth]{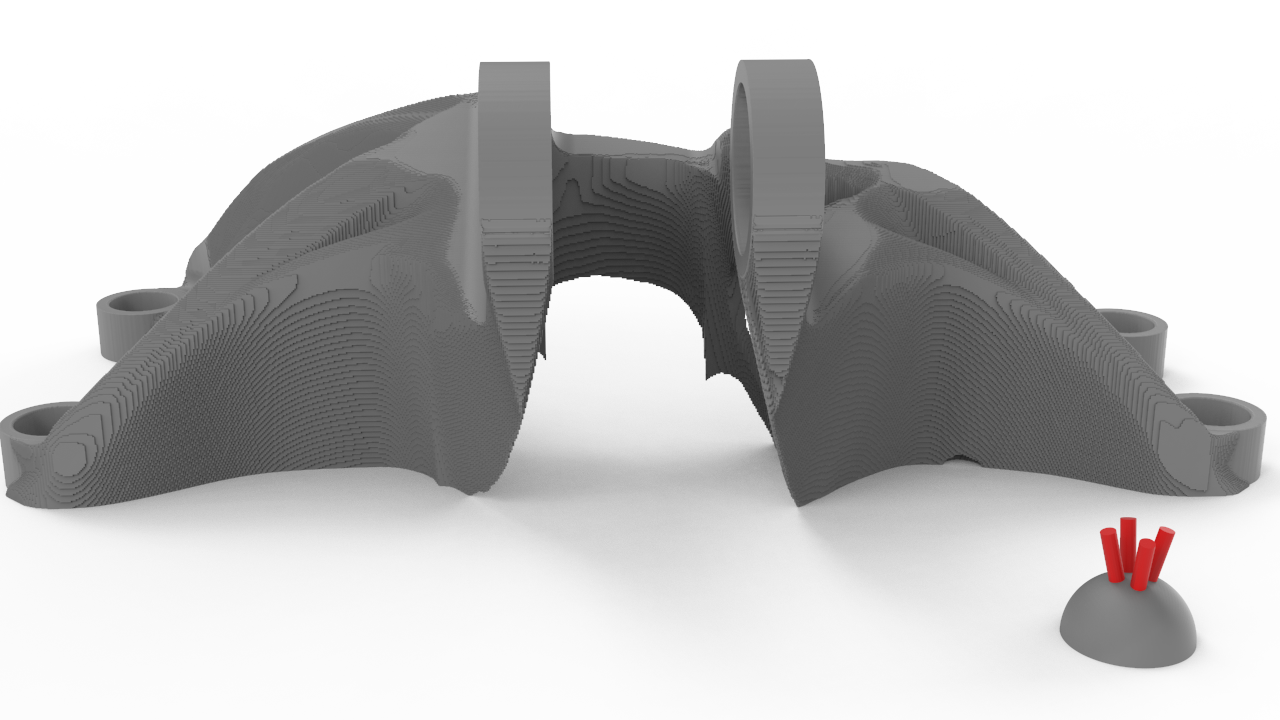}
    \end{subfigure}
    \caption{Design with four milling directions at an 14.4 degree angle. $C=\SI{6.09e7}{J}$.}
    \label{fig:bracketFourHigh}
\end{figure*}

\begin{figure}[htb]
    \centering
    \includegraphics[width=0.5\linewidth]{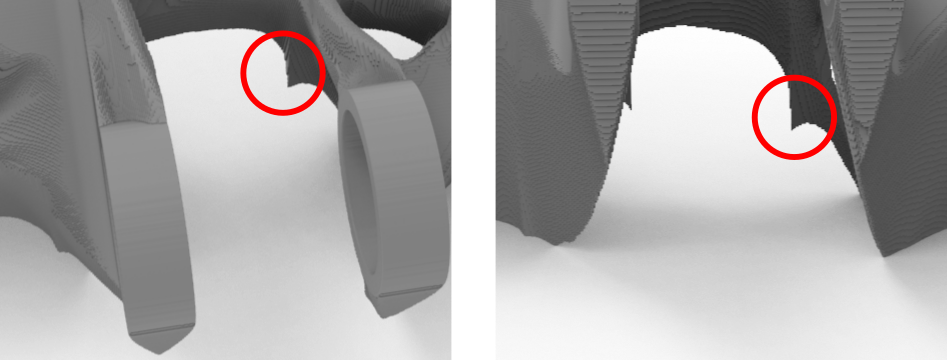}
    \caption{Zoom in on the wedge-like features from the design obtained with the four milling directions at a 14.4 degree angle, seen in \cref{fig:bracketFourHigh}.}
    \label{fig:bracketWedges}
\end{figure}

\begin{figure*}[htb]
    \centering
    \begin{subfigure}[t]{0.33\textwidth}
        \includegraphics[width=\linewidth]{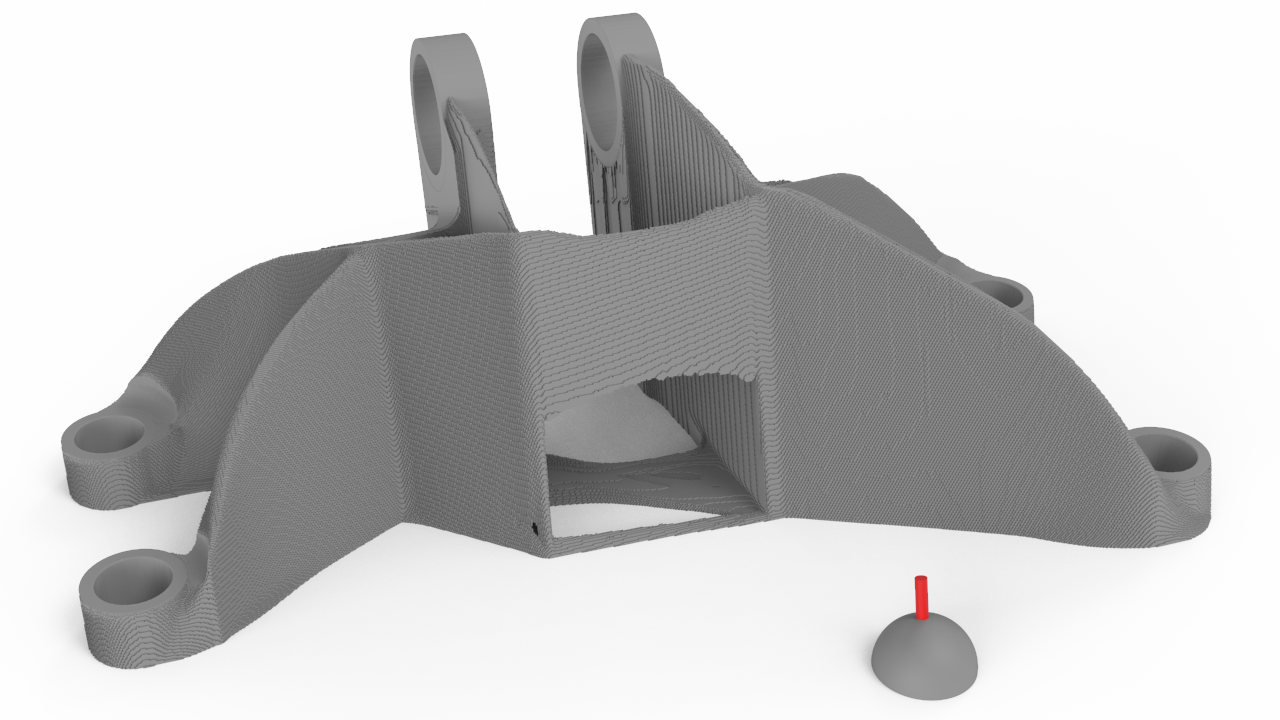}
    \end{subfigure}
    \begin{subfigure}[t]{0.33\textwidth}
        \includegraphics[width=\linewidth]{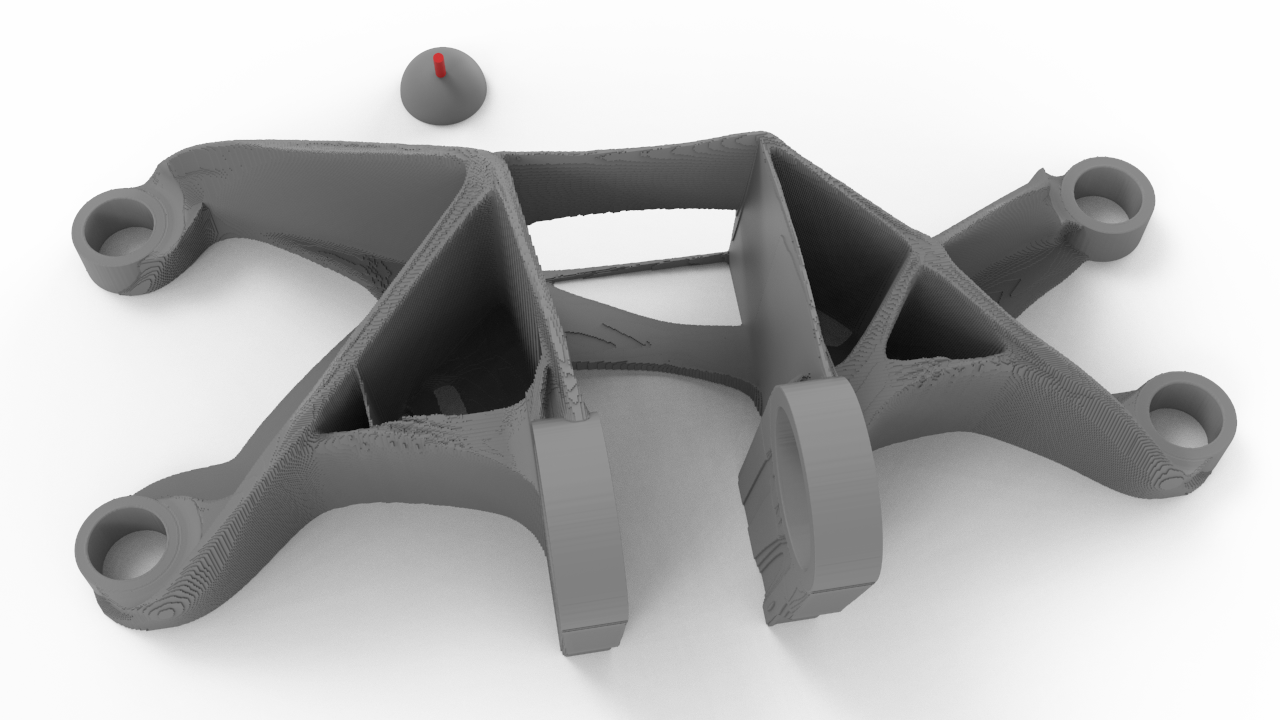}
    \end{subfigure}
    \begin{subfigure}[t]{0.33\textwidth}
        \includegraphics[width=\linewidth]{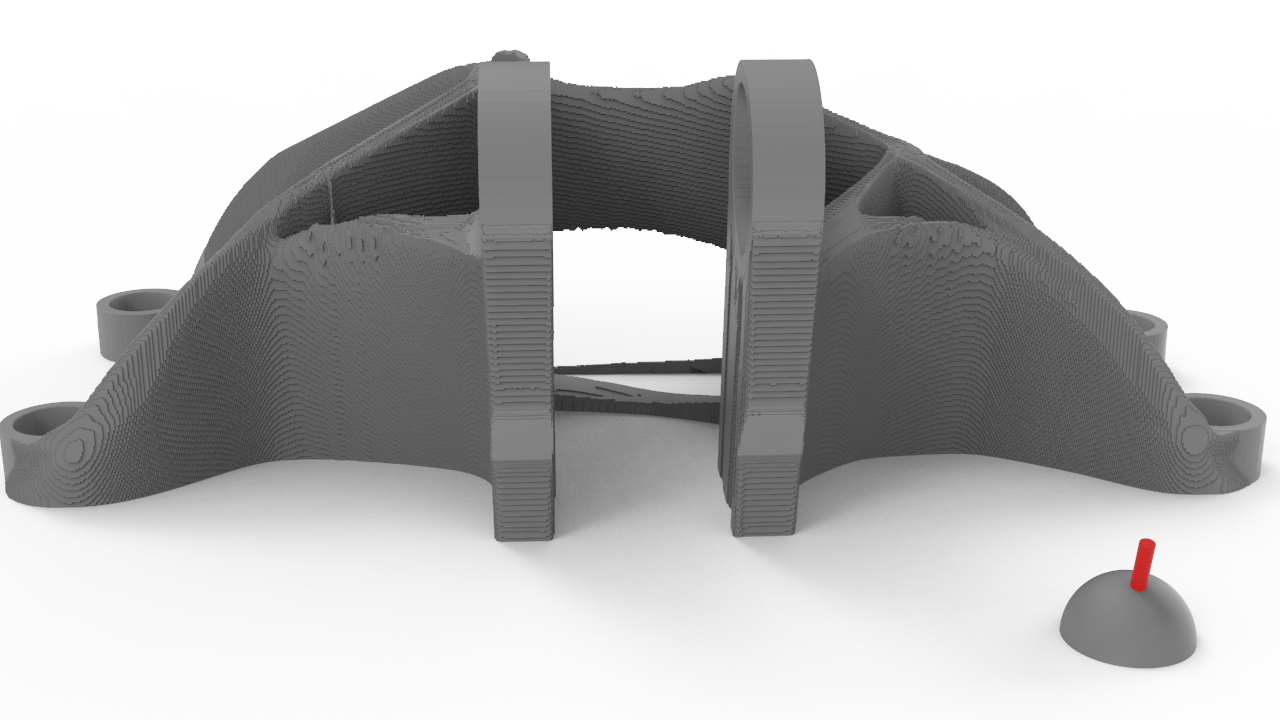}
    \end{subfigure}
    \caption{Design with one milling directions at an 14.4 degree angle. $C=\SI{9.52e7}{J}$.}
    \label{fig:bracketOneAlt}
\end{figure*}

The first example, which is performed using only one milling direction from the top is shown in \cref{fig:bracketOne}. It can be seen that the volume underneath the passive rings, which contain the rigid pin, is set to solid, due to the Dirichlet boundary condition on the shadowing step. This prescribes the use of an already small global volume fraction, leaving little material for the remaining structure. All four supporting bolts are connected by thin legs, which connect in a cross shaped structure. From the resulting compliance values, shown in \cref{tab:BracketCompare}, it is clear that the single direction performs poorly, resulting in approximately twice the compliance value of the reference case. This is somewhat expected, given the restriction of available material and severely restricted design freedom.

\begin{table}[htb]
    \centering
    \caption{Comparison of compliance of the GE jet engine bracket examples.}
    \label{tab:BracketCompare}
    \begin{tabular}{lrrr}
    \hline
        Figure  & Num. tools & $C$ &      $C/C_\mathit{ref}$\\\hline
        \ref{fig:bracketRef} & 0 & \SI{4.21e7}{J} & \num{1} \\ 
        \ref{fig:bracketOne} & 1 &\SI{9.28e7}{J} & \num{2.20} \\ 
        \ref{fig:bracketFive} & 5 &\SI{4.89e7}{J} & \num{1.16} \\ 
        \ref{fig:bracketFiveAlt} & 5 & \SI{4.91e7}{J} & \num{1.17} \\ 
        \ref{fig:bracketFour} & 4 &\SI{5.12e7}{J} & \num{1.22} \\ 
        \ref{fig:bracketFourHigh} & 4 & \SI{6.09e7}{J} & \num{1.45} \\ 
        \ref{fig:bracketOneAlt} & 1 &\SI{9.52e7}{J} & \num{2.26} \\ 
    \end{tabular}
\end{table}

Two additional bracket examples each using five mutually orthogonal milling directions, which are visualized by the red bars along with the resulting structures in \cref{fig:bracketFive,fig:bracketFiveAlt}. The structures both differ from the reference by being more compact almost truss-like, as opposed to the shell-like reference structure. The compliance value of both examples with five directions are very close to the reference value, considering the design restrictions, deviating with only 16\% and 17\%. From this, and similar cantilever examples with many tool directions, it can be seen that structures generated with milling constraints using many tool directions in general perform very well compared to reference designs, when taking into account the severe limitations in design freedom.

Additional numerical examples are also given with more constrained tool directions, to evaluate the milling approach itself, as structures under very limiting manufacturing constraints are generated. \Cref{fig:bracketFour} shows the bracket with four non cartesian tool directions. Again, an agreeable structure with only 22\% increase in compliance compared to the reference is obtained. If closely examined, the structure shows two 'spikes' in the center of the structure. These spikes do not carry any load, but cannot be removed by any of the given tool directions without also removing some vital load-carrying structural member. Therefore they can be compared to the phenomenon seen in \cref{fig:2D4dir}, where non-load carrying material is also present. 

An alternate version of the four directions is shown in \cref{fig:bracketFourHigh}, where all four directions have a 14.4 degree angle to the vertical axis and are aligned with one of the other axes. In this case, the resulting structure is forced to use a significant amount of material under the supporting rings in form of spikes, as can also be seen in the zoom into the corresponding area in \cref{fig:bracketWedges}. These spikes appear as there is no milling direction which allows the removal of this material. Similarly, any structural member which has material removed from under it needs to have a 'wedge'-like shape as the tools need to reach the void under the structure.

Finally, a last example with a single tool direction is shown again, but this time the tool direction is at a 14.4 degree angle with the vertical axis, like in the previous example, but only the direction coming from the side of the bolt flanges is considered. Again it can be seen that a thin-walled structure appears, but this time all of the walls have the expected angle. One curious artifact of this example is that the optimizer is able to circumvent the milling filter, and generate a hole in the bottom part of the central wall. This is understandable in a structural sense as the bottom part of the wall carries little load, but still a very undesirable artifact in the milling filter. This effect occurs due to an interesting interplay between the advection-diffusion filter and the Heaviside projection. As can be seen in \cref{fig:peinf4}, the advection-diffusion-filtered value can slowly decrease along the advection direction. This is partly due to the numerical diffusion due to the used upwind scheme, which is necessary for numerical stability. If the value gets below the Heaviside parameter $\eta$, which is \num{0.5} in this case, the Heaviside projection will project the design towards 0, rather than 1, allowing such a hole to appear. This can be solved by setting a lower $\eta$ value, although the example is included here for completeness.

\section{Conclusion}
\label{sec:conclusion}
The article presents a formulation for performing topology optimization of manufacturable structures through milling by using a PDE-based alternative to a cumulative summation. The proposed method has been shown to generate manufacturability by machining in a number of numerical examples in two and three dimensions, resulting in topologies where all void regions are reachable through a tool direction from outside the domain. The proposed method has also shown itself to scale well to larger topology optimization problems, as shown by the numerical examples consisting of 64 million elements. From the large-scale examples it can also be seen that the increased number of elements allows thin shell members in the structures to be resolved, notably in the bracket. This would not be possible on lower resolutions.

The proposed method for performing the cumulative summation by solving the advection-diffusion equation, can also be considered a significant simplification of the mapping process \cite{Langelaar2019}, since it is not necessary to keep track of multiple meshes, and their relative orientation. 

Some disadvantages of the PDE-based formulation can also be noted from the numerical examples. The advection-diffusion equation is unable to transport material through regions which have not been meshed, such as between the two flanges of the bracket example. When a milling direction across flanges is chosen, no coupling between the flanges occurs. 

The transport problem could in principle be solved by developing special boundary conditions, which couple the field values across the gap in the mesh. This would require a quite non-trivial effort in terms of mathematical and software development. An easier alternative, would be to use an auxiliary mesh of the bounding box of the domain, and solve the advection-diffusion problems on this mesh, somewhat similar to the approach used in \cite{Langelaar2019}. Solving the advection-diffusion equation in the bounding box is still expected to scale better with the number of elements, compared to performing a cumulative summation.

Another disadvantage is the lack of control in the tool-shape, which is possible when explicitly computing the cumulative summations \cite{Langelaar2019}. A further challenge is the diffusivity term of the advection-diffusion equation, which is needed for numerical stability. It can be seen from the presented results that the diffusive terms can be kept very small. 

While it is difficult to guarantee a tool shape based on the solution of the advection-diffusion equation, it could be possible to ensure a minimum member thickness of the features in the resulting design. This could be done by providing eroded, nominal, and dilated variations of the design field during Heaviside projection phase. These fields could then be used to provide a robust formulation based on \cite{Wang2011}. This extension of the milling formulation is left to further work, as it considered somewhat a separate issue from the advection-diffusion based approach.

It is found that the chosen milling directions have a large impact on the compliance of the resulting structures, especially when few milling directions are used. This is to be expected, as the milling filter poses a large restriction on the possible structures.  

\section*{Acknowledgments}
The authors would like to thank Assoc. Prof. Dr. Schousboe Andreasen, Assoc. Prof. Dr. Aage, and Prof. Dr. Sigmund for technical discussions, proof-reading, and supervising our PhDs. 

The authors acknowledge the support of the Villum Foundation for funding the aforementioned PhDs through the Villum Investigator Project InnoTop.

\appendix
\section{Finite Volume implementation with upwind scheme}
\label{app:FVMimplementation}

The advection-diffusion equation from \cref{eq:ad_concrete} is integrated over a control volume. After applying the Gauss divergence theorem, the finite volume form of the equation over a control volume $V$ with boundaries $S$ is obtained \cite{Ferziger2002}:
\begin{equation}
\label{eq:FV}
    \int_S n_i u_i^s \rhoshad dS - \int_S\frac{1}{\Pe}n_i\nabla \rhoshad dS=\int_V \rhoreg dV
\end{equation}
Finite difference schemes are used to describe the fluxes through the faces of the cell. The stencil for the respective face consists of the two cells adjacent to the face. The diffusive flux through the face is described using a Center Difference Scheme (CDS):
\begin{equation}
\label{eq:CDS}
    \left.n_i\nabla \rhoshad\right|_f=\left|\frac{\mathbf{d}}{\mathbf{d}\cdot\mathbf{A}}|\mathbf{A}|^2\right|\frac{\left.\rhoshad\right|_N-\left.\rhoshad\right|_P}{|\mathbf{d}|}
\end{equation}
where $\mathbf{A}$ is the normal area vector and $\mathbf{d}$ the distance between the cell center and the neighbor cell center \cite{Dilgen2018}. The subscript $f$ refers to the face value, $P$ to the cell and $N$ to the neighbor adjacent to the face.

In order to guarantee numerical stability, an upwind difference scheme is used for the advection coefficient in the hence, the interpolated value at the face is the one from the upstream cell:
\begin{equation}
\begin{split}
 \mathrm{if: }\;(n_iu_i^s)_f>0:\;&\left.\rhoshad\right|_f=\left.\rhoshad\right|_N\\
 \mathrm{if: }\;(n_iu_i^s)_f<0:\;&\left.\rhoshad\right|_f=\left.\rhoshad\right|_P
 \end{split}
\end{equation}
where $n_i$ is the outward pointing surface normal.

The Robin boundary conditions are introduced in \cref{eq:robinBC}, using a linear interpolation of the $\rhoshad$ value on the cell boundary and CDS \cref{eq:CDS} on the corresponding gradient, the following boundary condition contribution is found for the boundary face:
\begin{equation}
    \label{eq:discRobin}
    \left.\rhoshad\right|_N = -\left.\rhoshad\right|_P\frac{\frac{1}{2}-n_i\frac{1}{\scale\Pe}\frac{\mathbf{A}_d}{\mathbf{|d|}|\mathbf{A}|}}{\frac{1}{2}+n_i\frac{1}{\scale\Pe}\frac{\mathbf{A}_d}{\mathbf{|d|}|\mathbf{A}|}}
\end{equation}

Dirichlet boundary conditions are implemented adjacent to passive domains. Here, the surface contribution is given as:
\begin{equation}
    \rhoshad=1\quad x\in\partial\Omega
\end{equation}
the right hand side in these cells is corrected with:
\begin{equation}
    -2a_f
\end{equation}
where $a_f$ is the surface contribution of the boundary face.

\section{Sensitivity analysis}
\label{app:sensitivity}
The sensitivity of the full milling filter can be found by applying the chain rule on all operations, as seen in \cref{eq:chainrule}.

The derivative of the compliance w.r.t. the used density field can be found for each element as \cite{bendsoe}
\begin{equation}
     \frac{dc}{d\rhophysel} = -\mathbf{u}^\top \frac{d\mathbf{K}}{d\rhophysel} \mathbf{u} = -\mathbf{u^e}^\top \frac{d\mathbf{K^e}}{d\rhophysel} \mathbf{u^e}
\end{equation}
where the $e$ superscripts denote the element density, deformations, and local stiffness matrix.

The partial derivative of the Heaviside projection can be found analytically for every entry of the density field as
\begin{equation}
     \diff{\rhophys^e}{\rhoaggl^e} =  \frac{\beta \left(1 - \tanh^2(\beta(\rhoaggl^e-\eta)) \right)}{\tanh(\beta\eta) +\tanh(\beta(1-\eta)))}
\end{equation}

The partial derivative of the p-norm agglomeration for a given field $i$ can be found analytically for every element $e$ as
\begin{equation}
    \diff{\rhoaggl^e}{\rhoshad^e} = \left(\frac{1}{n} \sum_{s=1}^{n_s} (\rhoshad^e)^p \right)^{\frac{1}{p}-1} \frac{1}{n}(\rhoshad^e)^{p-1}
\end{equation}

The final partial derivative $\diff{\rhoreg}{\rhodesign}$ depends on the chosen density filtering strategy. In both cases applying the partial derivative corresponds to performing the density filtering on the accumulated sensitivity $\frac{dc}{d\rhoreg}$. 

\bibliographystyle{elsarticle-num-names}
\bibliography{bib.bib}   

\end{document}